\documentclass[reprint,superscriptaddress,amsmath,amssymb,aps,floatfix,longbibliography]{revtex4-1}

\usepackage[pdftex]{graphicx}
\usepackage{bm}
\usepackage{color}
\usepackage{times}
\usepackage{amsmath}
\usepackage{amssymb}

\begin{document}

\title{Large deformations of a soft porous material}

\author{Christopher W. MacMinn}
\email{christopher.macminn@eng.ox.ac.uk}
\affiliation{Department of Engineering Science, University of Oxford, Oxford, OX1 3PJ, UK}
\author{Eric R. Dufresne}
\affiliation{Department of Materials, ETH Zurich, CH-8093 Zurich, Switzerland}
\author{John S. Wettlaufer}
\affiliation{Yale University, New Haven, CT, 06520, USA}
\affiliation{The Mathematical Institute, University of Oxford, Oxford, OX1 3LB}
\affiliation{Nordita, Royal Institute of Technology and Stockholm University, SE-10691 Stockholm, Sweden}

\date{\today}


\begin{abstract}
Compressing a porous material will decrease the volume of the pore space, driving fluid out. Similarly, injecting fluid into a porous material can expand the pore space, distorting the solid skeleton. This poromechanical coupling has applications ranging from cell and tissue mechanics to geomechanics and hydrogeology. The classical theory of linear poroelasticity captures this coupling by combining Darcy's law with Terzaghi's effective stress and linear elasticity in a linearized kinematic framework. Linear poroelasticity is a good model for very small deformations, but it becomes increasingly inappropriate for moderate to large deformations, which are common in the context of phenomena such as swelling and damage, and for soft materials such as gels and tissues. The well-known theory of large-deformation poroelasticity combines Darcy's law with Terzaghi's effective stress and nonlinear elasticity in a rigorous kinematic framework. This theory has been used extensively in biomechanics to model large elastic deformations in soft tissues, and in geomechanics to model large elastoplastic deformations in soils. Here, we first provide an overview and discussion of this theory with an emphasis on the physics of poromechanical coupling. We present the large-deformation theory in an Eulerian framework to minimize the mathematical complexity, and we show how this nonlinear theory simplifies to linear poroelasticity under the assumption of small strain. We then compare the predictions of linear poroelasticity with those of large-deformation poroelasticity in the context of two uniaxial model problems: Fluid outflow driven by an applied mechanical load (the consolidation problem) and compression driven by a steady fluid throughflow. We explore the steady and dynamical errors associated with the linear model in both situations, as well as the impact of introducing a deformation-dependent permeability. We show that the error in linear poroelasticity is due primarily to kinematic nonlinearity, and that this error (i) plays a surprisingly important role in the dynamics of the deformation and (ii) is amplified by nonlinear constitutive behavior, such as deformation-dependent permeability. 
\end{abstract}

\maketitle

\section{Introduction}

In a deformable porous material, deformation of the solid skeleton is mechanically coupled to flow of the interstitial fluid(s). This poromechanical coupling has relevance to problems as diverse as cell and tissue mechanics~\citep[\textit{e.g.},][]{mow-jbiomecheng-1980, lai-jbiomecheng-1991, yang-jbiomech-1991, cowin-jbiomech-1999, franceschini-jmps-2006, skotheim-science-2005, charras-nature-2005, dumais-annrevfluidmech-2012, moeendarbary-natmaterials-2013}, magma/mantle dynamics~\citep[\textit{e.g.},][]{mckenzie-jpetrology-1984, fowler-geoastrofluiddynamics-1985, spiegelman-jfm-1993a, bercovici-jgr-2001a, katz-jpetrology-2008, hesse-geophysjint-2011}, and hydrogeology~\citep[\textit{e.g.},][]{bear-wrr-1981a, wang-princeton-2000, szulczewski-pnas-2012, rutqvist-geotechgeologeng-2012, chang-wrr-2013, verdon-pnas-2013, yarushina-geophysjint-2013, jha-wrr-2014, cai-geophysjint-2014, hewitt-jfm-2015}. These problems are notoriously difficult due to the inherently two-way nature of poromechanical coupling, where deformation drives flow and vice-versa.

In poroelasticity, the mechanics of the solid skeleton are described by elasticity theory. The theory of poroelasticity has a rich and interdisciplinary history~\citep{deboer-applmechrev-1996}. \citet{wang-princeton-2000} provides an excellent discussion of the historical roots of linear poroelasticy, which models poroelastic loading under infinitesimal deformations. Major touchstones in the development of linear poroelasticity include the works of M.~A.~Biot~\citep[\textit{e.g.},][]{biot-japplphys-1941, biot-jacoustsocam-1956a, biot-japplphys-1962}, who formalized the linear theory and provided a variety of analytical solutions through analogy with thermoelasticity, as well as that of \citet{rice-revgeophysspacephys-1976}, who reformulated the linear theory in terms of more tangible material parameters and derived solutions to several model problems.

When the deformation of the skeleton is not infinitesimal, poroelasticity must be cast in the framework of large-deformation elasticity~\citep[\textit{e.g.},][]{biot-indianaumathj-1972, coussy-wiley-2004}. Large-deformation poroelasticity has found extensive application over the past few decades in computational biomechanics for the study of soft tissues, which are porous, fluid-saturated, and can accommodate large deformations reversibly. Much of this effort has been directed at capturing the complex and varied structure and constitutive behavior of biological materials~\citep[\textit{e.g.},][]{holmes-jbiomech-1990, kwan-jbiomechanics-1990, simon-applmechrev-1992, argoubi-jbiomech-1996, federico-mechmater-2012, tomic-imajapplmath-2014, vuong-compmethapplmecheng-2015}. Large-deformation poroelasticity has also been applied in computational geomechanics for the study of soils and other geomaterials. Soils typically accommodate large deformations through plasticity (ductile failure or yielding) due to their granular and weakly cemented microstructure, and much of the effort has been directed at the challenges of large-deformation elastoplasticity~\citep[\textit{e.g.},][]{carter-intjnag-1979a, borja-compmethapplmecheng-1995, li-compmethapplmecheng-2004, uzuoka-intjnag-2011}. Large deformations can also occur through swelling~\citep[\textit{e.g.},][]{bennethum-intjengsci-1996, huyghe-intjengsci-1997}, which has attracted interest recently in the context of gels~\citep[\textit{e.g.},][]{hong-jmps-2008, duda-jmps-2010, chester-jmps-2010}.

Large-deformation (poro)elasticity is traditionally approached almost exclusively with computational tools, and these are based almost exclusively on the finite-element method and in a Lagrangian framework~\citep[\textit{e.g.,}][]{hong-jmps-2008, chester-jmps-2010}. A thorough treatment of the Lagrangian approach to large-deformation poroelasticity can be found in Ref.~\citep{coussy-wiley-2004}. Although powerful, these tools can be cumbersome from the perspective of developing physical insight. They are also poorly suited for studying nontrivial flow and solute transport through the pore structure. Here, we instead consider the general theory of large-deformation poroelasticity in an Eulerian framework. Although the Eulerian approach is well known \citep[\textit{e.g.},][]{coussy-wiley-2004}, it has very rarely been applied to practical problems and a clear and unified presentation is lacking. This approach is useful in the present context to emphasize the physics of poromechanical coupling.

In the first part of this paper, we review and discuss the well-known theory~(\S\ref{s:poroelasticity}--\ref{s:LinearPoroelasticity}). We first consider the exact description of the kinematics of flow and deformation, adopting a simple but nonlinear elastic response in the solid skeleton~(\S\ref{s:poroelasticity}). We then show how this theory reduces to linear poroelasticity in the limit of infinitesimal deformations~(\S\ref{s:LinearPoroelasticity}). 

In the second part of this paper, we compare the linear and large-deformation theories in the context of two uniaxial model problems~(\S\ref{s:rect}--\ref{s:flow_rect}). The primary goals of this comparison are to study the role of kinematic nonlinearity in large deformations, and to examine the resulting error in the linear theory. The two model problems are: (i)~Compression driven by an applied load (the consolidation problem) and (ii)~compression driven by a net fluid throughflow.  In the former, the evolution of the deformation is controlled by the rate at which fluid is squeezed through the material and out at the boundaries; as a result, fluid flow is central to the rate of deformation but plays no role in the steady state. In the latter, fluid flow is also central to the steady state since this is set by the steady balance between the gradient in fluid pressure and the gradient in stress within the solid skeleton. We show that, in both cases, the error in linear poroelasticity due to the missing kinematic nonlinearities plays a surprisingly important role in the dynamics of the deformation, and that this error is amplified by nonlinear constitutive behavior such as deformation-dependent permeability.

\section{Large-deformation poroelasticity}\label{s:poroelasticity}

Poroelasticity is a multiphase theory in that it describes the coexistence and interaction of multiple immiscible constituent materials, or phases~\citep[\textit{e.g.},][]{hassanizadeh-awr-1979}. Here, we restrict ourselves to two phases: A solid and a fluid. The solid phase is arranged into a porous and deformable macroscopic structure, ``the solid skeleton'' or ``the skeleton'', and the pore space of the solid skeleton is saturated with a single interstitial fluid. Deformation of the solid skeleton leads to rearrangement of its pore structure, with corresponding changes in the local volume fractions (see \S\ref{ss:VolumeFractions}). Throughout, we use the terms ``the solid grains'' or ``the solid'' to refer to the solid phase and ``the interstitial fluid'' or ``the fluid'' to refer to the fluid phase. Although the term ``grain'' is inappropriate in the context of porous materials with fibrous microstructure, such as textiles, polymeric gels, or tissues, we use it generically for convenience. 

We assume here that the two constituent materials are incompressible, meaning that the mass densities of the fluid, $\rho_f$, and of the solid, $\rho_s$, are constant. Note that this does not prohibit compression of the solid skeleton---variations in the macroscopic mass density of the porous material are enabled by changes in its pore volume. The assumption of incompressible constituents is standard in soil mechanics and biomechanics, where fluid pressures and solid stresses are typically very small compared to the bulk moduli of the constituent materials. However, some caution is merited in the context of deep aquifers where, at a depth of a few kilometers, the hydrostatic pressure and lithostatic stress themselves approach a few percent of the bulk moduli of water and mineral grains. In these cases, it may be appropriate to allow for fluid compressibility while retaining the incompressibility of the solid grains, in which case much of the theory discussed here still applies. The theory of poroelasticity can be generalized to allow for compressible constituents~\citep[\textit{e.g.},][]{wang-princeton-2000, coussy-wiley-2004, bennethum-jengmech-2006, jaeger-wiley-2007, gajo-rspa-2010}, but this is beyond the scope of this paper.

What follows is, in essence, a brief and superficial introduction to continuum mechanics in the context of a porous material. We have minimized the mathematical complexity where possible for the sake of clarity, and to preserve an emphasis on the fundamental physics. For the more mathematically inclined reader, excellent resources are available for further study~\citep[\textit{e.g.},][]{marsden-dover-1983, coussy-wiley-2004, gurtin-cambridge-2010}.

\subsection{Eulerian and Lagrangian reference frames}\label{ss:Euler_vs_Lagrange}

A core concept in continuum mechanics is the distinction between a Lagrangian reference frame (fixed to the material) and an Eulerian one (fixed in space). These two perspectives are rigorously equivalent, so the choice is purely a matter of convention and convenience. A Lagrangian frame is the natural and traditional choice in solid mechanics, where displacements are typically small and where the current state of stress is always tied to the reference (undeformed) configuration of the material through the current state of strain (displacement gradients). An Eulerian frame is the natural and traditional choice in fluid mechanics, where displacements are typically large and where the current state of stress depends only on the instantaneous rate of strain (velocity gradients). More complex materials such as viscoelastic solids and non-Newtonian fluids can have elements of both, with a dependence on both strain and strain rate~\citep[\textit{e.g.},][]{bird-wiley-1987a, bird-wiley-1987b}.

Problems involving fluid-solid coupling lead to a clear conflict between the Eulerian and Lagrangian approaches. In classical fluid-structure interaction problems, such as air flow around a flapping flag or blood flow through an artery, the fluid and the solid exist in separate domains that are coupled along a shared moving boundary. In a porous material, in contrast, the fluid and the solid coexist in the same domain and are coupled through bulk conservation laws. As a result, the entire problem must be posed either in a fixed Eulerian frame or in a Lagrangian frame attached to the solid skeleton. One major advantage of the latter approach is that it eliminates moving (solid) boundaries since the skeleton is stationary relative to a Lagrangian coordinate system; this feature is particularly powerful and convenient in the context of computation. However, the Eulerian approach leads to a simpler and more intuitive mathematical model in the context of fluid flow and transport, and it is straightforward and even advantageous when boundary motion is absent or geometrically simple. This conflict can be avoided altogether when the deformation of the skeleton is small, such that the distinction between an Eulerian frame and a Lagrangian one can be ignored, which is a core assumption of linear (poro)elasticity.

In the present context, we consider two model problems where the fluid and the skeleton are tightly coupled, the deformation of the skeleton is large, and there is a moving boundary. We pose these problems fully in an Eulerian frame and write all quantities as functions of an Eulerian coordinate $\mathbf{x}$, fixed in the laboratory frame. Accordingly, we adopt the notation
\begin{equation}
    \boldsymbol{\nabla}\equiv\hat{\boldsymbol{e}}_i \,\frac{\partial}{\partial{x_i}},
\end{equation}
where the $x_i$ are the components of the Eulerian coordinate system, $i=1,2,3$, with $\hat{\boldsymbol{e}}_i$ the associated unit vectors, and adopting the Einstein summation convention. For reference and comparison, we summarize the key aspects of the Lagrangian framework in Appendix~\ref{app:s:Lagrangian}.

\subsection{Volume fractions}\label{ss:VolumeFractions}

We denote the local volume fractions of fluid and solid by $\phi_f$ (the porosity, fluid fraction, or void fraction) and $\phi_s$ (the solid fraction), respectively. These are the \textit{true} volume fractions in the sense that they measure the current phase volume per unit current total volume, such that $\phi_f+\phi_s\equiv{}1$. As such, the true porosity is the relevant quantity for calculating flow and transport through the pore structure. However, changes in $\phi_f$ at a spatial point $\mathbf{x}$ reflect both deformation and motion of the underlying skeleton, so the relevant state of stress must be calculated with some care.

Alternatively, it is possible to define \textit{nominal} volume fractions that measure the current phase volume per unit reference total volume~\citep{coussy-wiley-2004}. These nominal quantities are convenient in a Lagrangian frame where, if the solid phase is incompressible, the nominal solid fraction is constant by definition and the nominal porosity is linearly related to the local volumetric strain. However, the nominal porosity is not directly relevant to flow and transport. In addition, the nominal volume fractions do not sum to unity; rather, they must sum to the Jacobian determinant $J$ (see \S\ref{ss:Kinematics_solid}). Here, we avoid these nominal quantities and work strictly with the true porosity. Note that \citet{coussy-wiley-2004} denotes the true porosity (``Eulerian porosity'') by $n$ and the nominal porosity (``Lagrangian porosity'') by $\phi$, whereas we denote the true porosity by $\phi_f$ and the nominal porosity by $\Phi_f$ (see Appendix~\ref{app:s:Lagrangian}).

\subsection{Kinematics of solid deformation}\label{ss:Kinematics_solid}

The most primitive quantity for calculating deformation is the displacement field, which is a map between the current configuration of the solid skeleton and its reference configuration. In other words, the displacement field measures the displacement of material points from their reference positions. In an Eulerian frame, the solid displacement field $\mathbf{u}_s(\mathbf{x},t)$ is given by
\begin{equation}\label{eq:udef}
    \mathbf{u}_s(\mathbf{x},t)=\mathbf{x}-\mathbf{X}(\mathbf{x},t),
\end{equation}
where $\mathbf{X}$ is the reference position of the material point that sits at position~$\mathbf{x}$ at time~$t$ (\textit{i.e.}, it is the Lagrangian coordinate in our Eulerian frame). We adopt the convention that $\mathbf{X}(\mathbf{x},0)=\mathbf{x}$ such that $\mathbf{u}_s(\mathbf{x},0)=0$; this is not required, but it simplifies the analysis.

The displacement field is not directly a measure of deformation because it includes rigid-body motions. The deformation-gradient tensor $\mathbf{F}$, which is the Jacobian of the deformation field, excludes translations by considering the spatial gradient of the displacement field. In an Eulerian frame, $\mathbf{F}$ is most readily defined through its inverse,
\begin{equation}\label{eq:Fdef}
    \mathbf{F}^{-1} =\boldsymbol{\nabla}\mathbf{X} =\mathbf{I}-\boldsymbol{\nabla}\mathbf{u}_s,
\end{equation}
where $\displaystyle(\cdot)^{-1}$ denotes the inverse and $\mathbf{I}$ is the identity tensor. The deformation-gradient tensor $\mathbf{F}$ still includes rigid-body rotations, but these can be excluded by multiplying $\mathbf{F}$ by its transpose. Hence, measures of strain are ultimately derived from the right~Cauchy-Green deformation tensor $\displaystyle\mathbf{C}=\mathbf{F}^\mathsf{T}\mathbf{F}$ or the left~Cauchy-Green deformation tensor $\displaystyle\mathbf{B}=\mathbf{F}\mathbf{F}^\mathsf{T}$, where $\displaystyle(\cdot)^\mathsf{T}$ denotes the transpose.

The eigenvalues $\lambda_i^2$ of $\mathbf{C}$ (or, equivalently, of $\mathbf{B}$) are the squares of the principal stretches $\lambda_i$, with $i=1,2,3$. The stretches measure the amount of elongation along the principal axes of the deformation, which are themselves related to the eigenvectors of $\mathbf{C}$ and $\mathbf{B}$: In the reference configuration, they are the normalized eigenvectors of $\mathbf{C}$; in the current configuration, they are the normalized eigenvectors of $\mathbf{B}$.

The Jacobian determinant $J$ measures the amount of local volume change during the deformation,
\begin{equation}\label{eq:F_to_J}
    J(\mathbf{x},t)=\det{\left(\mathbf{F}\right)} =\frac{1}{\det{\left(\mathbf{F}^{-1}\right)}}=\lambda_1\lambda_2\lambda_3,
\end{equation}
where $\det{(\cdot)}$ denotes the determinant. The Jacobian determinant is precisely the ratio of the current volume of the material at point $\mathbf{x}$ to its reference volume. For an incompressible solid skeleton, $J\equiv{}1$. For a compressible solid skeleton made up of incompressible solid grains, as considered here, deformation occurs strictly through rearrangement of the pore structure. The Jacobian determinant is then connected directly to the porosity,
\begin{equation}\label{eq:J_to_phi_true}
    J(\mathbf{x},t)=\frac{1-\phi_{f,0}(\mathbf{x},t)}{1-\phi_f(\mathbf{x},t)},
\end{equation}
where $\phi_{f,0}(\mathbf{x},t)\equiv\phi_f\big(\mathbf{x}-\mathbf{u}_s(\mathbf{x},t),0\big)$ is the reference porosity field. In an Eulerian frame, the reference porosity field depends on $\mathbf{u}_s$ because it refers to the initial porosity of the material that is currently located at $\mathbf{x}$ but that was originally located at $\mathbf{x}-\mathbf{u}_s$. Note that $\phi_{f,0}(\mathbf{x},t)\neq{}\phi_f(\mathbf{x},0)$ unless $\phi_f(\mathbf{x},0)$ is spatially uniform, in which case $\phi_f(\mathbf{x},0)=\phi_{f,0}$ is simply a constant and this distinction is unimportant.

Lastly, local continuity for the incompressible solid phase is written
\begin{equation}\label{eq:continuity_solid}
    \frac{\partial{\phi_s}}{\partial{t}} +\boldsymbol{\nabla}\cdot\big(\phi_s\mathbf{v}_s\big)=0 \quad\mathrm{or}\quad \frac{\partial{\phi_f}}{\partial{t}} -\boldsymbol{\nabla}\cdot\big[(1-\phi_f)\mathbf{v}_s\big]=0,
\end{equation}
where $\mathbf{v}_s(\mathbf{x},t)$ is the solid velocity field. The solid velocity is the material derivative of the solid displacement,
\begin{equation}\label{eq:us_to_vs}
    \mathbf{v}_s =\frac{\mathrm{D}\mathbf{u}_s}{\mathrm{D}t} \equiv\frac{\partial{\mathbf{u}_s}}{\partial{t}}+\mathbf{v}_s\cdot{}\boldsymbol{\nabla}\mathbf{u}_s =\frac{\partial{\mathbf{u}_s}}{\partial{t}}\cdot\mathbf{F}.
\end{equation}
Equations~\eqref{eq:udef}--\eqref{eq:us_to_vs} provide an exact kinematic description of the deformation of the solid skeleton, assuming only that the solid phase is incompressible. This description is valid for arbitrarily large deformations and, because it is simply a geometric description of the changing pore space, it \textit{makes no assumptions about the fluid that occupies the pore space}. This description remains rigorously valid when the fluid phase is compressible, and in the presence of multiple fluid phases. Further, this description \textit{makes no additional assumptions about the constitutive behavior of the solid skeleton}---it remains rigorously valid for any elasticity law, and in the presence of viscous dissipation or plasticity.

\subsection{Kinematics of fluid flow}\label{ss:Kinematics_fluid}

We assume that the pore space of the solid skeleton is saturated with a single fluid phase. For a compressible fluid phase, local continuity is written
\begin{equation}\label{eq:continuity_fluid_compressible}
    \frac{\partial}{\partial{t}}\big(\rho_f\phi_f\big) +\boldsymbol{\nabla}\cdot\big(\rho_f\phi_f\mathbf{v}_f\big)=0,
\end{equation}
where $\mathbf{v}_f(\mathbf{x},t)$ is the fluid velocity field. This expression remains valid for multiple fluid phases if $\rho_f$ and $\mathbf{v}_f$ are calculated as fluid-phase-averaged quantities, in which case Eq.~\eqref{eq:continuity_fluid_compressible} must also be supplemented by a conservation law for each of the individual fluid phases. For simplicity, we focus here on the case of a single, incompressible fluid phase, for which we have
\begin{equation}\label{eq:continuity_fluid}
    \frac{\partial{\phi_f}}{\partial{t}} +\boldsymbol{\nabla}\cdot\big(\phi_f\mathbf{v}_f\big)=0.
\end{equation}
There is no need to introduce a fluid displacement field since we assume below that the fluid is Newtonian. The constitutive law for a Newtonian fluid, and also for many non-Newtonian fluids, depends only on the fluid velocity.

\subsection{Constitutive laws for fluid flow}\label{ss:Darcy}

We assume that the fluid flows relative to the solid skeleton according to Darcy's law. For a single Newtonian fluid, Darcy's law can be written
\begin{equation}\label{eq:darcy}
    \phi_f(\mathbf{v}_f-\mathbf{v}_s) =-\frac{k(\phi_f)}{\mu}\left(\boldsymbol{\nabla}p-\rho_f\mathbf{g}\right),
\end{equation}
where $k(\phi_f)$ is the permeability of the solid skeleton, which we have taken to be an isotropic function of porosity only, $\mu$ is the dynamic viscosity of the fluid, $\mathbf{g}$ is the body force per unit mass due to gravity, and we have neglected body forces other than gravity.

Darcy's law is an implicit statement about the continuum-scale form of the mechanical interactions between the fluid and solid~\citep[\textit{e.g.}, \S{}3.3.1 of][]{coussy-wiley-2004}. We simply adopt it here as a phenomenological model for flow of a single fluid through a porous material. In the presence of multiple fluid phases, Eq.~\eqref{eq:darcy} can be replaced by the classical multiphase extension of Darcy's law~\citep[\textit{e.g.},][]{pinder-wiley-2008}. For a single but non-Newtonian fluid phase, Eq.~\eqref{eq:darcy} must be modified accordingly~\citep[\textit{e.g.},][]{wu-advporousmed-1996}.

Generally, the permeability will change with the pore structure as the skeleton deforms, although this dependence is neglected in linear poroelasticity, where it is assumed that deformations are infinitesimal. The simplest representation of this dependence is to take the permeability to be a function of the porosity, as we have done above, and here again the true porosity is the relevant quantity. A common choice is the Kozeny-Carman formula, one form of which is
\begin{equation}\label{eq:kozeny-carman}
    k(\phi_f)=\frac{d^2}{180}\,\frac{\phi_f^3}{(1-\phi_f)^2},
\end{equation}
where $d$ is the typical pore or grain size. Although derived from experimental measurements in beds of close-packed spheres, this formula is commonly used for a wide range of materials. One reason for this wide use is that the Kozeny-Carman formula respects two physical limits that are important for poromechanics: The permeability vanishes as the porosity vanishes, and diverges as the porosity approaches unity. The former requirement ensures that fluid flow cannot drive the porosity below zero, and the latter prevents the flow from driving the porosity above unity.

We use a normalized Kozeny-Carman formula here,
\begin{equation}\label{eq:kozeny-carman-k0}
    k(\phi_f) =k_0\,\frac{(1-\phi_{f,0})^2}{\phi_{f,0}^3}\,\frac{\phi_f^3}{(1-\phi_f)^2},
\end{equation}
where $k(\phi_{f,0})=k_0$ is the relaxed or undeformed permeability. This expression preserves the qualitative characteristics of the original relationship while allowing the initial permeability and the initial porosity to be imposed independently. Clearly, it is straightforward to design other permeability laws that have the same characteristics. Note that the particular choice of permeability law will dominate the flow and mechanics in the limit of vanishing permeability since the pressure gradient, which is coupled with the solid mechanics, is inversely proportional to the permeability.

Since porosity is strictly volumetric, writing $k=k(\phi_f)$ neglects the impacts of rotation and shear. This form of deformation dependence is overly simplistic for materials with inherently anisotropic permeability fields, the axes of which would rotate under rigid-body rotation and would be distorted in shear. It is also possible that permeability anisotropy could emerge through anisotropic deformations, or through other effects creating orthotropic structure. We neglect these effects here for simplicity.

\subsection{Nonlinear flow equation}\label{ss:ade}

One convenient way of combining Eqs.~\eqref{eq:continuity_solid}, \eqref{eq:continuity_fluid}, and \eqref{eq:darcy} is by defining the total volume flux $\mathbf{q}$ as
\begin{equation}
    \mathbf{q}\equiv{}\phi_f\mathbf{v}_f+(1-\phi_f)\mathbf{v}_s.
\end{equation}
This measures the total volume flow per unit total cross-sectional area per unit time. The total flux can also be viewed as a phase-averaged, composite, or bulk velocity. From this, it is then straightforward to derive
\begin{subequations}\label{eq:NonlinearADEEulerian}
    \begin{align}
        \frac{\partial{\phi_f}}{\partial{t}} +\boldsymbol{\nabla}\cdot\bigg[\phi_f\mathbf{q}-(1-\phi_f)\frac{k(\phi_f)}{\mu}\big(&\boldsymbol{\nabla}p-\rho_f\mathbf{g}\big)\bigg] =0~,\label{eq:NonlinearADEEulerian_pde} \\
       \textrm{and}\qquad &\boldsymbol{\nabla}\cdot\mathbf{q} =0~,\label{eq:NonlinearADEEulerian_divq}
    \end{align}
\end{subequations}
with
\begin{subequations}\label{eq:vf_vs_nonlinear}
    \begin{align}
        \mathbf{v}_f &=\mathbf{q}-\left(\frac{1-\phi_f}{\phi_f}\right)\frac{k(\phi_f)}{\mu}\left(\boldsymbol{\nabla}p-\rho_f\mathbf{g}\right)~, \label{eq:vf_nonlinear} \\
      \quad\textrm{and}\quad \mathbf{v}_s &=\mathbf{q}+\frac{k(\phi_f)}{\mu}\left(\boldsymbol{\nabla}p-\rho_f\mathbf{g}\right)~. \label{eq:vs_nonlinear}
    \end{align}
\end{subequations}
Equations~\eqref{eq:NonlinearADEEulerian} and \eqref{eq:vf_vs_nonlinear} embody Darcy's law and the kinematics of the deformation, describing the coupled relative motion of the fluid and the solid skeleton. It remains to enforce mechanical equilibrium, and to provide a constitutive relation between stress and deformation within the solid skeleton.

\subsection{Mechanical equilibrium}\label{ss:Equilibrium}

Mechanical equilibrium requires that the fluid and the solid skeleton must jointly support the local mechanical load, and this provides the fundamental poromechanical coupling. The total stress $\boldsymbol{\sigma}$ is the total force supported by the two-phase system per unit area, and can be written
\begin{equation}\label{eq:sig_s,sig_f}
    \boldsymbol{\sigma}=(1-\phi_f)\boldsymbol{\sigma}_s +\phi_f\boldsymbol{\sigma}_f,
\end{equation}
where $\boldsymbol{\sigma}_s$ and $\boldsymbol{\sigma}_f$ are the solid stress and the fluid stress, respectively. The solid stress is the force supported by the solid per unit solid area, and $(1-\phi_f)\boldsymbol{\sigma}_s$ is then the force supported by the solid per unit total area. Similarly, the fluid stress is the force supported by the fluid per unit fluid area, and $\phi_f\boldsymbol{\sigma}_f$ is then the force supported by the fluid per unit total area. Note that it is implicitly assumed here and elsewhere that the phase area fractions are equivalent to the phase volume fractions.

Any stress tensor can be decomposed into isotropic (volumetric) and deviatoric (shear) components without loss of generality. For a fluid within a porous solid, it can be shown that the shear component of the stress is negligible relative to the volumetric component at the continuum scale~\citep[\textit{e.g.}, \S{}3.3.1 of][]{coussy-wiley-2004}, so that $\boldsymbol{\sigma}_f=-p\mathbf{I}$, where $p\equiv-(1/3)\mathrm{tr}(\boldsymbol{\sigma}_f)$ is the fluid pressure with $\mathrm{tr}(\cdot)$ the trace. Note that we have adopted the sign convention from solid mechanics that tension is positive and compression negative. The opposite convention is usually used in soil mechanics, rock mechanics, and geomechanics since geomaterials are almost always in compression.

Because the fluid permeates the solid skeleton, the solid stress must include an isotropic and compressive component in response to the fluid pressure. This component is present even when the fluid is at rest, and/or when the skeleton carries no external load, but this component cannot contribute to deformation unless the solid grains are compressible. Subtracting this component from the solid stress leads to Terzaghi's effective stress $\boldsymbol{\sigma}^\prime$~\citep{terzaghi-procsmfe-1936},
\begin{equation}\label{eq:sig_prime_def}
    \boldsymbol{\sigma}^\prime\equiv{}(1-\phi_f)(\boldsymbol{\sigma}_s+p\mathbf{I}),
\end{equation}
which is the force per unit total area supported by the solid skeleton through deformation~\citep[\textit{e.g.},][]{terzaghi-procsmfe-1936, biot-japplphys-1941, wang-princeton-2000, coussy-wiley-2004}. We can then rewrite Eq.~\eqref{eq:sig_s,sig_f} in its more familiar form,
\begin{equation}\label{eq:totalstress}
    \boldsymbol{\sigma}=\boldsymbol{\sigma}^\prime-p\mathbf{I},
\end{equation}
which can be modified to allow for compressibility of the solid grains~\citep[\textit{e.g.}, ][]{nur-jgr-1971, gajo-rspa-2010}.

Neglecting inertia, and in the absence of body forces other than gravity, mechanical equilibrium then requires that
\begin{equation}\label{eq:equilibrium}
    \boldsymbol{\nabla}\cdot\boldsymbol{\sigma} =\boldsymbol{\nabla}\cdot\boldsymbol{\sigma}^\prime-\boldsymbol{\nabla}p =-\rho\,\mathbf{g},
\end{equation}
where $\rho\equiv{}\phi_f\rho_f+(1-\phi_s)\rho_s$ is the phase-averaged, composite, or bulk density. A useful but nonrigorous physical interpretation of Eq.~\eqref{eq:equilibrium} is that the fluid pressure gradient acts as a body force within the solid skeleton.

The stress tensors in Eq.~\eqref{eq:equilibrium} are Cauchy or \textit{true} stresses. These are Eulerian quantities, and Eq.~\eqref{eq:equilibrium} is an Eulerian statement: \textit{The current forces on current areas in the current configuration must balance.}

\subsection{Constitutive law for the solid skeleton}\label{ss:HenckyElasticity}

We assume that the solid skeleton is an elastic material, for which the state of stress depends on the displacement of material points from a relaxed reference state. This behavior distinguishes the theory of poroelasticity from, for example, the poroviscous framework traditionally used in magma/mantle dynamics, where the skeleton is assumed to behave as a viscous fluid over geophysical time scales~\citep[\textit{e.g.},][]{bercovici-jgr-2001a}. This assumption greatly simplifies the mathematical framework for large deformations by eliminating any dependence on displacement, but we cannot take advantage of it here.

The constitutive law for an elastic solid skeleton typically links the effective stress to the solid displacement via an appropriate measure of strain or strain energy. For large deformations, elastic behavior is nonlinear for two reasons. First, the kinematics is inherently nonlinear because the geometry of the body evolves with the deformation (kinematic nonlinearity). Second, most materials harden or soften under large strains as their internal microstructure evolves---that is, the material properties change with the deformation (material or constitutive nonlinearity).

To capture the kinematic nonlinearities introduced by the evolving geometry, relevant measures of finite strain are typically derived from one of the Cauchy-Green deformation tensors. A wide variety of finite-strain measures exist, each of which is paired with an appropriate measure of stress through a stress-strain constitutive relation that includes at least two elastic parameters. In modern hyperelasticity theory, this constitutive relation takes the form of a strain-energy density function. Selection of an appropriate constitutive law and subsequent tuning of the elastic parameters can ultimately match a huge variety of material behaviors, but our focus here is simply on capturing kinematic nonlinearity. For this purpose, we consider a simple hyperelastic model known as Hencky elasticity.

The key idea in Hencky elasticity is to retain the classical strain-energy density function of linear elasticity, but replacing the infinitesimal strain with the Hencky strain~\citep{anand-japplmech-1979, xiao-actamechanica-2002}. Hencky strain, also known ``natural strain'' or ``true strain'', is an extension to three dimensions of the one-dimensional concept of logarithmic strain. Hencky elasticity is a generic model in that it does not account for material-specific constitutive nonlinearity, but it captures the full geometric nonlinearity of large deformations and thus provides a good model for the elastic behavior of a wide variety of materials under moderate to large deformations~\citep{anand-japplmech-1979, anand-jmps-1986}. It is also very commonly used in large-deformation plasticity. Hencky strain has some computational disadvantages~\citep{bazant-jengmatertechnol-1998}, but these are not relevant here.

Hencky elasticity can be written
\begin{subequations}\label{eq:hencky_elasticity}
    \begin{align}
        J\boldsymbol{\sigma}^\prime &=\Lambda\,\mathrm{tr}(\mathbf{H})\mathbf{I} +(\mathcal{M}-\Lambda)\mathbf{H}, \label{eq:hencky_stress} \quad\mathrm{and} \\
        \mathbf{H} &=\frac{1}{2}\ln(\mathbf{F}\mathbf{F}^\mathsf{T}),
    \end{align}
\end{subequations}
where $\mathbf{H}$ is the Hencky strain tensor and the $J$ on the left-hand side of Eq.~\eqref{eq:hencky_stress} accounts for volume change during the deformation. Hencky elasticity reduces to linear elasticity for small strains and, conveniently, it uses the same elastic parameters as linear elasticity (see \S\ref{s:LinearPoroelasticity}). For compactness, we work in terms of the oedometric or $p$-wave modulus $\mathcal{M}=\mathcal{K}+\frac{4}{3}\mathcal{G}$ and Lam\'{e}'s first parameter $\Lambda=\mathcal{K}-\frac{2}{3}\mathcal{G}$, where $\mathcal{K}$ and $\mathcal{G}$ are the bulk modulus and shear modulus of the solid skeleton, respectively. Note that Lam\'{e}'s first parameter is often denoted $\lambda$, but we use $\Lambda$ here to avoid confusion with the principal stretches $\lambda_i$. All of these elastic moduli are ``drained'' properties, meaning that they are mechanical properties of the solid skeleton alone and must be measured under quasistatic conditions where the fluid is allowed to drain (leave) or enter freely.

\subsection{Boundary conditions}

Poromechanics describes flow and deformation within a porous material, so the boundaries of the spatial domain typically coincide with the boundaries of the solid skeleton. These boundaries may move as the skeleton deforms; in an Eulerian framework, this constitutes a moving-boundary problem. This feature is the primary disadvantage of working in an Eulerian framework as it can be analytically and numerically inconvenient. One noteworthy exception is in infinite or semi-infinite domains, in which case suitable far-field conditions are applied; this situation is common in geophysical problems, which are often spatially extensive.

To close the model presented above, we require kinematic and dynamic boundary conditions for the fluid and the skeleton. Kinematic conditions are straightforward: For the fluid, the most common kinematic conditions are constraints on the flux through the boundaries; for the solid, kinematic conditions typically enforce that the boundaries of the domain are \textit{material} boundaries, meaning that they move with the skeleton.

The simplest dynamic conditions are an imposed total stress, an imposed effective stress, or an imposed fluid pressure. At a permeable boundary, any two of these three quantities can be imposed. At an unconstrained permeable boundary, for example, the normal component of the total stress will come from the fluid pressure and the shear component must vanish; this then implies that both the normal and shear components of the effective stress must vanish. At an impermeable boundary, in contrast, only the total stress can be imposed---the decomposition of the load into fluid pressure and effective stress within the domain will arise naturally through the solution of the problem (although imposed shear stress can only be supported by the solid skeleton, via effective stress). Some care is required with more complex dynamic conditions that provide coupling with a non-Darcy external flow~\citep[\textit{e.g.,}][]{beavers-jfm-1967, shavit-tipm-2009, mosthaf-wrr-2011}, but this is beyond the scope of this paper.

\section{Linear poroelasticity}\label{s:LinearPoroelasticity}

We now briefly derive the theory of linear poroelasticity by considering the limit of infinitesimal deformations. For a deformation characterized by typical displacements of size $\delta\sim{}||\mathbf{u}_s||$ varying over spatial scales of size $\mathcal{L}\sim||\mathbf{x}||\sim{}||\mathbf{X}||$, the characteristic strain is of size $\epsilon\equiv{}\delta/\mathcal{L}\sim{}||\boldsymbol{\nabla}\mathbf{u}_s||$. The assumption of infinitesimal deformations requires that $\epsilon\ll{}1$. We develop the well-known linear theory by retaining terms to first order in $\epsilon$, neglecting terms of order $\epsilon^2$ and higher. Note that the deformation itself enters at first order by definition.

\subsection{Linear flow equation}

We have from Eqs.~\eqref{eq:Fdef} and \eqref{eq:F_to_J} that
\begin{equation}\label{eq:phi_to_u_linear}
    \frac{\phi_f-\phi_{f,0}}{1-\phi_{f,0}} \approx \boldsymbol{\nabla}\cdot\mathbf{u}_s\sim{}\epsilon.
\end{equation}
This result motivates rewriting Eq.~\eqref{eq:NonlinearADEEulerian_pde} in terms of the normalized change in porosity, $\tilde{\phi}_f\equiv{}(\phi_f-\phi_{f,0})/(1-\phi_{f,0})\sim\epsilon$,
\begin{equation}
    \frac{\partial{\tilde{\phi}_f}}{\partial{t}} +\boldsymbol{\nabla}\cdot\left[\tilde{\phi}_f\mathbf{q}-(1-\tilde{\phi}_f)\frac{k(\phi_f)}{\mu}\left(\boldsymbol{\nabla}p-\rho_f\mathbf{g}\right)\right]=0,
\end{equation}
where we have taken the initial porosity field to be uniform. We then eliminate $\mathbf{q}$ in favor of $\mathbf{v}_s$ using Eq.~\eqref{eq:vs_nonlinear},
\begin{equation}\label{eq:pre_LDE}
    \frac{\partial{\tilde{\phi}_f}}{\partial{t}} +\boldsymbol{\nabla}\cdot\left[\tilde{\phi}_f\mathbf{v}_s-\frac{k(\phi_f)}{\mu}\left(\boldsymbol{\nabla}p-\rho_f\mathbf{g}\right)\right]=0.
\end{equation}
Equation~\eqref{eq:us_to_vs} implies that $||\mathbf{v}_s||\sim{}\delta$, and therefore that $||\boldsymbol{\nabla}\cdot(\tilde{\phi}_f\mathbf{v}_s)||\sim{}\epsilon^2$. Simplifying Eq.~\eqref{eq:pre_LDE} accordingly, we arrive at one form of the well-known linear poroelastic flow equation:
\begin{equation}\label{eq:LinearDiffusionEulerian}
    \frac{\partial{\phi_f}}{\partial{t}} -\boldsymbol{\nabla}\cdot\left[(1-\phi_{f,0})\frac{k_0}{\mu}\left(\boldsymbol{\nabla}p-\rho_f\mathbf{g}\right)\right]\approx{}0,
\end{equation}
where $k_0=k(\phi_{f,0})$ is the relaxed/undeformed permeability, and where we have reverted from $\tilde{\phi}_f$ to $\phi_f$.

Comparing Eq.~\eqref{eq:LinearDiffusionEulerian} with Eqs.~\eqref{eq:NonlinearADEEulerian} highlights the fact that exact kinematics render the model nonlinear and also introduce a fundamentally different mathematical character: Equation~\eqref{eq:LinearDiffusionEulerian} can be written as a linear diffusion equation after introducing linear elasticity in the solid skeleton, whereas Eqs.~\eqref{eq:NonlinearADEEulerian} feature an additional, advection-like term related to the divergence-free total flux.

\subsection{Linear elasticity}\label{ss:LinearElasticity}

It is straightforward to show that Hencky elasticity (\S\ref{ss:HenckyElasticity}) reduces to classical linear elasticity at leading order in $\epsilon$, as do many other (but not all) finite-deformation elasticity laws. Linear elasticity can be written as
\begin{subequations}\label{eq:linear_elasticity}
    \begin{align}
        \boldsymbol{\sigma}^\prime &=\Lambda\,\mathrm{tr}(\boldsymbol{\varepsilon})\mathbf{I} +(\mathcal{M}-\Lambda)\boldsymbol{\varepsilon}, \quad\mathrm{and} \\
        \boldsymbol{\varepsilon} &=\frac{1}{2}\left[\boldsymbol{\nabla}\mathbf{u}_s +(\boldsymbol{\nabla}\mathbf{u}_s)^{\mathsf{T}}\right],
    \end{align}
\end{subequations}
where $\boldsymbol{\varepsilon}$ is the infinitesimal (``small'') strain tensor. By linearizing the strain in the displacement ($\mathbf{H}\approx{}\boldsymbol{\varepsilon}$) and the stress in the strain ($J\boldsymbol{\sigma}^\prime\approx{}\boldsymbol{\sigma}^\prime$), linear elasticity neglects both kinematic nonlinearity and constitutive nonlinearity as well as the distinction between the deformed configuration and the reference configuration.

\subsection{Discussion}\label{ss:LPE_Discussion}

A closed linear theory is provided by combining the linear flow equation (Eq.~\ref{eq:LinearDiffusionEulerian}) with mechanical equilibrium (Eq.~\ref{eq:equilibrium}), linear elasticity (Eqs.~\ref{eq:linear_elasticity}), and the linearized statement of volumetric compatibility (Eq.~\ref{eq:phi_to_u_linear}). The resulting model is valid to first order in $\epsilon$. A discussion of the various forms of the linear theory commonly used in hydrology, hydrogeology, and petroleum engineering can be found in Ref.~\citep{wang-princeton-2000}, and reviews of numerous classical results in linear poroelasticity can be found in Refs.~\citep{wang-princeton-2000} and \citep{rice-revgeophysspacephys-1976}.

Note that variations in permeability do not enter at this order because Eqs.~\eqref{eq:equilibrium} and \eqref{eq:linear_elasticity} together imply that $||\boldsymbol{\nabla}p/(\mathcal{M}/\mathcal{L})||=||(\boldsymbol{\nabla}\cdot\boldsymbol{\sigma}^\prime)/(\mathcal{M}/\mathcal{L})||\sim{}\epsilon$. This latter scaling should also be viewed as a constraint: Imposing pressure or stress gradients of size approaching $\mathcal{M}/\mathcal{L}$ will drive a deformation that violates the assumption $\epsilon\ll{}1$, invalidating the linear theory.

The linear theory can alternatively be derived from a Lagrangian perspective~(Ref.~\citep{coussy-wiley-2004} and Appendix~\ref{app:s:Lagrangian}). This must necessarily result in the same model, but in terms of the Lagrangian coordinate $\mathbf{X}$ instead of the Eulerian coordinate $\mathbf{x}$. These coordinates themselves differ at first order, $||(\mathbf{x}-\mathbf{X})/\mathcal{L}||=||\mathbf{u}_s/\mathcal{L}||\sim{}\epsilon$, but all quantities related to the deformation are also first order and this implies, for example, that $p(\mathbf{X},t)=p(\mathbf{x},t)-(\boldsymbol{\nabla}p)\cdot\mathbf{u}_s+\ldots\approx{}p(\mathbf{x},t)$. As a result, replacing $\mathbf{x}$ with $\mathbf{X}$ in Eqs.~\eqref{eq:equilibrium}, \eqref{eq:phi_to_u_linear}, \eqref{eq:LinearDiffusionEulerian}, and \eqref{eq:linear_elasticity} will result in a Lagrangian interpretation of the linear model that is still valid to first order in $\epsilon$. These two models are equivalent in the limit of $\epsilon\to0$, but they will always differ at order $\epsilon^2$ and diverge from each other as the deformation grows. This conceptual ambiguity is one awkward aspect of linear (poro)elasticity (see Appendix~\ref{app:s:strainambiguity}).

Here, our interest is in the behavior of the linear theory as the deformation becomes non-negligible. We next consider two model problems involving uniaxial flow and deformation, using these as a convenient setting for comparing the predictions of linear poroelasticity with the large-deformation theory.

\section{Models for uniaxial flow \\ and deformation}\label{s:rect}

\begin{figure}
    \centering
    \includegraphics[width=8.6cm]{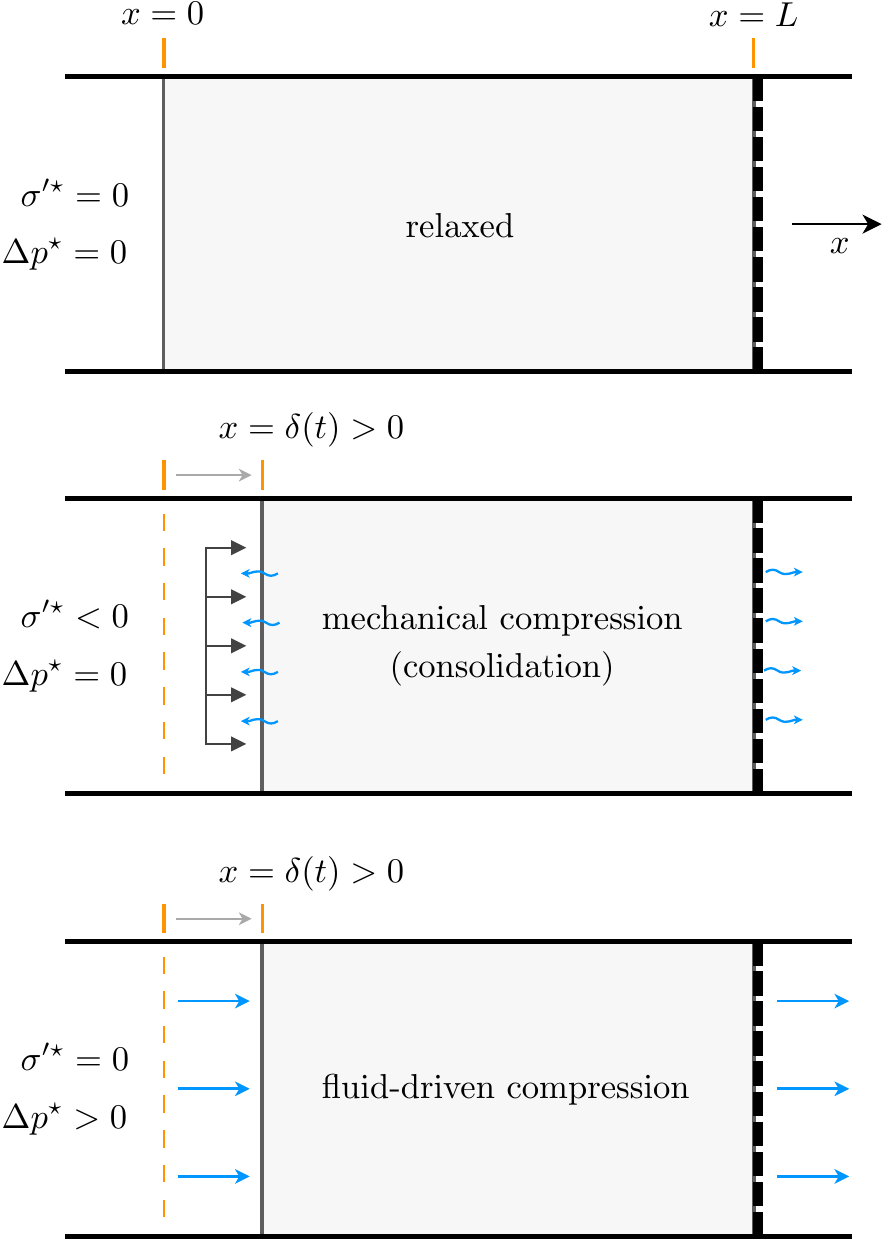}
    \caption{We consider the uniaxial deformation of a soft porous material by an applied effective stress and/or an applied fluid pressure drop. (a)~The solid is laterally confined and has relaxed length $L$. Its right edge is attached to a rigid permeable barrier ($x=L$, thick dashed black line), but the rest is free to move. We denote the instantaneous position of the left edge by $x=\delta(t)$, taking $\delta(0)=0$ (dashed orange line). The material can be compressed against the barrier ($\delta>0$) by (b)~an applied effective stress $\sigma^{\prime\star}<0$ (dark gray arrows), in which case the rate of deformation is set by the rate of fluid outflow (wiggly blue arrows) and/or by (c)~an applied fluid pressure drop $\Delta{p}^\star>0$, in which case the deformation is driven by a net flow from left to right (straight blue arrows). \label{fig:rect} }
\end{figure}

We now consider the uniaxial deformation of a deformable porous material, as shown schematically in Fig.~\ref{fig:rect}.
Provided that the material properties are uniform in the lateral directions, both the flow and the deformation will be restricted to one spatial dimension,
\begin{subequations}
    \begin{align}
        \mathbf{v}_f &=v_f(x,t)\hat{\mathbf{e}}_x, \\
        \mathbf{v}_s &=v_s(x,t)\hat{\mathbf{e}}_x, \\
        \mathbf{u}_s &=u_s(x,t)\hat{\mathbf{e}}_x, \label{eq:usrect}
    \end{align}
    and
    \begin{equation}
        \phi_f =\phi_f(x,t).
    \end{equation}
\end{subequations}
As a result, the analysis is tractable even when the deformation is large, which allows for the exploration of a variety of complex material models~\citep{barry-jbiomech-1990, barry-intjnonlinmech-1991, preziosi-intjmultiphaseflow-1996} and loading scenarios, including mechanical compression~\citep{holmes-jbiomecheng-1986, dawson-japplmech-2008}, forced infiltration~\citep{preziosi-intjmultiphaseflow-1996, sommer-jfm-1996}, and spontaneous imbibition~\citep{anderson-physfluids-2005, siddique-physfluids-2009}. Here, we consider two canonical problems: Mechanical compression (the consolidation problem) and fluid-driven compression. These differ only in the boundary conditions, so we develop a single model that applies to both cases. We assume that gravity is unimportant.

For the solid, a one-dimensional displacement field implies that the material is either laterally confined or laterally infinite, otherwise the Poisson effect would lead to lateral expansion or contraction. Our model and results are independent of the shape and size of the $y$-$z$ cross section as long as the lateral boundaries are rigid, frictionless, and impermeable. For example, the material could be a rectangular slab within a duct~~\citep[\textit{e.g.},][]{beavers-jfluidseng-1981a, beavers-jfluidseng-1981b} or a cylinder within a tube~\citep[\textit{e.g.},][]{parker-japplmech-1987}. Although we focus here on compression, our models and solutions remain valid if we reverse the sign of the effective stress and/or the pressure gradient; this will reverse the direction of the displacement and/or the flow, stretching the skeleton to the left in a state of tension.

\subsection{Five models}\label{ss:models}

Poromechanical phenomena are highly coupled. In order to highlight the nonlinear interactions between the various physical mechanisms at play, as well the qualitative and quantitative behavior of the error introduced by linearizing these mechanisms, we consider five different models below: A fully linear model~(\S\ref{s:LinearPoroelasticity}), two fully nonlinear models, and two intermediate models. The nonlinear models combine rigorous large-deformation kinematics with Hencky elasticity~(\S\ref{ss:HenckyElasticity}) and one of two permeability laws: Constant ($k=k_0$) or deformation-dependent ($k=k(\phi_f)$) via the normalized Kozeny-Carman formula, Eq.~\eqref{eq:kozeny-carman-k0}. The intermediate models are the same as the nonlinear models, but replacing Hencky elasticity with linear elasticity~(\S\ref{ss:LinearElasticity}) while retaining all other nonlinearity. We refer to these models as:
\begin{enumerate}
    \item ``linear'': Linear poroelasticity;
    \item ``nonlinear-$k_0$'': Nonlinear kinematics with Hencky elasticity and constant permeability;
    \item ``nonlinear-$k_\mathrm{KC}$'': Nonlinear kinematics with Hencky elasticity and deformation-dependent permeability;
    \item ``intermediate-$k_0$'': Nonlinear kinematics with linear elasticity and constant permeability; and
    \item ``intermediate-$k_\mathrm{KC}$'': Nonlinear kinematics with linear elasticity and deformation-dependent permeability.
\end{enumerate}
Note that although the intermediate approach retains most of the kinematic nonlinearity of the fully nonlinear model, it is not kinematically rigorous because the nonlinearity of Hencky elasticity is also kinematic in origin. The intermediate approach should also be considered with caution because it is asymptotically mixed, which can lead to nonphysical behavior at large deformations. However, it is useful for illustration.

We derive and discuss below the fully nonlinear models (\S\ref{ss:rect_large}) and the linear model (\S\ref{ss:rect_linear}), but we present results from all five models~\cite{see_supp}. We adopt the shorthand names given above for conciseness.

\subsection{Large-deformation poroelasticity}\label{ss:rect_large}

We first consider the exact kinematics of flow and deformation with a Hencky-elastic response in the solid skeleton. The results from this section provide the nonlinear-$k_0$ and nonlinear-$k_\mathrm{KC}$ models by introducing the appropriate permeability function, and can be readily modified to provide the intermediate-$k_0$ and intermediate-$k_\mathrm{KC}$ models by replacing Hencky elasticity with linear elasticity in any steps involving the elasticity law.

\subsubsection{Kinematics and flow}

We assume that the porosity in the initial state is spatially uniform and given by $\phi_f(x,0)=\phi_{f,0}$, where $\phi_{f,0}$ is a known constant, thereby giving
\begin{equation}\label{eq:J_to_phi}
    J(x,t) =\frac{1-\phi_{f,0}}{1-\phi_f(x,t)}.
\end{equation}
The deformation-gradient tensor can be written as (\textit{c.f.}, Eq.~\ref{eq:Fdef})
\begin{equation}\label{eq:F_large_rect}\renewcommand{\arraystretch}{1.5}
	\mathbf{F}=\left[
	\begin{array}{ccc}
		J  &  0  &  0 \\
		0  &  1  &  0 \\
		0  &  0  &  1
	\end{array}\right],
\end{equation}
where the Jacobian determinant is
\begin{equation}\label{eq:J_rect}
    J=\det{\left(\mathbf{F}\right)} =\left(1-\frac{\partial{u_s}}{\partial{x}}\right)^{-1}.
\end{equation}
The displacement field is linked to the porosity field via Eq.~\eqref{eq:J_to_phi},
\begin{equation}\label{eq:phi_to_u}
    \frac{\phi_f-\phi_{f,0}}{1-\phi_{f,0}}=\frac{\partial{u_s}}{\partial{x}}.
\end{equation}
For uniaxial flow, Eqs.~\eqref{eq:NonlinearADEEulerian}, \eqref{eq:vf_vs_nonlinear}, and \eqref{eq:equilibrium} become
\begin{subequations}\label{eq:fluidmech_rect}
    \begin{align}
        \frac{\partial{\phi_f}}{\partial{t}}+ \frac{\partial}{\partial{x}}\bigg[\phi_f{}q(t)-(1-\phi_f)\frac{k(\phi_f)}{\mu}\frac{\partial{p}}{\partial{x}}\bigg] &=0, \label{eq:fluidmech_rect_pde}
    \end{align}
with
    \begin{align}
        v_f &=q(t)-\left(\frac{1-\phi_f}{\phi_f}\right)\frac{k(\phi_f)}{\mu}\frac{\partial{p}}{\partial{x}}, \label{eq:fluidmech_rect_vf} \\
        v_s &=q(t)+\frac{k(\phi_f)}{\mu}\frac{\partial{p}}{\partial{x}}, \label{eq:fluidmech_rect_vs}
    \end{align}
and
    \begin{align}
        \frac{\partial{p}}{\partial{x}}&=\frac{\partial{\sigma^\prime_{xx}}}{\partial{x}}, \label{eq:equilibrium_rect}
    \end{align}
\end{subequations}
where the total volume flux $q(t)=\phi_fv_f+(1-\phi_f)v_s$ is a function of time only. Equations~\eqref{eq:phi_to_u} and \eqref{eq:fluidmech_rect} constitute a kinematically exact model for any constitutive behavior in the solid skeleton. This model has been derived previously~\citep[\textit{e.g.}, Eq.~(44) of Ref.][]{preziosi-intjmultiphaseflow-1996}.

\subsubsection{Hencky elasticity}\label{ss:mech_rect_large}

We take the constitutive response of the solid skeleton to be Hencky elastic, in which case the associated effective stress is
\begin{equation}\label{eq:sigma_large_rect}\renewcommand{\arraystretch}{1.5}
	\boldsymbol{\sigma}^{\prime}=\left[
	\begin{array}{ccc}
		\mathcal{M}\displaystyle\frac{\ln{J}}{J} &  0  &  0 \\
		0  &  \Lambda\displaystyle\frac{\ln{J}}{J}  &  0 \\
		0  &  0  &  \Lambda\displaystyle\frac{\ln{J}}{J}
	\end{array}\right].
\end{equation}
Although the displacement and the strain are uniaxial, the stress has three nontrivial components due to the Poisson effect under lateral confinement. If the material were laterally unconfined, the stress would be uniaxial and the strain would have three nontrivial components.

We link the mechanics of the skeleton with those of the fluid by combining Eq.~\eqref{eq:sigma_large_rect} with Eqs.~\eqref{eq:phi_to_u} and \eqref{eq:equilibrium_rect} to obtain
\begin{equation}\label{eq:p_to_phi_large_rect}
    \begin{split}
        \frac{\partial{p}}{\partial{x}} =\frac{\partial{\sigma^{\prime}_{xx}}}{\partial{x}} &=\frac{\partial}{\partial{x}}\left[\mathcal{M}\,\frac{\ln{J}}{J}\right] \\
        &=\frac{\partial}{\partial{x}}\left[ \mathcal{M}\,\left(\frac{1-\phi_f}{1-\phi_{f,0}}\right)\ln\left(\frac{1-\phi_{f,0}}{1-\phi_f}\right)\right].
    \end{split}
\end{equation}
With appropriate boundary conditions, Eqs.~\eqref{eq:phi_to_u}--\eqref{eq:p_to_phi_large_rect} provide a closed model for the evolution of the porosity.

For uniaxial deformation, the Hencky stress and strain depend only on $J$ and can therefore be written directly in terms of $\phi_f$. In fact, this is the case for any constitutive law since $\mathbf{F}$ itself depends only on $J$---that is, the deformation can be completely characterized by the local change in porosity. This is a special feature of uniaxial deformation: The effective stress can be written exclusively as a function of porosity, $\boldsymbol{\sigma}^\prime=\boldsymbol{\sigma}^\prime(\phi_f)$, \textit{for any constitutive law}. As a result, the framework of large-deformation elasticity can be avoided in a uniaxial setting by simply positing or measuring the function $\sigma^\prime_{xx}(\phi_f)$~\citep[\textit{e.g.},][]{beavers-japplmech-1975}. This approach is simple and appealing, but has the obvious disadvantage that it cannot be readily generalized to more complicated loading scenarios. It also has the more subtle disadvantage that even in the uniaxial case it is unable to provide answers to basic questions about the 3D state of stress within the material. For example: How much stress does the material apply to the lateral confining walls? What is the maximum shear stress within the material?

\subsubsection{Boundary conditions}\label{sss:rect_large_bcs}

The left and right boundaries of the solid skeleton are located at $x=\delta(t)$ and $x=L$, respectively, and we take $\delta(0)=0$ without loss of generality (Fig.~\ref{fig:rect}). We then have four kinematic boundary conditions for the skeleton from the fact that the left and right edges are material boundaries: Two on displacement,
\begin{equation}\label{eq:u_bc}
    u_s(\delta,t)=\delta \quad\mathrm{and}\quad u_s(L,t)=0,
\end{equation}
and two on velocity,
\begin{equation}\label{eq:vs_bc}
    v_s(\delta,t)=\dot{\delta}\equiv\frac{\mathrm{d}\delta}{\mathrm{d}t} \quad\mathrm{and}\quad v_s(L,t)=0.
\end{equation}
We use the former pair in calculating the displacement field from the porosity field, and the latter pair in deriving boundary conditions for porosity.

We take the pressure drop across the material to be imposed and equal to $\Delta{p}\equiv{}p(\delta,t)-p(L,t)$, and without loss of generality we then write
\begin{equation}\label{eq:dp_bc}
    p(\delta,t)=\Delta{p} \quad\mathrm{and}\quad p(L,t)=0.
\end{equation}
We further assume that a mechanical load is applied to the left edge in the form of an imposed effective stress $\sigma^{\prime\star}$. The effective stress at the right edge can then be derived by integrating Eq.~\eqref{eq:equilibrium_rect} from $\delta$ to $L$ to arrive at $\sigma_{xx}(\delta,t)=\sigma_{xx}(L,t)$, which is simply a statement of macroscopic force balance in the absence of inertia or body forces. From result and the pressures at $\delta$ and $L$, we then have that
\begin{equation}\label{eq:sig_bc}
    \sigma_{xx}^\prime(\delta,t)=\sigma^{\prime\star} \quad\mathrm{and}\quad \sigma_{xx}^\prime(L,t)=\sigma^{\prime\star}-\Delta{p}.
\end{equation}
Since the effective stress is directly related to the porosity in this geometry (see \S\ref{ss:mech_rect_large}), Eqs.~\eqref{eq:sig_bc} provide $\phi_f(\delta,t)$ and $\phi_f(L,t)$ and constitute Dirichlet conditions. For Hencky elasticity, these values can be readily calculated from
\begin{equation}\label{eq:sig_to_phi}
    \phi_f(x,t) =1+(1-\phi_{f,0}) \frac{\sigma_{xx}^\prime/\mathcal{M}}{\mathrm{W}(-\sigma_{xx}^\prime/\mathcal{M})}
\end{equation}
where $\mathrm{W}(\,\cdot\,)$ denotes the Lambert~W function ($y=\mathrm{W}(x)/x$ solves $x=-\ln(y)/y$).

When the pressure drop is imposed, the volume flux $q(t)$ through the material will vary in time and this appears explicitly in Eqs.~\eqref{eq:fluidmech_rect}. One approach to deriving an expression for $q(t)$ is to rearrange and integrate Eq.~\eqref{eq:fluidmech_rect_vs} \citep[\textit{c.f.}, Eqs.~(21)--(23) of][]{preziosi-intjmultiphaseflow-1996},
\begin{equation}
    q(t) =\frac{\Delta{p}^\star+\displaystyle\int_\delta^L\,\frac{\mu}{k(\phi_f)}\,v_s\,\mathrm{d}x} {\displaystyle\int_\delta^L\,\frac{\mu}{k(\phi_f)}\,\mathrm{d}x},
\end{equation}
but this is awkward in practice since it requires explicit calculation of $v_s$ from $u_s$ via Eq.~\eqref{eq:us_to_vs}, which is otherwise unnecessary. Alternatively, we can evaluate Eq.~\eqref{eq:fluidmech_rect_vs} at $x=L$ to obtain
\begin{equation}\label{eq:q}
    q(t) =-\left[\frac{k(\phi_f)}{\mu} \frac{\partial{p}}{\partial{x}}\right]\bigg|_{x=L} =-\left[\frac{k(\phi_f)}{\mu}\,\frac{\mathrm{d}\sigma^\prime_{xx}}{\mathrm{d}\phi_f}\,\frac{\partial{\phi_f}}{\partial{x}}\right]\bigg|_{x=L},
\end{equation}
which we supplement with the Dirichlet condition above on $\phi_f(L,t)$. Equation~\eqref{eq:q} is straightforward to implement.

For fluid-driven deformation, an imposed pressure drop $\Delta{p}^\star$ will eventually lead to a steady state in which the solid is stationary, the fluid flow is steady, and the volume flux is constant, $q(t)\to{}q^\star$. Imposing instead this same flux $q^\star$ from the outset and allowing the pressure drop to vary must eventually lead to precisely the same steady state, in which $\Delta{p}(t)\to{}\Delta{p}^\star$. As a result, the only difference between these two conditions is in the dynamic approach to steady state. We focus on the pressure-driven case below, but we provide analytical and numerical solutions that are valid for both cases~\cite{see_supp} and we explore the relationship between $q^\star$ and $\Delta{p}^\star$ at steady state. Note that, for an imposed flux, the pressure at $x=\delta$ is unknown and the Dirichlet condition at $x=L$ must be replaced by the Neumann condition that $\phi_f(L,t)v_f(L,t)=q^\star$.

For an incompressible solid skeleton, conservation of solid volume requires that
\begin{equation}\label{eq:Vol_rect}
    \int_\delta^L\,(1-\phi_f)\,\mathrm{d}x=(1-\phi_{f,0})\,L,
\end{equation}
and it is straightforward to confirm that this is identically satisfied by Eqs.~\eqref{eq:phi_to_u} and \eqref{eq:u_bc}. If any of these relationships are approximated, the resulting model will no longer be volume conservative. Conservation of mass or volume is typically not a primary concern in solid mechanics because most engineering materials are only slightly compressible and typically experience very small deformations. It becomes more important in poromechanics because porous materials are much more compressible than nonporous ones since the skeleton can deform through rearrangement of the solid grains. This rearrangement allows for large volume changes through large changes in the pore volume, which are then strongly coupled to the fluid mechanics.

\subsection{Linear poroelasticity}\label{ss:rect_linear}

We now derive the linear model. We do this by linearizing the nonlinear model above, so we write the results in terms of the Eulerian coordinate $x$. As described in \S\ref{ss:LPE_Discussion}, however, the spatial coordinate in the linear model is ambiguous: Simply replacing the Eulerian coordinate $x$ with the Lagrangian coordinate $X$ in the expressions below will result in a model that is still accurate to leading order in $\delta/L$. Whereas the Eulerian interpretation of this model (with $x$) will satisfy the boundary conditions at $x=\delta$ only at first order, the resulting Lagrangian interpretation (with $X$) will satisfy them exactly at $X=0$. However, the Eulerian interpretation will respect the relationship between porosity and displacement exactly since this relationship is linear in the Eulerian coordinate~(\textit{c.f.}, Eq.~\ref{eq:phi_to_ux_linear}), whereas the Lagrangian interpretation will respect this relationship only at first order.

\subsubsection{Kinematics and flow}

Adopting the assumption of infinitesimal deformations and linearizing in the strain, Eq.~\eqref{eq:phi_to_u_linear} becomes
\begin{equation}\label{eq:phi_to_ux_linear}
    \frac{\phi_f-\phi_{f,0}}{1-\phi_{f,0}}=\frac{\partial{u_s}}{\partial{x}}.
\end{equation}
Note that this is identical to Eq.~\eqref{eq:phi_to_u}, and is therefore exact. This is another special feature of uniaxial deformation: The exact relationship between $\phi_f$ and $u_s$ is linear. This does not hold for even simple biaxial deformations.

From Eqs.~\eqref{eq:equilibrium} and \eqref{eq:LinearDiffusionEulerian}, we further have
\begin{subequations}\label{eq:fluidmech_rect_linear}
    \begin{align}
        \frac{\partial{\phi_f}}{\partial{t}}- \frac{\partial}{\partial{x}}\left[(1-\phi_{f,0})\frac{k_0}{\mu}\frac{\partial{p}}{\partial{x}}\right]&\approx{}0, \label{eq:fluidmech_rect_pde_linear}\\
        \frac{\partial{p}}{\partial{x}}&=\frac{\partial{\sigma^\prime_{xx}}}{\partial{x}}. \label{eq:equilibrium_rect_linear}
    \end{align}
\end{subequations}
Comparing Eq.~\eqref{eq:fluidmech_rect_pde_linear} with Eq.~\eqref{eq:fluidmech_rect_pde} again highlights the fundamentally different mathematical character of the linear model as compared to the nonlinear model.

\subsubsection{Linear elasticity}

We take the constitutive response of the solid skeleton to be linear elastic, in which case the associated effective stress tensor is
\begin{equation}\label{eq:sigma_small_rect}\renewcommand{\arraystretch}{1.5}
	\boldsymbol{\sigma}^{\prime}=\left[
	\begin{array}{ccc}
		\mathcal{M} \displaystyle\frac{\partial{u_s}}{\partial{x}}  &  0  &  0 \\
		0  &  \Lambda \displaystyle\frac{\partial{u_s}}{\partial{x}}  &  0 \\
		0  &  0  &  \Lambda \displaystyle\frac{\partial{u_s}}{\partial{x}}
	\end{array}\right].
\end{equation}
Combining this with Eqs.~\eqref{eq:phi_to_ux_linear} and \eqref{eq:equilibrium_rect_linear}, we obtain
\begin{equation}\label{eq:p_to_phi_small_rect}
    \frac{\partial{p}}{\partial{x}} =\frac{\partial{\sigma^{\prime}_{xx}}}{\partial{x}} =\frac{\partial}{\partial{x}}\left[\mathcal{M} \frac{\partial{u_s}}{\partial{x}}\right] =\frac{\partial}{\partial{x}} \left[\mathcal{M}\,\left(\frac{\phi_f-\phi_{f,0}}{1-\phi_{f,0}}\right)\right].
\end{equation}
With appropriate boundary conditions, Eqs.~\eqref{eq:phi_to_ux_linear}--\eqref{eq:p_to_phi_small_rect} provide a closed linear model for the evolution of the porosity.

\subsubsection{Boundary conditions}

The kinematic conditions on the solid displacement (Eqs.~\ref{eq:u_bc}) become
\begin{equation}\label{eq:u_bc_linear}
    u_s(0,t)\approx{}\delta \quad\mathrm{and}\quad u_s(L,t)=0,
\end{equation}
where the distinction between $u_s(\delta,t)$ and $u_s(0,t)$ does not enter at first order. The latter condition is used when calculating the displacement field from the porosity field, and the former then provides an expression for $\delta(t)$. Neither is necessary when solving for the porosity field itself.

For an imposed pressure drop, the dynamic conditions on the pressure and the stress become
\begin{equation}\label{eq:sig_bc_linear}
    \sigma_{xx}^\prime(0,t)\approx{}\sigma^{\prime\star} \quad\mathrm{and}\quad \sigma_{xx}^\prime(L,t)=\sigma^{\prime\star}-\Delta{p}^\star,
\end{equation}
and these again provide Dirichlet conditions on the porosity via the elasticity law,
\begin{equation}\label{eq:dp_bc_linear}
    \phi_f(x,t)=\phi_{f,0} +(1-\phi_{f,0})\frac{\sigma_{xx}^\prime}{\mathcal{M}}.
\end{equation}
With these conditions on porosity, the linear model is fully specified. It is not necessary to calculate the total flux because it does not appear explicitly in the linear conservation law, but the flux can be calculated at any time from
\begin{equation}\label{eq:q_bc_linear}
    v_s(L,t)=0 \quad\to\quad q(t)\approx{}-\frac{1}{(1-\phi_{f,0})}\,\frac{k_0\mathcal{M}}{\mu}\,\frac{\partial{\phi_f}}{\partial{x}}\Big|_{x=L}.
\end{equation}
When the flux is imposed instead of the pressure drop, Eq.~\eqref{eq:q_bc_linear} can be rearranged to provide a Neumann condition at $x=L$ that replaces the Dirichlet condition above. The pressure drop $\Delta{p}(t)$ is then unknown, and must be calculated by rearranging Eqs.~\eqref{eq:sig_bc_linear} and \eqref{eq:dp_bc_linear}.

\subsection{Scaling}

We consider the natural scaling
\begin{equation}
    \begin{split}
        \tilde{t}=\frac{t}{T_\mathrm{pe}}\,,\quad{}\tilde{x}&=\frac{x}{L}\,, \quad{}\tilde{k}=\frac{k}{k_0}\,, \\
        \quad\tilde{p}&=\frac{p}{\mathcal{M}}\,,\quad\tilde{\sigma}_{xx}^\prime=\frac{\sigma_{xx}^\prime}{\mathcal{M}}\,,\quad{}\tilde{u}_s=\frac{u_s}{L},
    \end{split}
\end{equation}
where the characteristic permeability is $k_0=k(\phi_{f,0})$ and the classical poroelastic time scale is $T_\mathrm{pe}=\mu{}L^2/(k_0\mathcal{M})$. The problem is then controlled by one of two dimensionless groups that measure the strength of the driving stresses relative to the stiffness of the skeleton: $\tilde{\sigma}^{\prime\star}\equiv{}\sigma^{\prime\star}/\mathcal{M}$ for deformation driven by an applied mechanical load, or $\Delta{\tilde{p}}^\star\equiv{}\Delta{p}^\star/\mathcal{M}$ for deformation driven by fluid flow with a constant pressure drop $\Delta{p}^\star$. For an imposed flux $q^\star$, the relevant dimensionless group is instead $\tilde{q}^\star\equiv{}\mu{}q^\star{}L/(k_0\mathcal{M})$.

The problem also depends on the initial porosity $\phi_{f,0}$. When the permeability is constant, $\phi_{f,0}$ can be scaled out by working instead with the normalized change in porosity,
\begin{equation}
    \tilde{\phi}_f=\frac{\phi_f-\phi_{f,0}}{1-\phi_{f,0}}.
\end{equation}
When the permeability is allowed to vary, the initial value $\phi_{f,0}$ cannot be eliminated because the permeability must depend on the current porosity rather than on the change in porosity.

The discussion below uses dimensional quantities for expository clarity, but we present the results in terms of dimensionless parameter combinations to emphasize this scaling.

\subsection{Summary}

Each of the models described above can ultimately be written as a single parabolic conservation law for $\phi_f$; this will be linear and diffusive for the linear model, and nonlinear and advective diffusive for the intermediate and nonlinear models. The boundary condition at the left is a Dirichlet condition for all of the cases considered here, and the boundary condition at the right is either Dirichlet for flow driven by a imposed pressure drop or Neumann for flow driven by an imposed fluid flux. For the nonlinear and intermediate models, we must also solve for the unknown position of the free left boundary. Below, we study these models dynamically and at steady state in the context of two model problems.

\section{Mechanical compression: The consolidation problem}\label{s:mech_rect}

We now consider the uniaxial mechanical compression of a porous material~(Fig.~\ref{fig:rect}b), in which an effective stress $\sigma^{\prime\star}$ is suddenly applied to the left edge of the material at $t=0^+$ and the fluid pressure at both edges is held constant and equal to the ambient pressure, $p(\delta,t)=p(L,t)=0$. The process by which the material relaxes under this load, squeezing out fluid as the pore volume decreases, is known as \textit{consolidation}. The consolidation problem is a classical one, with direct application to the engineering of foundations; it has been studied extensively in that context and others~\citep[\textit{e.g.},][]{carter-intjnag-1979a, oloyede-clinicalbiomech-1991, murad-tipm-1995, lancellotta-intjengsci-1997, wang-princeton-2000, franceschini-jmps-2006, jaeger-wiley-2007}.

Force balance requires that the total stress everywhere in the material must immediately support the applied load, $\sigma_{xx}(x,t)=\sigma^\prime_{xx}(x,t)-p(x,t)=\sigma^{\prime\star}$ for $t>0$. However, the effective stress can only contribute through strain in the solid skeleton, and the solid skeleton can only deform by displacing fluid, and this is not instantaneous. As a result, the fluid pressure must immediately jump to support the entire load: $p(x,0^+)=-\sigma^{\prime\star}$. In soil and rock mechanics, this is known as an \textit{undrained} response: The mechanical response of a fluid-solid mixture under conditions where the fluid content is fixed. Over time, this high pressure relaxes as fluid flows out at the boundaries, and the effective stress supports an increasing fraction of the load as the material is compressed. When the process is finished, the effective stress will support the entire load and the fluid pressure will have returned to its ambient value. This is classical consolidation theory.

\subsection{Steady state}\label{ss:consolidation_ss}

When the consolidation process is finished, the solid and fluid are both stationary, $v_s(x)=v_f(x)=0$, and the fluid pressure is uniform, $p(x)=0$. As a result, the steady state is determined entirely by the boundary conditions and the elastic response of the skeleton; the fluid plays no role. In soil and rock mechanics, this is known as the \textit{drained} response of the material.

\begin{figure*}
    \centering
    \includegraphics[width=17.2cm]{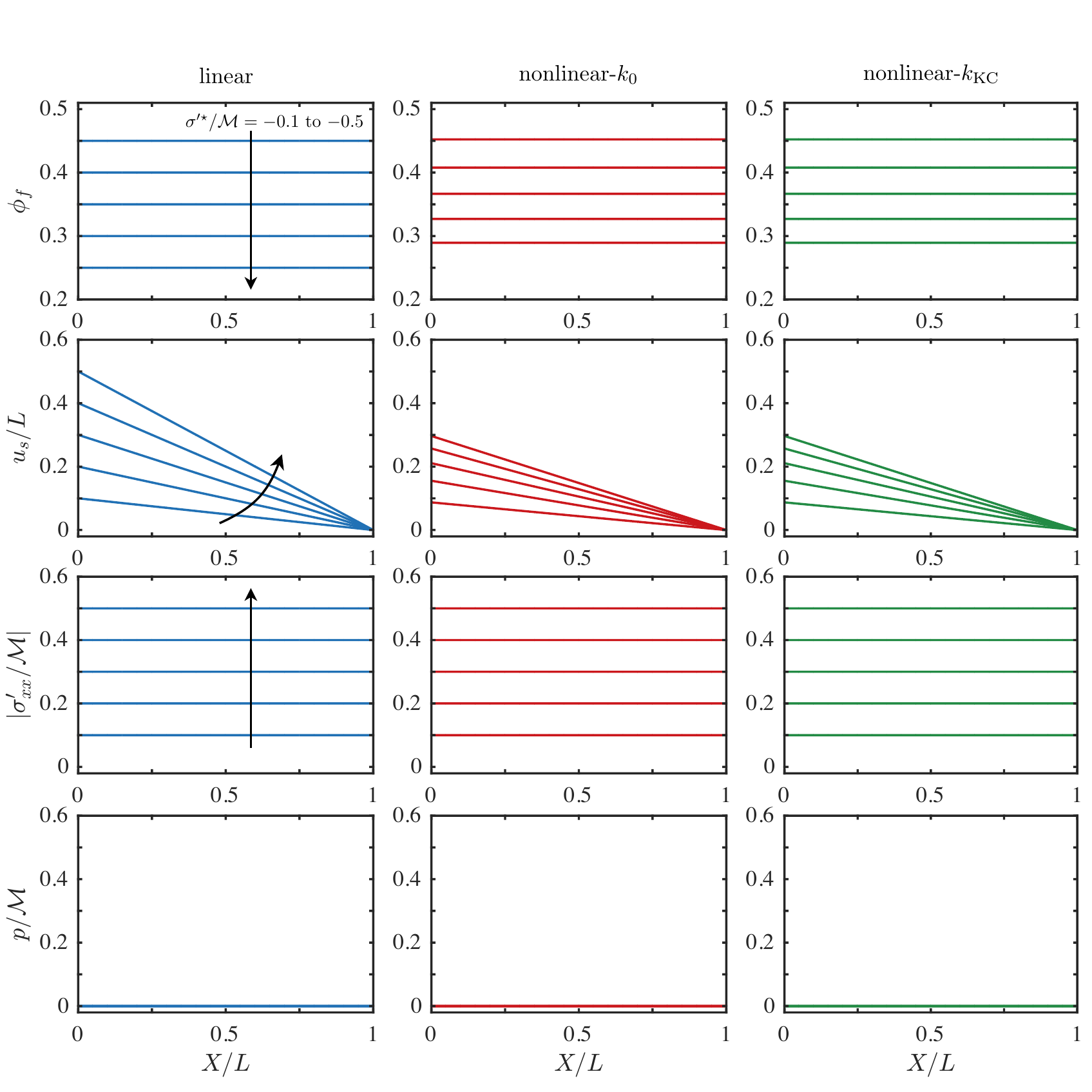}
    \caption{Steady state in the consolidation of a soft porous material under an applied effective stress $\sigma^{\prime\star}<0$, here for $\sigma^{\prime\star}/\mathcal{M}=-0.1$, $-0.2$, $-0.3$, $-0.4$, and $-0.5$, as indicated. We show the porosity (taking $\phi_{f,0}=0.5$; first row), displacement (second row), effective stress (third row), and pressure (last row) for the linear model (left column, blue), the nonlinear-$k_0$ model (middle column, red), and the nonlinear-$k_\mathrm{KC}$ model (right column, green) (see \S\ref{ss:models}). For the nonlinear models, we plot these results against the Lagrangian coordinate $X=x-u_s(x,t)$ for clarity; for the linear model, we adopt a Lagrangian interpretation and simply replace $x$ with $X$ in the relevant expressions (see \S\ref{ss:LPE_Discussion} and \S\ref{ss:mech_rect_dy}). In all cases, the displacement is linear and the porosity, stress, and pressure are uniform. Fluid flow plays no role in the steady state, so the middle and right columns are identical. \label{fig:mech_rect_ss_v_x} }
\end{figure*}

Without a fluid pressure gradient, mechanical equilibrium implies that the effective stress and the porosity must be uniform, $\sigma^\prime_{xx}(x)=\sigma^{\prime\star}$ and $\phi_f(x)=\phi_f^\star$ (Eqs.~\ref{eq:p_to_phi_large_rect} and \ref{eq:p_to_phi_small_rect}). Since the fluid plays no role, the nonlinear-$k_0$ and nonlinear-$k_\mathrm{KC}$ models are identical at steady state. For both of these, we have that
\begin{subequations}\label{eq:mech_rect_large}
    \begin{align}
        \frac{\phi_f^\star-\phi_{f,0}}{1-\phi_{f,0}} &=1-\frac{1}{J^\star}, \\
        \frac{u_s(x)}{L} &=-\left(1-\frac{1}{J^\star}\right)\left(1-\frac{x}{L}\right), \\
        \frac{\delta^\star}{L} &=1-J^\star,
    \end{align}
\end{subequations}
where the Jacobian determinant $J^\star$ is found by inverting $\sigma^\prime_{xx}(J^\star)=\sigma^{\prime\star}$ with the aid of Eq.~\ref{eq:sig_to_phi},
\begin{equation}
    J^\star =-\frac{\mathrm{W}(-\sigma^{\prime\star}/\mathcal{M})}{\sigma^{\prime\star}/\mathcal{M}},
\end{equation}
and the deflection $\delta^\star/L$ is the change in length per unit reference length, usually known as the ``engineering'' or \textit{nominal} strain. For the linear model, we instead have that
\begin{subequations}\label{eq:mech_rect_small}
    \begin{align}
        \frac{\phi_f^\star-\phi_{f,0}}{1-\phi_{f,0}} &\approx{}\frac{\sigma^{\prime\star}}{\mathcal{M}}, \label{eq:mech_rect_large_phi} \\
        \frac{u_s(x)}{L} &\approx{}-\frac{\sigma^{\prime\star}}{\mathcal{M}}\left(1-\frac{x}{L}\right), \,\,\mathrm{and} \\
        \frac{\delta^\star}{L} &\approx{}-\frac{\sigma^{\prime\star}}{\mathcal{M}}.
    \end{align}
\end{subequations}
We compare these results in Fig.~\ref{fig:mech_rect_ss_v_x},
showing the linear model (Lagrangian interpretation), the nonlinear-$k_0$ model, and the nonlinear-$k_\mathrm{KC}$ model (see \S\ref{ss:models}). We include the latter for completeness but, as mentioned above, it is identical to the nonlinear-$k_0$ model at steady state since there is no flow.

\begin{figure*}
    \centering
    \includegraphics[width=17.2cm]{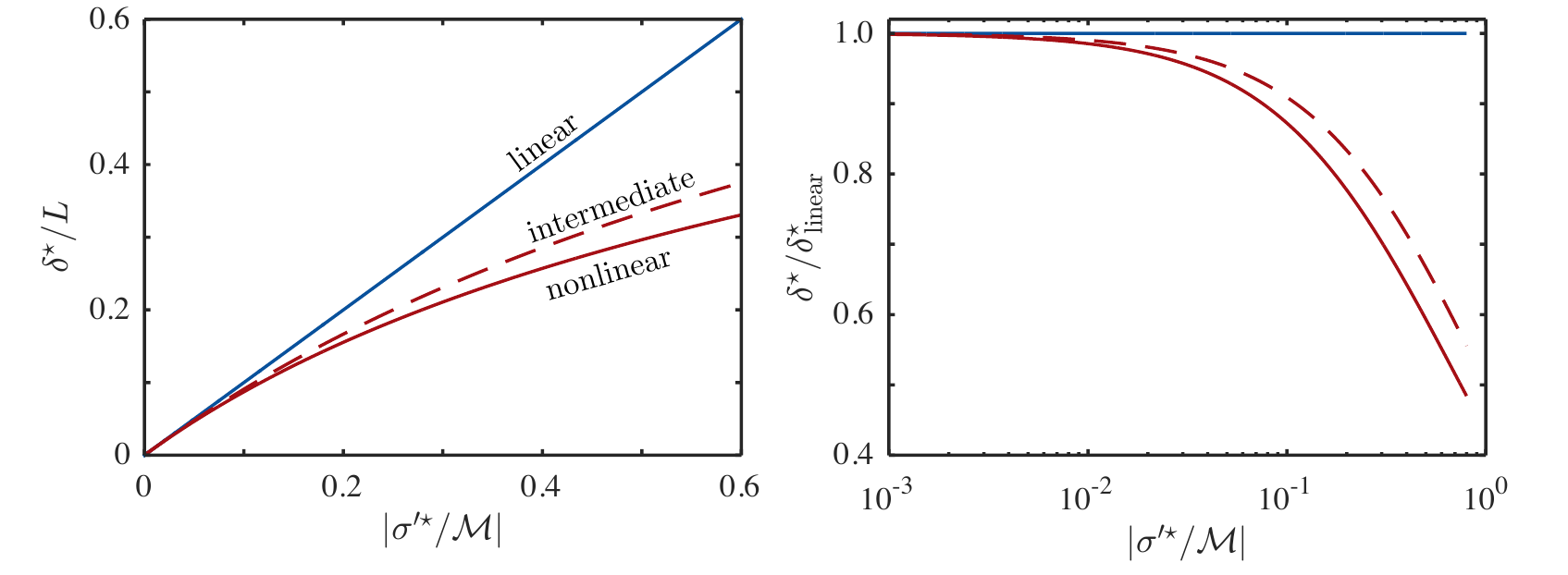}
    \caption{The linear model overpredicts the final deflection in consolidation under an applied effective stress, and this error is primarily kinematic. Here we plot (left)~the final deflection $\delta^\star/L$ against the applied effective stress $\sigma^{\prime\star}/\mathcal{M}$ for the linear model (blue), the nonlinear model (solid red), and the intermediate model (dashed red) (see \S\ref{ss:models}). We also show (right)~the ratio of these predictions to the linear one, $\delta^\star/\delta^\star_{\,\mathrm{linear}}$, on a semilogarithmic scale. The nonlinear and intermediate models both exhibit much stiffer behavior than the linear model in compression, and the nonlinear model is stiffer than the intermediate model. The relative error in both the linear and intermediate models is of size $\delta^\star/L$, which is consistent with the assumptions of linear (poro)elasticity. \label{fig:mech_rect_ss_v_sig} }
\end{figure*}

In all cases, the only nontrivial component of the deformation is the displacement, and this is simply linear in $x$. The difference between the models lies in the amount of deformation that results from a given load: The nonlinear and intermediate models deform much less than the linear model, and increasingly so for larger compressive loads~(Fig.~\ref{fig:mech_rect_ss_v_sig}).
The relative error between the linear and nonlinear models is of size $\delta^\star/L$, which is consistent with the assumptions of linear (poro)elasticity. To highlight the origin of this error, we further compare these two models with the intermediate model, in which we replace Hencky elasticity with linear elasticity in the nonlinear kinematic framework~(see \S\ref{ss:models}; Fig.~\ref{fig:mech_rect_ss_v_sig}). This comparison illustrates the fact that the majority of the error associated with the linear model results in this case from the kinematics of the deformation, and not from nonlinearity in the elasticity law. One source of kinematic nonlinearity at steady state is the cumulative nature of strain, where increments of displacement correspond to increasingly larger increments of strain as the material is compressed because the overall length decreases. The opposite occurs in tension: The nonlinear model deforms much more than the linear model because increments of displacement correspond to increasingly smaller increments of strain as the material is stretched. Another source of kinematic nonlinearity is the moving boundary, since the linear model satisfies the boundary conditions there only at leading order in $\delta^\star/L$.

The nonlinear model implies that the material can support an arbitrarily large compressive stress, with $\delta^\star/L$ approaching unity (\textit{i.e.}, the length of the deformed solid approaching zero) as the compressive stress diverges. Closer inspection reveals that the porosity will vanish when the deflection $\delta^\star/L$ reaches $\phi_{f,0}$, which occurs at a finite compressive stress. One would expect the stiffness of the skeleton to change relatively sharply across the transition from compressing pore space to compressing solid grains, and significant microstructural damage would likely occur \textit{en route} (\textit{e.g.}, grain crushing)---A material-specific constitutive model would be necessary to capture this behavior. This behavior is also important and problematic from the perspective of the fluid mechanics, which can become nonphysical unless the permeability law accounts appropriately for the changing porosity (see \S\ref{ss:Darcy} above).

\subsection{Dynamics}\label{ss:mech_rect_dy}

To explore the dynamics of consolidation, we solve the nonlinear and intermediate models numerically using a finite-volume method with an adaptive grid (Appendix~\ref{app:s:FV_rect} and \cite{see_supp}), and we solve the linear model analytically via separation of variables. The well-known analytical solution can be written
\begin{widetext}
    \begin{subequations}\label{eq:consolidation_solution}
        \begin{align}
            \phi_f(x,t) &= \phi_f^\star -(\phi_f^\star-\phi_{f,0})\sum_{n=1}^\infty\,\frac{2}{n\pi}\Big[1+(-1)^{n+1}\Big]e^{-\frac{(n\pi)^2t}{T_\mathrm{pe}}}\sin\left(\frac{n\pi{}x}{L}\right) \quad\mathrm{and} \\
            \frac{u_s(x,t)}{L} &=-\left(\frac{\phi_f^\star-\phi_{f,0}}{1-\phi_{f,0}}\right)\left\{1-\frac{x}{L}+\sum_{n=1}^\infty \frac{2}{(n\pi)^2}\Big[1+(-1)^{n+1}\Big]e^{-\frac{(n\pi)^2t}{T_\mathrm{pe}}}\left[(-1)^n-\cos\left(\frac{n\pi{}x}{L}\right)\right]  \right\},
        \end{align}
    \end{subequations}
\end{widetext}
where $\phi_f(0,t)=\phi_f(L,t)=\phi_f^\star=\phi_{f,0}+(1-\phi_{f,0})(\sigma^{\prime\star}/\mathcal{M})$, as in Eq.~\eqref{eq:mech_rect_large_phi}, and all other quantities of interest can readily be calculated from the porosity and displacement fields. Note that, as in the steady state, the Eulerian interpretation of Eqs.~\eqref{eq:consolidation_solution} (as written) satisfies the boundary conditions at the moving boundary only to leading order in $\delta^\star/L$. The Lagrangian interpretation (replacing $x$ with $X$) rigorously satisfies the boundary conditions at $X=0$, but at the expense of exact conservation of mass (Eq.~\ref{eq:phi_to_ux_linear}). However, both interpretations predict the same deflection, which is often the quantity of primary interest in engineering applications (Eulerian: $\delta^\star\approx{}u_s(x=0,t)$; Lagrangian: $\delta^\star=u_s(X=0,t)$).
\begin{figure*}
    \centering
    \includegraphics[width=17.2cm]{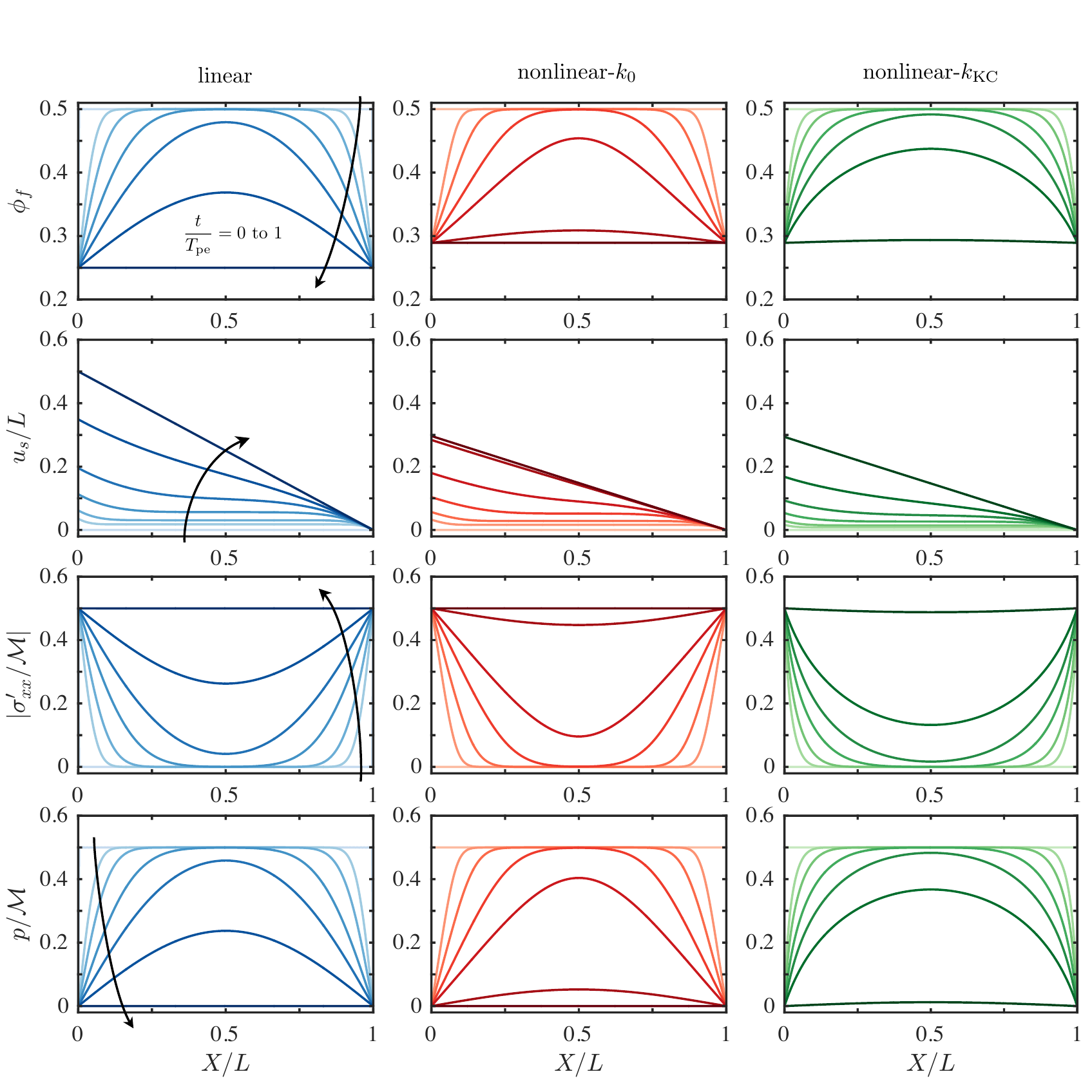}
    \caption{Dynamics of the consolidation process for a soft porous material under an applied effective stress of $\sigma^{\prime\star}/\mathcal{M}=-0.5$. We show the porosity (first row), displacement (second row), effective stress (third row), and pressure (last row) at $t/T_\mathrm{pe}=0$, 0.001, 0.003, 0.01, 0.03, 0.1, and 1, as indicated (light to dark colors), for the linear model (left column, blue), the nonlinear-$k_0$ model (middle column, red), and the nonlinear-$k_\mathrm{KC}$ model (right column, green). For the nonlinear models, we plot these results against the Lagrangian coordinate $X=x-u_s(x,t)$ for clarity; for the linear model, we again adopt a Lagrangian interpretation and simply replace $x$ with $X$ in the relevant expressions (see \S\ref{ss:LPE_Discussion} and \S\ref{ss:mech_rect_dy}). These results are for $\phi_{f,0}=0.5$. \label{fig:mech_rect_dy_v_x} }
\end{figure*}

In Fig.~\ref{fig:mech_rect_dy_v_x}, we compare the dynamics of consolidation for the linear model (Lagrangian interpretation), the nonlinear-$k_0$ model, and the nonlinear-$k_\mathrm{KC}$ model. In all cases, the skeleton is initially relaxed in the middle and very strongly deformed at the edges, from which the fluid can easily escape. The deformation propagates inward toward the middle from both ends over time, and the pressure decays as the skeleton supports an increasing fraction of the total stress. The nonlinear-$k_\mathrm{KC}$ model exhibits a more rounded deformation profile than either the linear model or the nonlinear-$k_0$ model, which is a result of the fact that the reduced permeability in the compressed outer regions slows and spreads the relaxation of the pressure field. The two nonlinear models ultimately arrive at the same steady state, which is determined strictly by the elasticity law (\textit{c.f.}, Fig.~\ref{fig:mech_rect_ss_v_x}). The nonlinear models deform much less than the linear model overall.

\begin{figure*}
    \centering
    \includegraphics[width=17.2cm]{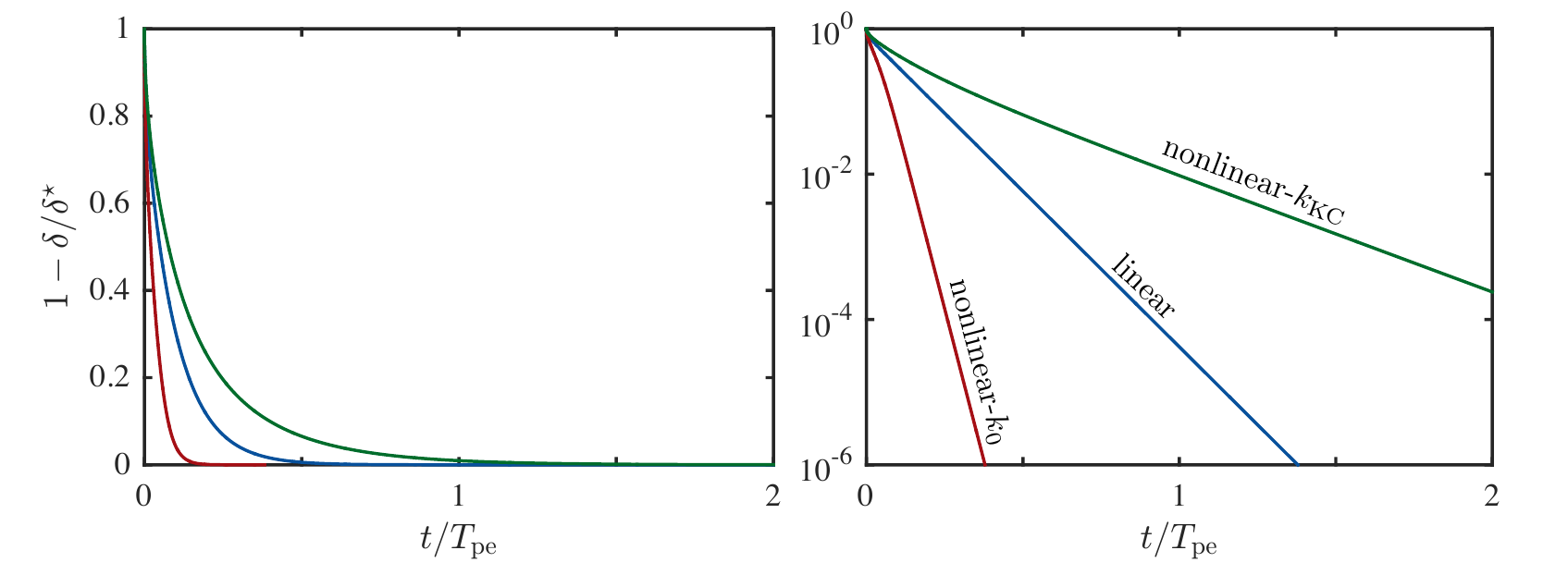}
    \caption{Relaxation of the deflection $\delta$ toward its steady-state value $\delta^\star$ during consolidation for $\sigma^{\prime\star}/\mathcal{M}=-0.5$ on a linear scale (left) and on a semilogarithmic scale (right). We again show the linear model (blue), the nonlinear-$k_0$ model (red), and the nonlinear-$k_\mathrm{KC}$ model (green). All three models relax exponentially, but the nonlinear-$k_0$ model relaxes about 4 times faster than the linear model, whereas the nonlinear-$k_\mathrm{KC}$ model relaxes at less than half the rate of the linear model. These results are for $\phi_{f,0}=0.5$. \label{fig:mech_rect_dy_v_t} }
\end{figure*}
\begin{figure*}
    \centering
    \includegraphics[width=17.2cm]{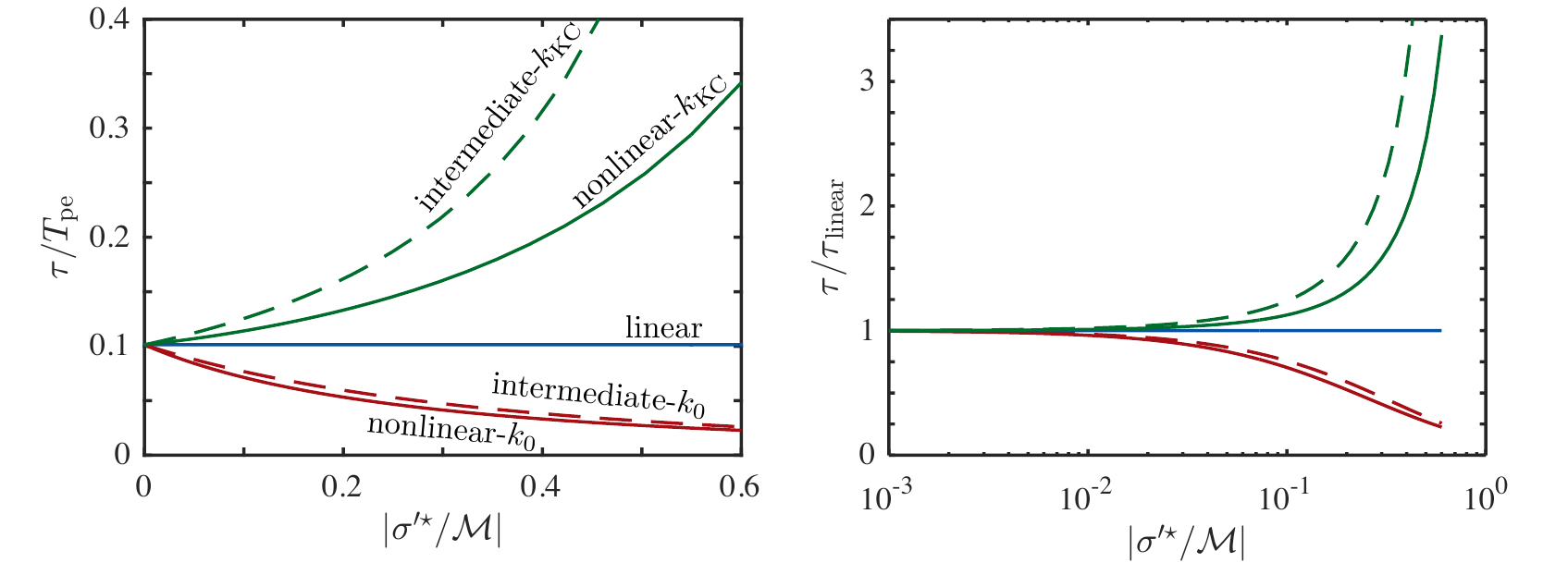}
    \caption{The consolidation time scale is constant and equal to $\pi^{-2}$ for the linear model, but depends strongly on the magnitude of the applied effective stress for the nonlinear models. Here, we plot (left) the time scale $\tau/T_\mathrm{pe}$ against the magnitude of the applied effective stress $|\sigma^{\prime\star}/\mathcal{M}|$ on a linear scale for the linear model (solid blue), the nonlinear-$k_0$ model (solid red), the nonlinear-$k_\mathrm{KC}$ model (solid green), the intermediate-$k_0$ model (dashed red), and the intermediate-$k_\mathrm{KC}$ model (dashed green). We also compare the relaxation time scales of all models with that of the linear model by plotting (right) $\tau/\tau_{\,\mathrm{linear}}$ against $|\sigma^{\prime\star}/\mathcal{M}|$ on a semilogarithmic scale. The nonlinear and intermediate models with constant permeability always relax much faster than the linear model, whereas those with deformation-dependent permeability always relax much more slowly than the linear model. These results are for $\phi_{f,0}=0.5$. \label{fig:mech_rect_dy_v_sig} }
\end{figure*}

We examine the rate of deformation in Fig.~\ref{fig:mech_rect_dy_v_t}. All three models relax exponentially toward their respective steady states, but the rate of relaxation depends very strongly on the magnitude of the applied effective stress and on the nonlinearities of the model. Specifically, the nonlinear-$k_0$ model relaxes much faster than the linear model, whereas the nonlinear-$k_\mathrm{KC}$ model relaxes much more slowly than the linear model. The relaxation time scale $\tau$, which is the characteristic time associated with the decaying exponentials shown in Fig.~\ref{fig:mech_rect_dy_v_t}, is constant for the linear model, but decreases with $|\sigma^{\prime\star}|$ for the nonlinear-$k_0$ model and increases strongly with $|\sigma^{\prime\star}|$ for the nonlinear-$k_\mathrm{KC}$ model~(Fig.~\ref{fig:mech_rect_dy_v_sig}). The time scales of the nonlinear models differ from that of the linear model by severalfold for moderate strain.

\section{Fluid-driven deformation}\label{s:flow_rect}

We now consider the uniaxial deformation of a porous material driven by a net fluid flow through the material from left to right (Fig.~\ref{fig:rect}c), which compresses the material against the rigid right boundary. This problem has attracted interest since the 1970s for applications in filtration and the manufacturing of composites~\citep[\textit{e.g.},][]{beavers-japplmech-1975, beavers-jfluidseng-1981a, beavers-jfluidseng-1981b, sommer-jfm-1996, preziosi-intjmultiphaseflow-1996}, in tissue mechanics~\citep[\textit{e.g.},][]{barry-jbiomech-1990, barry-intjnonlinmech-1991}, and as a convenient model problem in poroelasticity~\citep[\textit{e.g.},][]{parker-japplmech-1987, lanir-japplmech-1990, sobac-mecind-2011, hewitt-pre-2016}.

We assume that a pressure drop $\Delta{p}^\star$ is suddenly applied across the material at $t=0^+$, and we write this as $p(\delta,t)=\Delta{p}^\star$ and $p(L,t)=0$ without loss of generality. We also assume for simplicity that the left edge is unconstrained, $\sigma^\prime(\delta,t)=0$, but our models and solutions do not require this. Force balance then leads to $\sigma^\prime(L,t)=-\Delta{p}^\star$, implying that the right edge of the skeleton is compressed against the right boundary.

As in the consolidation problem, the deformation will evolve toward a state in which the solid is stationary. Unlike in the consolidation problem, fluid flow is central to this steady state because the flow drives the deformation. The resulting deformation field is highly nonuniform because it must balance the internal pressure gradient. As discussed in \S\ref{sss:rect_large_bcs} above, the same steady state can be achieved when the flow is instead driven by an imposed fluid flux $q^\star$; we focus on the case of an applied pressure drop here for simplicity, but our models and solutions are general and can also be used for the case of an imposed flux.

\subsection{Steady state}\label{ss:flow_rect_ss}

The deformation will eventually reach a state in which the flow is steady ($q(t)\to{}q^\star$ and $v_f(x,t)\to{}v_f(x)$) and the solid is stationary ($v_s\to{}0$ and $\phi_f{}v_f\to{}q^\star$). We present in Appendix~\ref{app:s:proc_steady_rect} a general procedure for constructing steady-state solutions to the kinematically exact model for arbitrary elasticity and permeability laws, and we provide the key results for the two nonlinear models and the two intermediate models in Appendix~\ref{app:s:ss_Integrals}. Below, we discuss the results for the nonlinear-$k_0$ model and the linear model.

For the nonlinear-$k_0$ model, the pressure and effective stress fields can be calculated by integrating Eq.~\eqref{eq:fluidmech_rect_vf} or \eqref{eq:fluidmech_rect_vs} with \eqref{eq:equilibrium_rect},
\begin{subequations}
    \begin{align}
        \frac{p(x)}{\mathcal{M}} &=\frac{\mu{}q^\star{}L}{k_0\mathcal{M}}\left(1-\frac{x}{L}\right), \\
        \frac{\sigma^{\prime}_{xx}(x)}{\mathcal{M}} &=-\frac{\mu{}q^\star{}L}{k_0\mathcal{M}}\left(\frac{x}{L}-\frac{\delta^\star}{L}\right).
    \end{align}
\end{subequations}
Since the permeability is constant, the pressure drops linearly from $p(\delta^\star)=\Delta{p}^\star$ to $p(L)=0$. The effective stress must then also vary linearly in $x$, rising in magnitude from $\sigma^\prime_{xx}(\delta^\star)=0$ to $\sigma^\prime_{xx}(L)=-\Delta{p}^\star$. The total stress is uniform and equal to $\sigma_{xx}(x)=\sigma_{xx}^\prime(x)-p(x)=-\Delta{p}^\star$, and this is supported entirely by the fluid at the left and entirely by the skeleton at the right.

The unknown flux $q^\star$ can be calculated directly from (see Appendix~\ref{app:s:proc_steady_rect})
\begin{equation}\label{eq:flow_rect_ss_large_J(L)_to_q}
	\frac{\mu{}q^\star{}L}{k_0\mathcal{M}} =\frac{1}{4}\left(\frac{1}{J(L)^2}-1\right)-\left(\frac{1}{2J(L)}\right)\,\frac{\ln{}J(L)}{J(L)},
\end{equation}
where $J(L)=(1-\phi_{f,0})/(1-\phi_f(L))$ is the Jacobian determinant at $x=L$, which is readily calculated by inverting $\sigma^\prime_{xx}\big(\phi_f(L)\big)=-\Delta{p}^\star$ using the elasticity law (Eq.~\ref{eq:sig_to_phi}). For an imposed flux, Eq.~\eqref{eq:flow_rect_ss_large_J(L)_to_q} should instead be solved for $J(L)$, which will then provide $\Delta{p}^\star$.

The unknown deflection $\delta^\star$ can then be calculated by evaluating the pressure at $x=\delta^\star$ or the effective stress at $x=L$, both of which lead to
\begin{equation}
    \frac{\delta^\star}{L} = 1-\left(\frac{\Delta{p}^\star}{\mathcal{M}}\right)\left(\frac{k_0\mathcal{M}}{\mu{}q^\star{}L}\right).
\end{equation}
We can then calculate the Jacobian determinant field $J(x)$ from the effective stress field using Eq.~\eqref{eq:sig_to_phi},
\begin{equation}
	J(x)=-\left[\frac{\mu{}q^\star{}L}{k_0\mathcal{M}}\left(\frac{x}{L}-\frac{\delta^\star}{L}\right)\right]^{-1}\mathrm{W}\left[-\frac{\mu{}q^\star{}L}{k_0\mathcal{M}}\left(\frac{x}{L}-\frac{\delta^\star}{L}\right)\right],
\end{equation}
where $\mathrm{W}(\,\cdot\,)$ is again the Lambert~W function. The porosity field $\phi_f(x)$ is again given by,
\begin{equation}
    \phi_f(x)=1-\frac{1-\phi_{f,0}}{J(x)}
\end{equation}
and, finally, the displacement field is
\begin{equation}
    \begin{split}
        \frac{u_s(x)}{L} =\frac{\delta^\star}{L}-\frac{k_0\mathcal{M}}{\mu{}q^\star{}L}\bigg[&\frac{1}{4}\left(\frac{1}{J(x)^2}-1\right) \\
        &-\frac{1}{2}\left(\frac{1}{J(x)}-2\right)\,\frac{\ln{}J(x)}{J(x)}\bigg].
    \end{split}
\end{equation}

The linear model is, of course, much simpler. The pressure and effective stress fields are similar to those for the nonlinear-$k_0$ model,
\begin{subequations}
    \begin{align}
        \frac{p(x)}{\mathcal{M}} &\approx{}\frac{\mu{}q^\star{}L}{k_0\mathcal{M}}\left(1-\frac{x}{L}\right), \\
        \frac{\sigma^{\prime}_{xx}(x)}{\mathcal{M}} &\approx{}-\frac{\mu{}q^\star{}L}{k_0\mathcal{M}}\,\left(\frac{x}{L}\right).
    \end{align}
\end{subequations}
Evaluating the pressure at $x\approx{}0$ or the effective stress at $x=L$ immediately provides the relationship between the flux and the pressure drop,
\begin{equation}
    \frac{\mu{}q^\star{}L}{k_0\mathcal{M}} \approx{}\frac{\Delta{p}^\star}{\mathcal{M}}.
\end{equation}
The porosity field is calculated from the effective stress field and linear elasticity,
\begin{equation}
    \frac{\phi_f(x)-\phi_{f,0}}{1-\phi_{f,0}} \approx{}-\frac{\mu{}q^\star{}L}{k_0\mathcal{M}}\,\left(\frac{x}{L}\right),
\end{equation}
and the displacement field is calculated by integrating the porosity field,
\begin{equation}
    \frac{u_s(x)}{L} \approx{}\frac{\mu{}q^\star{}L}{2k_0\mathcal{M}}\left(1-\frac{x}{L}\right)^2.
\end{equation}
Since the stress and the strain increase linearly from left to right, the displacement is quadratic. Finally, the deflection is simply given by $\delta^\star\approx{}u_s(0)$,
\begin{equation}
    \frac{\delta^\star}{L} \approx{}\frac{1}{2}\,\frac{\mu{}q^\star{}L}{k_0\mathcal{M}}.
\end{equation}

\begin{figure*}
    \centering
    \includegraphics[width=17.2cm]{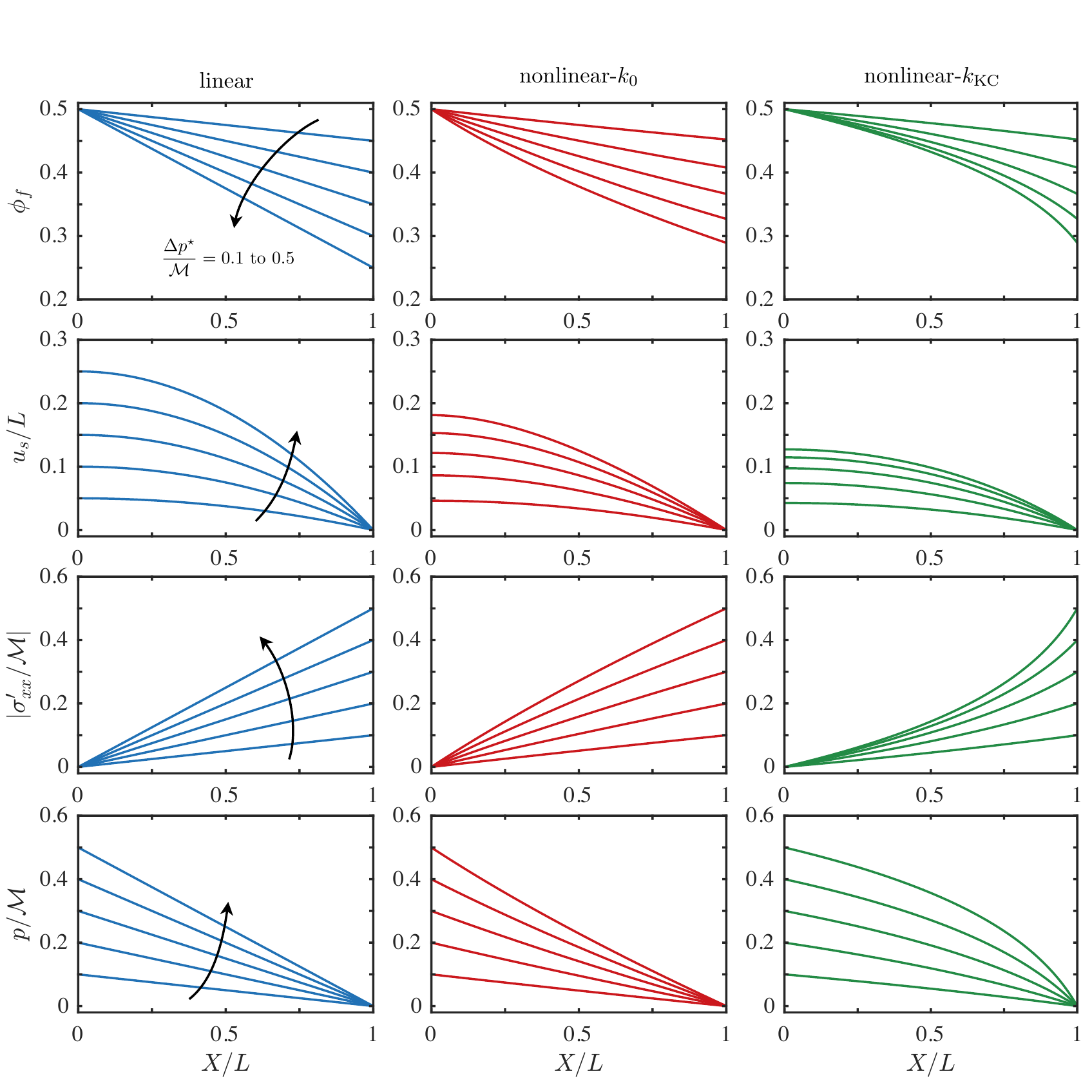}
    \caption{Steady-state in fluid-driven deformation of a soft porous material, where fluid flow through the material from left to right is driven by an imposed pressure drop $\Delta{p}^\star$. Here, we show the results for $\Delta{p}^\star/\mathcal{M}$ increasing from $0.1$ to $0.5$, as indicated. We show the porosity (first row), displacement (second row), effective stress (third row), and pressure (last row) for the linear model (left column, blue), the nonlinear-$k_0$ model (middle column, red), and the nonlinear-$k_\mathrm{KC}$ model (right column, green). For the nonlinear models, we again plot these results against the Lagrangian coordinate $X=x-u_s(x,t)$ for clarity; for the linear model, we again adopt a Lagrangian interpretation of the spatial coordinate. The nonlinear models deform less than the linear model in all cases, with the nonlinear-$k_\mathrm{KC}$ model deforming the least but exhibiting the most strongly nonlinear behavior. These results are for $\phi_{f,0}=0.5$. \label{fig:flow_rect_ss_v_x} }
\end{figure*}

\begin{figure*}
    \centering
    \includegraphics[width=17.2cm]{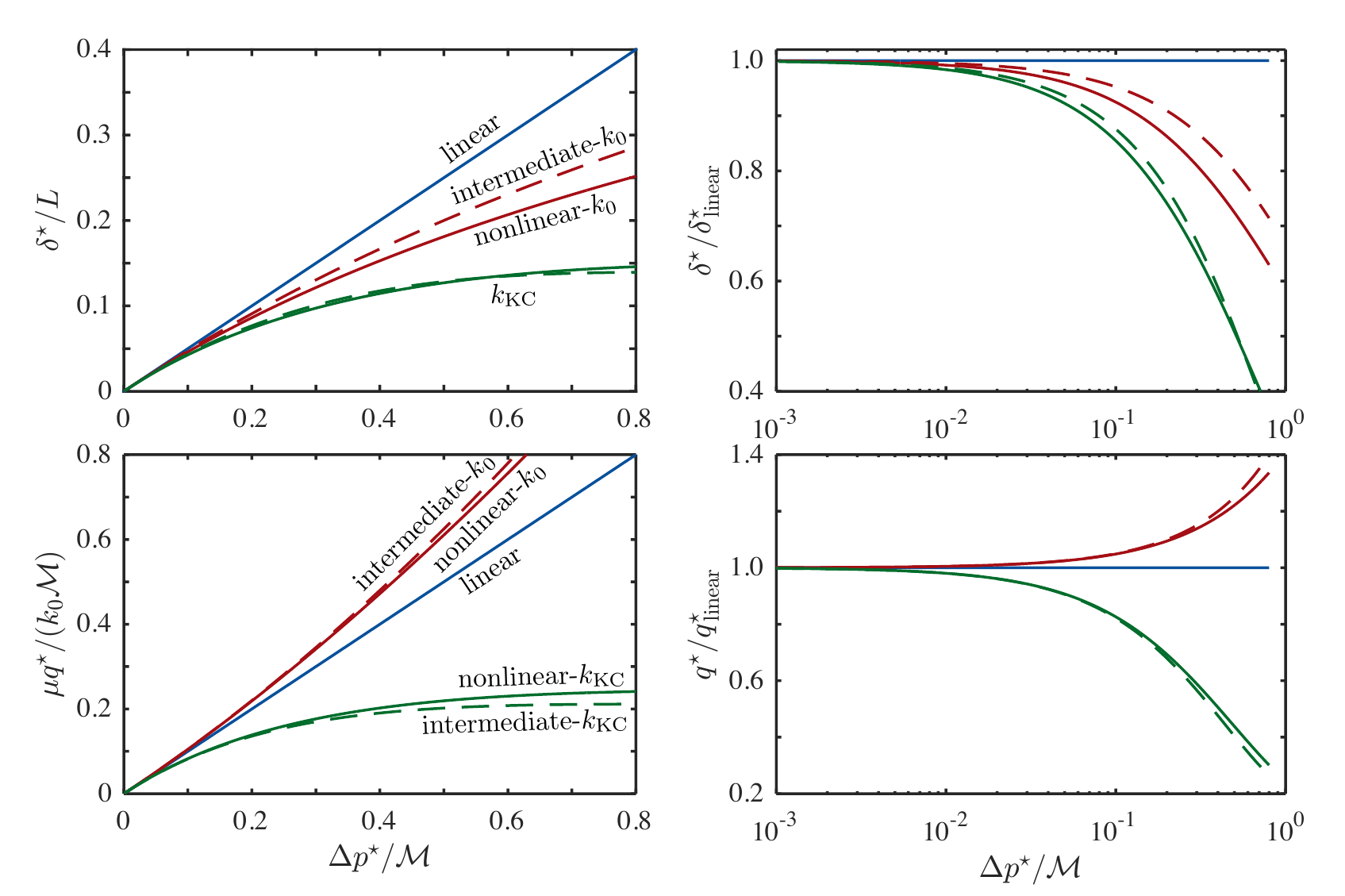}
    \caption{The linear model overpredicts the steady-state deformation relative to the nonlinear models during fluid-driven deformation. For flow driven by an applied pressure drop $\Delta{p}^\star$, we plot here the steady-state deflection $\delta^\star$ (top row) and the steady-state fluid flux $q^\star$ (bottom row) against $\Delta{p}^\star$ on a linear scale (left column) for the linear model (solid blue), the nonlinear-$k_0$ model (solid red), the nonlinear-$k_\mathrm{KC}$ model (solid green), the intermediate-$k_0$ model (dashed red), and the intermediate-$k_\mathrm{KC}$ model (dashed green). We also compare (right) the deflection and flux for all models with that of the linear model, $\delta^\star/\delta_{\,\mathrm{linear}}$ and $q^\star/q_{\,\mathrm{linear}}$, on a semilogarithmic scale. The nonlinear-$k_0$ and intermediate-$k_0$ models predict a somewhat higher flux than the linear model due to kinematic nonlinearity, but the nonlinear-$k_\mathrm{KC}$ and intermediate-$k_\mathrm{KC}$ models predict a much lower flux than the linear model because the permeability decreases strongly as the pressure drop increases, leading to a much lower flux for a given pressure drop. \label{fig:flow_rect_ss_v_dp} }
\end{figure*}

We compare these predictions qualitatively in Fig.~\ref{fig:flow_rect_ss_v_x}, including also the results for the nonlinear-$k_\mathrm{KC}$ model.
As with consolidation, the nonlinear models deform less than the linear model in all cases. Unlike with consolidation, the permeability law has a strong impact on the steady state: The nonlinear-$k_\mathrm{KC}$ model deforms less than the nonlinear-$k_0$ model, and exhibits more strongly nonlinear behavior. We compare the predictions for the final deflection $\delta^\star$ and the resulting flux $q^\star$ in Fig.~\ref{fig:flow_rect_ss_v_dp}, including also the two intermediate models.
Although all of the nonlinear and intermediate models predict a much smaller deflection than the linear model, the nonlinear-$k_0$ and intermediate-$k_0$ models predict a larger steady-state flux than the linear model, whereas the nonlinear-$k_\mathrm{KC}$ and intermediate-$k_\mathrm{KC}$ models predict a much smaller steady-state flux. This occurs because the steady-state flux results from two competing physical effects. As the driving pressure drop increases, we expect the deflection to increase. As the deflection increases, the overall length of the skeleton decreases and, since the pressure drop is fixed, the pressure gradient across the material increases. As a result, we expect from Darcy's law that the flux will scale like $q^\star\sim{}(k/\mu)\Delta{p}^\star/(L-\delta^\star)$. For constant permeability, we then expect the flux to increase faster than linearly with $\Delta{p}^\star$, and this is indeed what we see for the nonlinear-$k_0$ and intermediate-$k_0$ cases. The changing length is a kinematic nonlinearity that is neglected in the linear model, so $q^\star_{\,\mathrm{linear}}$ is simply proportional to $\Delta{p}^\star$ despite the fact that $\delta^\star_{\,\mathrm{linear}}$ is actually larger than the nonlinear or intermediate predictions. However, these models ignore the fact that the porosity decreases as the deformation increases. When the permeability is deformation dependent, this decreases very strongly with the porosity and overwhelms the effect of the changing length, leading to a strongly slower-than-linear growth of $q^\star$ with $\Delta{p}^\star$, and this is indeed what we see for the nonlinear-$k_\mathrm{KC}$ and intermediate-$k_\mathrm{KC}$ models.

\subsubsection{Dynamics}\label{sss:flow_rect_dynamics}

\begin{figure*}
    \centering
    \includegraphics[width=17.2cm]{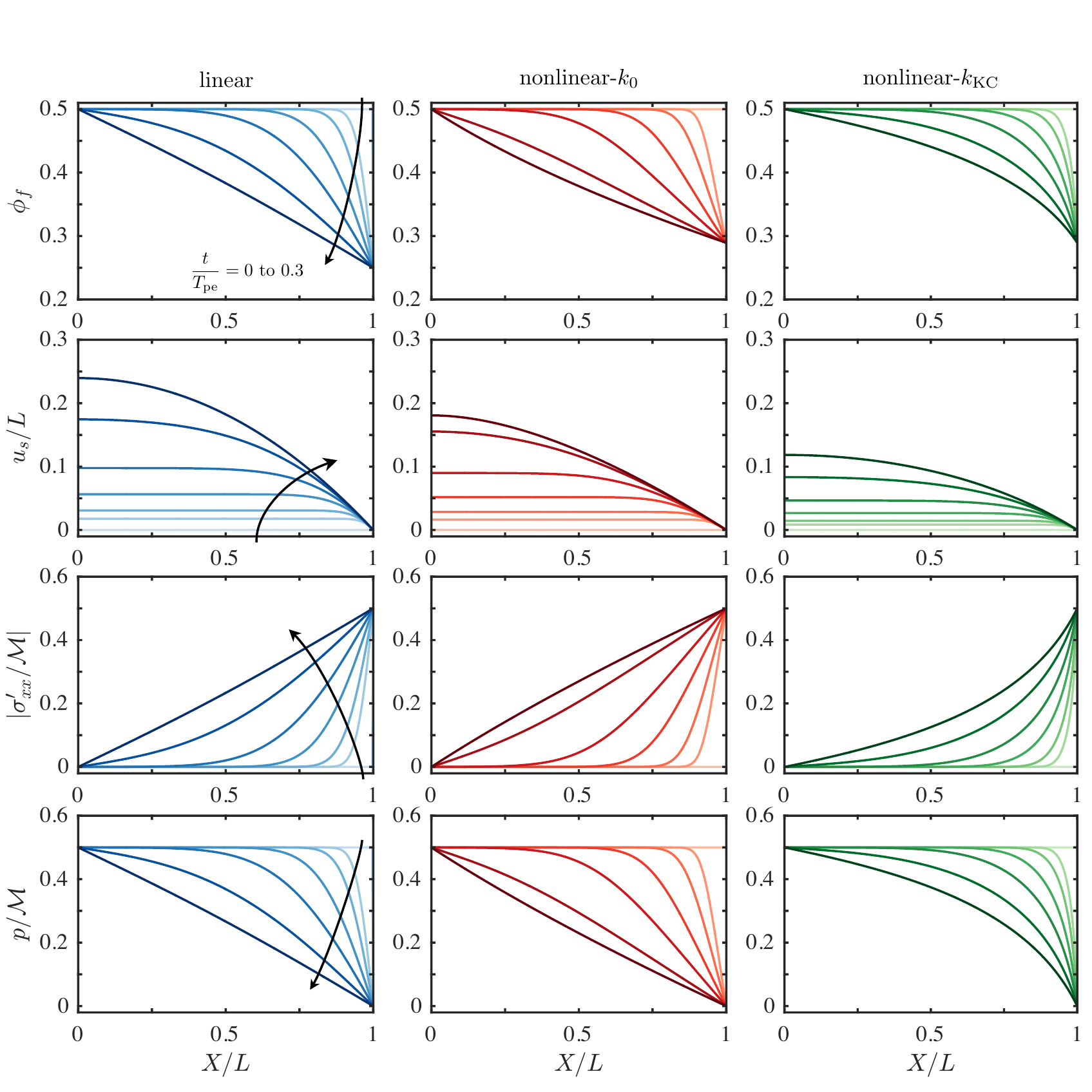}
    \caption{Dynamics of fluid-driven deformation for a soft porous material, where the net flow from left to right is driven by an applied pressure drop $\Delta{p}^\star=0.5$. We show the porosity (first row), displacement (second row), effective stress (third row), and pressure (last row) at $t/T_\mathrm{pe}=0$, 0.001, 0.003, 0.01, 0.03, 0.1, and 0.3, as indicated (light to dark colors), for the linear model (left column, blue), the nonlinear-$k_0$ model (middle column, red), and the nonlinear-$k_\mathrm{KC}$ model (right column, green). For the nonlinear models, we plot these results against the Lagrangian coordinate $X=x-u_s(x,t)$ for clarity; for the linear model, we again adopt a Lagrangian interpretation and simply replace $x$ with $X$ in the relevant expressions (see \S\ref{ss:LPE_Discussion} and \S\ref{ss:mech_rect_dy}). These results are for $\phi_{f,0}=0.5$. \label{fig:flow_rect_dy_v_x} }
\end{figure*}

We next focus on the dynamic evolution of the deformation. We again solve the nonlinear and intermediate models numerically (Appendix~\ref{app:s:FV_rect} and \cite{see_supp}), and we again solve the linear model analytically via separation of variables. The analytical solution can be written
\begin{widetext}
    \begin{subequations}
        \begin{align}
            \phi_f(x,t) &= \phi_{f,0} + (\phi_f^\star-\phi_{f,0})\left\{\frac{x}{L} +\sum_{n=1}^\infty\,\frac{2}{n\pi}\Big[(-1)^n\Big]e^{-\frac{(n\pi)^2t}{T_\mathrm{pe}}}\sin\left(\frac{n\pi{}x}{L}\right)\right\}, \\
            \frac{u_s(x,t)}{L} &=-\left(\frac{\phi_f^\star-\phi_{f,0}}{1-\phi_{f,0}}\right)\left\{\frac{1}{2}\left[1-\left(\frac{x}{L}\right)^2\right]-\sum_{n=1}^\infty \frac{2}{(n\pi)^2}\Big[(-1)^n\Big]e^{-\frac{(n\pi)^2t}{T_\mathrm{pe}}}\left[(-1)^n-\cos\left(\frac{n\pi{}x}{L}\right)\right]  \right\},
        \end{align}
    \end{subequations}
\end{widetext}
where $\phi_f(0,t)=\phi_{f,0}$ and $\phi_f(L,t)=\phi_f^\star=\phi_{f,0}-(1-\phi_{f,0})(\Delta{p}^\star/\mathcal{M})$. We compare these solutions qualitatively in Fig.~\ref{fig:flow_rect_dy_v_x}, including also the results for the nonlinear-$k_\mathrm{KC}$ case. Note once again that the spatial coordinate in the linear model is ambiguous, and we again adopt a Lagrangian interpretation.

When the flow starts, the fluid and the solid initially travel together to the right. The pressure remains uniform throughout most of the skeleton since there is no net flux of fluid \textit{through} the skeleton, but there is a very sharp pressure gradient at the right edge where the solid is necessarily stationary. The motion of the solid toward the right boundary gradually compresses the right edge of the skeleton against the boundary, and this motion slows over time as the effective stress builds from right to left. The motion of the solid eventually stops and the deformation reaches steady state when the strain in the skeleton is such that the gradient in effective stress balances the gradient in pressure. In this steady state, the skeleton remains completely relaxed at the left edge and is the most compressed at the right edge. From left to right, there is a gradual increase in deformation and magnitude of effective stress, and a gradual decrease in pressure and porosity.

Both here and in the consolidation problem, the deformation evolves with a classic boundary-layer structure that may be susceptible to a matched asymptotic approach with $t/T_\mathrm{pe}$ the small parameter. The prospect of more accurately capturing the kinematic nonlinearity while retaining some degree of analytical tractability is a promising one for future work.

\begin{figure*}
    \centering
    \includegraphics[width=17.2cm]{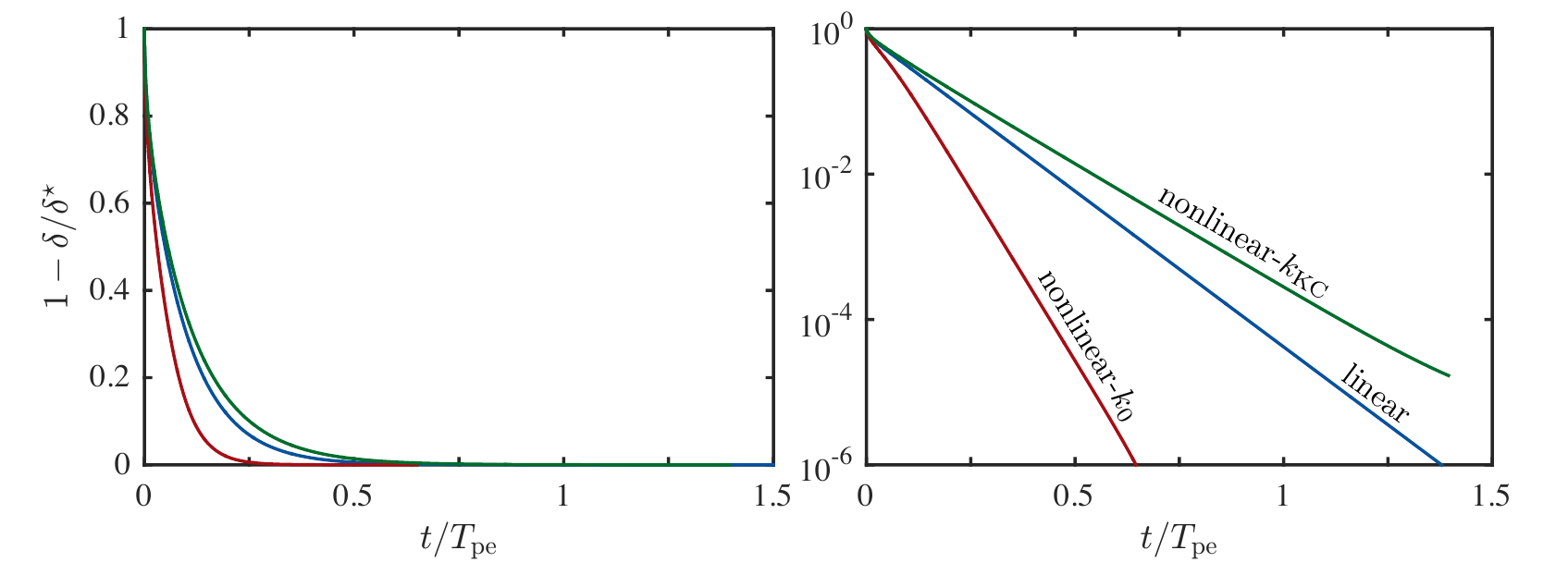}
    \caption{Relaxation toward steady state during fluid-driven deformation with $\Delta{p}^\star/\mathcal{M}=0.5$ on a linear scale (left) and on a semilogarithmic scale (right). We plot the relative difference between $\delta(t)$ and its final value $\delta^\star$ as a proxy for the global approach to steady state. We show again the linear (blue), nonlinear-$k_0$ (red), and nonlinear-$k_\mathrm{KC}$ (green) models. The final deflection $\delta^\star$ is largest for the linear model, followed by the nonlinear-$k_0$ model, followed by the nonlinear-$k_\mathrm{KC}$ model ($\delta^\star=0.25$, 0.181, and 0.127, respectively; \textit{c.f.}, Fig.~\ref{fig:flow_rect_ss_v_dp}). As with consolidation, however, the nonlinear-$k_0$ model relaxes more quickly than linear model (at more than twice the rate), whereas the nonlinear-$k_\mathrm{KC}$ model relaxes more slowly than the linear model (at about 80\% of the rate; \textit{c.f.}, Fig.~\ref{fig:mech_rect_dy_v_t}). These results are for $\phi_{f,0}=0.5$. \label{fig:flow_rect_dy_v_t} }
\end{figure*}

\begin{figure*}
    \centering
    \includegraphics[width=17.2cm]{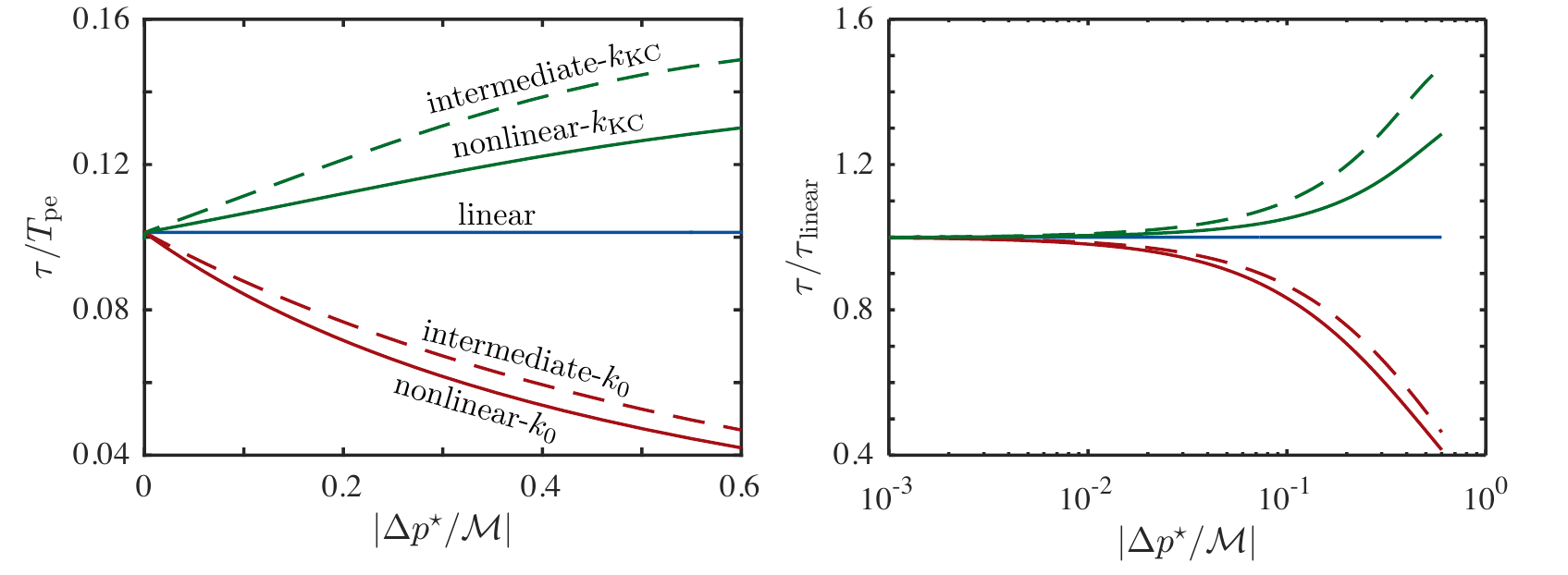}
    \caption{The deformation approaches steady state more quickly for larger $\Delta{p}^\star$ when the permeability is constant, and more slowly for larger $\Delta{p}^\star$ when the permeability is deformation dependent. Here, we plot (left) the relaxation time scale $\tau/T_\mathrm{pe}$ against $\Delta{p}^\star$ for the linear (solid blue), nonlinear-$k_0$ (solid red), nonlinear-$k_\mathrm{KC}$ (solid green), intermediate-$k_0$ (dashed red), and intermediate-$k_\mathrm{KC}$ (dashed green) models. The time scale is constant and equal to $\pi^{-2}$ for the linear model, but increases with $\Delta{p}^\star$ for the two $k_0$ models and decreases with $\Delta{p}^\star$ for the two $k_\mathrm{KC}$ models. We also compare the relaxation time scales of all models with that of the linear model by plotting (right) $\tau/\tau_{\,\mathrm{linear}}$ on a semilogarithmic scale. These results are for $\phi_{f,0}=0.5$. \label{fig:flow_rect_dy_v_dp} }
\end{figure*}

To examine the time scale of the deformation, we plot the evolution of the deflection toward its final value as a proxy for the global approach to steady state (Fig.~\ref{fig:flow_rect_dy_v_t}).
As for consolidation, we find that the deflection approaches steady state exponentially in all cases, and that the nonlinear-$k_0$ and nonlinear-$k_\mathrm{KC}$ models evolve more quickly and more slowly than the linear model, respectively. We also investigate the impact of $\Delta{p}^\star$ on the time scale (Fig.~\ref{fig:flow_rect_dy_v_dp}).
We find that the general trend is the same as in consolidation (\textit{c.f.}, Fig.~\ref{fig:mech_rect_dy_v_sig}), but that the magnitude of the effect is smaller---That is, the time scale during fluid-driven deformation depends less strongly on $\Delta{p}^\star$ than the time scale during consolidation depends on $\sigma^{\prime\star}$. This is most likely due to the fact that the steady state is uniform in consolidation, with potentially large compression throughout the entire material, whereas the steady state in fluid-driven deformation is highly nonuniform, with completely relaxed material at the left and highly compressed material at the right.

\section{Discussion \& Conclusions}

We have provided an overview and discussion of a complete Eulerian framework for the arbitrarily large deformation of a porous material. In doing so, our main goals were to (a)~elucidate the key aspects of the rigorous model, (b)~provide physical insight into the subtleties of poromechanical coupling, and (c)~investigate the qualitative and quantitative nature of the error introduced by linearizing this model. These points are often obscured by the powerful mathematical and computational machinery that is typically brought to bear on these problems. We intend that our approach here can serve as a concise, coherent, and approachable introduction to a large body of work in classical continuum and poromechanics. We believe that this overview now provides a rostrum to facilitate further theoretical advances and new applications in soil mechanics, hydrogeology, biophysics, and biomedical engineering.

We have also applied this theory to two canonical model problems in uniaxial deformation, one in which deformation drives fluid flow and one in which fluid flow drives deformation. In the former, the consolidation problem, an applied effective stress squeezes fluid from a porous material. Although the steady state is simple and controlled entirely by the solid mechanics, the evolution of the deformation is controlled by the rate at which fluid can flow through the material and out at the boundaries; we showed that the resulting rate of relaxation is impacted strongly by kinematic nonlinearity and even more strongly by deformation-dependent permeability. In the latter problem, fluid-driven deformation, a net throughflow compresses the material against a rigid permeable boundary. The steady state is highly nonuniform, controlled by the steady balance between the gradient in pressure and the gradient in stress. We showed that both the evolution of the deformation and the deflection and fluid flux at steady state are impacted strongly by kinematic nonlinearity and, again, even more strongly by deformation-dependent permeability.

In the interest of emphasizing the nonlinear kinematics of large deformations, we have avoided complex, material-specific constitutive models. Hencky elasticity captures the full kinematic nonlinearity of large deformations in a very simple form, and we believe that it provides a reasonable compromise between rigor and complexity for moderate deformations. However, real materials will always behave in a complex, material-specific way when subject to sufficiently large strains, and the framework considered here is fully compatible with other constitutive models. Similarly, we have considered one specific case of deformation-dependent permeability: The normalized Kozeny-Carman formula. We have shown that this typically amplifies the importance of kinematic nonlinearity and has striking qualitative and quantitative impacts on poromechanical behavior. Although this example captures the key qualitative features of the coupling between deformation and permeability, material-specific relationships will be needed to provide quantitative predictions for real materials.

In describing the kinematics of the solid skeleton, we have adopted the single assumption that the constituent material is incompressible. This has clear relevance to soil mechanics, biophysics, and any other situation where the pressure and stress are small compared to the bulk modulus of the solid grains (\textit{e.g.}, about 30--40~GPa for quartz sand). This assumption can be relaxed, although doing so substantially complicates the large-deformation theory~\citep[\textit{e.g.,}][]{coussy-wiley-2004, bennethum-jengmech-2006, gajo-rspa-2010}. We have also focused on the case of a single, incompressible pore fluid, but the theory is readily generalized to a compressible or multiphase fluid system~\citep[\textit{e.g.},][]{coussy-wiley-2004, uzuoka-intjnag-2011}.

Uniaxial deformations have provided a convenient testbed for our purposes here, but they are unusual in several respects that do not readily generalize to multiaxial scenarios. First, a uniaxial deformation can be fully characterized by the change in porosity, $\sigma_{xx}^\prime=\sigma_{xx}^\prime(\phi_f)$; this simplifies the analysis, but it is not the case for even simple biaxial deformations. Second, the cross section normal to the flow does not deform or rotate, which greatly simplifies the nonlinearity of poromechanical coupling. Finally, the exact relationship between displacement and porosity is linear; this is again not the case for even simple biaxial deformations. We expect kinematic nonlinearity to play an even stronger role for multiaxial deformations.



The authors gratefully acknowledge support from the Yale Climate \&~Energy Institute. ERD acknowledges support from the National Science Foundation (Grant No.~CBET-1236086). JSW acknowledges support from Yale University, the Swedish Research Council (Vetenskapsr\r{a}det grant 638-2013-9243), and a Royal Society Wolfson Research Merit Award.

\appendix
\section{Large-deformation poroelasticity in a Lagrangian framework}\label{app:s:Lagrangian}

Here we briefly summarize the Lagrangian approach to large-deformation poroelasticity, a thorough discussion and derivation of which is provided by \citet{coussy-wiley-2004}. In a Lagrangian frame, it is natural to work with so-called \textit{nominal} quantities, which measure the current stresses, fluxes, \textit{etc.} acting on or through the reference areas or volumes. For example, the nominal porosity $\Phi_f$ measures the current fluid volume per unit reference total volume, and is related to the true porosity via $\Phi_f=J\phi_f$. We denote the gradient and divergence operators in the Lagrangian coordinate system by $\mathrm{grad}(\cdot)$ and $\mathrm{div}(\cdot)$, respectively, to distinguish them from the corresponding operators in the Eulerian coordinate system. The Lagrangian displacement field is
\begin{equation}
    \mathbf{U}_s=\mathbf{x}(\mathbf{X},t)-\mathbf{X},
\end{equation}
where $\mathbf{X}$ is the Lagrangian (material) coordinate and $\mathbf{x}(\mathbf{X},t)$ is the current position of the skeleton that was initially at position $\mathbf{X}$. The corresponding deformation-gradient tensor is
\begin{equation}
    \mathbf{F} =\mathrm{grad}(\mathbf{x}) =\mathbf{I}+\mathrm{grad}(\mathbf{U}_s).
\end{equation}
The Jacobian determinant is then related to $\Phi_f$ by
\begin{equation}
    J = \det{\left(\mathbf{F}\right)}=1+\Phi_f-\Phi_{f,0},
\end{equation}
where $\Phi_{f,0}(\mathbf{X})$ is the reference porosity field, which we again take to be undeformed. Continuity requires that
\begin{equation}\label{eq:ContinuityLL}
    \frac{\partial{\Phi_f}}{\partial{t}} +\mathrm{div}\left(\mathbf{W}_f\right)=0,
\end{equation}
where $\mathbf{W}_f$ is the nominal flux of fluid through the solid skeleton. The nominal flux is related to the pressure gradient via Darcy's law,
\begin{equation}\label{eq:DarcyLL}
    \mathbf{W}_f=-J\mathbf{F}^{-1}\mathbf{F}^{-\mathsf{T}} \cdot\frac{k(\phi_f)}{\mu}\,\mathrm{grad}(p),
\end{equation}
where the $J\mathbf{F}^{-1}$ portion of the prefactor converts the true flux to the nominal flux, and the remaining factor of $\mathbf{F}^{-\mathsf{T}}$ converts the Eulerian gradient to the Lagrangian one. Mechanical equilibrium requires that
\begin{equation}
    \mathrm{div}(\mathbf{s})=0,
\end{equation}
where $\mathbf{s}$ is the nominal total stress, which is related to the true total stress via
\begin{equation}
    \mathbf{s}=J\boldsymbol{\sigma}\,\mathbf{F}^{-\mathsf{T}}.
\end{equation}
The nominal effective stress $\mathbf{s}^\prime$ is then given by
\begin{equation}\label{eq:StressLL}
    \mathbf{s}^\prime=\mathbf{s}+J\mathbf{F}^{-\mathsf{T}}p.
\end{equation}
Combining Eqs.~\eqref{eq:ContinuityLL}--\eqref{eq:StressLL}, we finally have
\begin{subequations}\label{eq:Lagrangian}
    \begin{align}
        \frac{\partial{\Phi_f}}{\partial{t}} -\mathrm{div}\left[J\mathbf{F}^{-1}\mathbf{F}^{-\mathsf{T}}\cdot\frac{k(\phi_f)}{\mu}\,\mathrm{grad}(p)\right] &=0 \quad\mathrm{and} \\
        \mathrm{div}(\mathbf{s}^\prime) &=\mathrm{div}(J\mathbf{F}^{-\mathsf{T}}p).
    \end{align}
\end{subequations}
Supplemented with a constitutive law for the solid skeleton (relating $\mathbf{s}^\prime$ to $\mathbf{U}_s$) and appropriate boundary conditions, Eqs.~\eqref{eq:Lagrangian} constitute a complete formulation of poroelasticity in a Lagrangian framework~\citep{coussy-wiley-2004}. The Lagrangian formulation is more suitable for computation than the Eulerian formulation since the domain is fixed, but the underlying physical structure is substantially more opaque. Note that the permeability must remain a function of the true porosity, $k=k(\phi_f)=k(\Phi_f/J)$.

Linearizing Eqs.~\eqref{eq:Lagrangian} in the strain and reverting from the nominal porosity to the true porosity leads to 
\begin{subequations}
    \begin{align}
        \frac{\partial{\phi_f}}{\partial{t}} -\mathrm{div}\left[(1-\phi_{f,0})\frac{k_0}{\mu}\,\mathrm{grad}(p)\right] &\approx{}0 \quad\mathrm{and} \\
        \mathrm{div}(\mathbf{\sigma}^\prime) &\approx{}\mathrm{grad}(p),
    \end{align}
\end{subequations}
which coincide with Eqs.~\eqref{eq:LinearDiffusionEulerian} and \eqref{eq:equilibrium}, respectively, but replacing $\mathbf{x}$ with $\mathbf{X}$. Note that the nominal porosity and the true porosity differ \textit{at leading order}:
\begin{equation}
    \Phi_f-\Phi_{f,0}=\frac{\phi_f-\phi_{f,0}}{1-\phi_f} \approx{}\frac{\phi_f-\phi_{f,0}}{1-\phi_{f,0}},
\end{equation}
where the reference fields are always precisely equivalent, $\Phi_{f,0}\equiv\phi_{f,0}$, because they must necessarily refer to the same reference state.

\section{Eulerian and Lagrangian interpretations of linear elasticity}\label{app:s:strainambiguity}

The Eulerian (Eulerian-Almansi) and Lagrangian (Green-Lagrange) finite-strain tensors are
\begin{equation}
    \mathbf{e} =\frac{1}{2}\left(\mathbf{I}-\mathbf{B}^{-1}\right) =\frac{1}{2}\left(\mathbf{I}-\mathbf{F}^{-\mathsf{T}}\mathbf{F}^{-1}\right) 
\end{equation}
and
\begin{equation}
    \mathbf{E} =\frac{1}{2}\left(\mathbf{C}-\mathbf{I}\right) =\frac{1}{2}\left(\mathbf{F}^{\mathsf{T}}\mathbf{F}-\mathbf{I}\right),
\end{equation}
respectively. Linear elasticity, as described above, is effectively a linearized Eulerian constitutive law, where stress is linear in the Eulerian infinitesimal strain
\begin{equation}\label{eq:LinearizedEulerianElasticity}
    \boldsymbol{\varepsilon}_e =\frac{1}{2}\left[\frac{\partial{\mathbf{u}_s}}{\partial{\mathbf{x}}} +\left(\frac{\partial{\mathbf{u}_s}}{\partial{\mathbf{x}}}\right)^{\mathsf{T}}\right]=\mathbf{I}-\frac{1}{2}\big(\mathbf{F}^{-1}+\mathbf{F}^{-\mathsf{T}}\big).
\end{equation}
However, it is equally valid to write a linearized Lagrangian constitutive law, where stress is linear in the Lagrangian infinitesimal strain
\begin{equation}\label{eq:LinearizedLagrangianElasticity}
    \boldsymbol{\varepsilon}_E =\frac{1}{2}\left[\frac{\partial{\mathbf{u}_s}}{\partial{\mathbf{X}}} +\left(\frac{\partial{\mathbf{u}_s}}{\partial{\mathbf{X}}}\right)^{\mathsf{T}}\right]=\frac{1}{2}\big(\mathbf{F}+\mathbf{F}^{\mathsf{T}}\big)-\mathbf{I}.
\end{equation}
The former quantity is nonlinear in a Lagrangian frame whereas the latter quantity is nonlinear in an Eulerian frame. We used the linearized Eulerian law above, but in a Lagrangian frame it would be more appropriate to use the linearized Lagrangian law. The results are equivalent at leading order in the strain ($\partial{u}/\partial{x}\approx\partial{u}/\partial{X}$), but they diverge as strains become non-negligible.

\section{Finite-volume method with a moving boundary}\label{app:s:FV_rect}

To solve Eq.~\eqref{eq:fluidmech_rect_pde} numerically, we formulate a finite-volume method on an adaptive grid. We provide a reference implementation in the Supplemental~Material~\cite{see_supp}. At any time $t$, the domain extends from $x=\delta(t)$ to $x=L$. We divide this domain into $N$ cells of equal width $\Delta{x}(t)=[L-\delta(t)]/N$, where cell $i$ has center $x_i(t)=\delta(t)+(i-1/2)\Delta{x}(t)$ and we denote its left and right edges by $x_{i-1/2}(t)=x_i(t)-\Delta{x}(t)/2$ and $x_{i+1/2}(t)=x_i(t)+\Delta{x}(t)/2$, respectively. Making use of the expressions
\begin{subequations}
    \begin{align}
        \frac{\partial}{\partial{t}}\,\Delta{x} &=-\frac{1}{N}\frac{\mathrm{d}\delta}{\mathrm{d}t} =-\frac{\Delta{x}}{L-\delta}\,\dot{\delta} \quad\mathrm{and} \\
        \frac{\partial}{\partial{t}}\,x_i &=\frac{L-x_i}{L-\delta}\,\dot{\delta},
    \end{align}
\end{subequations}
we formulate a finite-volume method in the standard way~\citep[\textit{e.g.},][]{leveque-cambridge-2004} by integrating Eq.~\eqref{eq:fluidmech_rect_pde} over cell $i$,
\begin{equation}
    \int_{x_{i-1/2}}^{x_{i+{1/2}}}\,\left\{ \frac{\partial{\phi_f}}{\partial{t}}+ \frac{\partial}{\partial{x}}\bigg[\phi_fv_f\bigg]\right\}\,\mathrm{d}x=0,
\end{equation}
where
\begin{equation}
    \phi_fv_f=\phi_f\,q(t)-(1-\phi_f)\frac{k(\phi_f)}{\mu}\,\frac{\partial{p}}{\partial{x}},
\end{equation}
as derived above. We arrive at
\begin{equation}
    \int_{x_{i-1/2}}^{x_{i+{1/2}}}\,\frac{\partial{\phi_f}}{\partial{t}}\,\mathrm{d}x +\bigg[\phi_fv_f\bigg]\bigg|_{x_{i-1/2}}^{x_{i+1/2}}=0.
\end{equation}
After manipulating the first term using the Leibniz integral rule and regrouping, we have
\begin{equation}
    \frac{\partial}{\partial{t}}\,\int_{x_{i-1/2}}^{x_{i+1/2}}\,\phi_f\,\mathrm{d}x +\bigg[-\left(\frac{L-x}{L-\delta}\right)\phi_f\,\dot{\delta} +\phi_fv_f\bigg]\bigg|_{x_{i-1/2}}^{x_{i+1/2}}=0.
\end{equation}
Defining $\phi_{f,i}$ to be the average of $\phi_f$ within cell $i$,
\begin{equation}
    \phi_{f,i}\equiv{}\frac{1}{\Delta{x}}\,\int_{x_{i-1/2}}^{x_{i+1/2}}\,\phi_f\,\mathrm{d}x,
\end{equation}
we finally have
\begin{equation}\label{eq:fluidmech_rect_pde_fv}
    \begin{split}
        \frac{\partial{\phi_{f,i}}}{\partial{t}} &-\left(\frac{\dot{\delta}}{L-\delta}\right)\phi_{f,i} \\
        &+\frac{1}{\Delta{x}}\,\bigg[-\left(\frac{L-x}{L-\delta}\right)\phi_f\,\dot{\delta} +\phi_fv_f\bigg]\bigg|_{x_{i-1/2}}^{x_{i+1/2}}=0.
    \end{split}
\end{equation}
We discretize the quantity in square brackets using upwinding for the advective components and central differencing for the diffusive components. Simultaneously, we must also solve an evolution equation for the position of the moving boundary. This comes from Eq.~\eqref{eq:fluidmech_rect_vs} and \eqref{eq:vs_bc},
\begin{equation}
    \dot{\delta} =q(t)+\left(\frac{k(\phi_f)}{\mu}\frac{\partial{p}}{\partial{x}}\right)\bigg|_{x=\delta}.
\end{equation}
This system can be written directly in terms of the porosity once a stress-strain constitutive law is specified, at which point we have a closed set of equations in $\phi_f$ and $\delta$. We integrate these equations in time using an explicit Runge-Kutta scheme.

\section{Steady-state solutions: General procedure}\label{app:s:proc_steady_rect}

At steady state, it is possible to construct (usually implicit) analytical solutions for any combination of elasticity and permeability law. We outline the general procedure below and provide a reference implementation in the Supplemental Material~\cite{see_supp}. \citet{barry-jaustralmathsocb-1993} suggested a somewhat similar procedure for axisymmetric geometries \citep[\textit{c.f.}, \S5 of Ref.][]{barry-jaustralmathsocb-1993}.

Of the four quantities $\sigma^{\prime\star}$, $q^\star$, $\Delta{p}^\star$, and $\delta^\star$, two must be known in advance. We assume here that these are $\sigma^{\prime\star}$ and either $q^\star$ or $\Delta{p}^\star$, but it is straightforward to adapt or invert this procedure for other pairs. This procedure degenerates when there is no flow at steady state, $q^\star=0\,\leftrightarrow\,\Delta{p}^\star=0$, as in consolidation. The solution then depends only on the solid mechanics, and is very straightforward to derive directly from the mechanics of uniaxial strain.

\medskip
\noindent\textbf{1.} We begin by formulating, and evaluating if possible, two dimensionless indefinite integrals:
\begin{subequations}\label{app:I1_I2}
    \begin{align}
        \mathcal{I}_1(\phi_f) &\equiv{}\frac{1}{k_0\mathcal{M}}\,\int\,k(\phi_f)\,\frac{\mathrm{d}{\sigma^\prime_{xx}}}{\mathrm{d}{\phi_f}}\,\mathrm{d}\phi_f \quad\mathrm{and}\\ \noalign{\medskip}
        \mathcal{I}_2(\phi_f) &\equiv{}\frac{1}{k_0\mathcal{M}}\,\int\,\left(\frac{\phi_f-\phi_{f,0}}{1-\phi_{f,0}}\right)\,k(\phi_f)\frac{\mathrm{d}\sigma^\prime_{xx}}{\mathrm{d}\phi_f}\,\mathrm{d}\phi_f.
    \end{align}
\end{subequations}
This relies on the fact that the effective stress can always be written directly in terms of the porosity in this geometry, $\sigma^\prime_{xx}=\sigma^\prime_{xx}(\phi_f)$ (see the discussion at the end of \S\ref{ss:mech_rect_large}). For the elasticity and permeability laws considered above, we provide the results in Appendix~\ref{app:s:ss_Integrals}.

\medskip
\noindent\textbf{2.} At steady state, we have that $v_s=0$ and $\phi_f{}v_f=q^\star$. The former statement with Eq.~\eqref{eq:fluidmech_rect_vs}, or the latter statement with Eq.~\eqref{eq:fluidmech_rect_vf}, leads to
\begin{equation}\label{eq:proc_ss_q_dsigdx}
    q^\star +\frac{k(\phi_f)}{\mu}\frac{\partial{\sigma^\prime_{xx}}}{\partial{x}}=0,
\end{equation}
where we have replaced the pressure gradient with the effective stress gradient using Eq.~\eqref{eq:equilibrium_rect}. Equation~\eqref{eq:proc_ss_q_dsigdx} can be rearranged and integrated to give
\begin{equation}\label{eq:proc_ss_q_I1}
    -\frac{\mu{}q^\star{}L}{k_0\mathcal{M}} \left(\frac{x}{L}-\frac{\delta^\star}{L}\right) =\mathcal{I}_1\big(\phi_f(x)\big) -\mathcal{I}_1\big(\phi_f(\delta^\star)\big),
\end{equation}
where $\phi_f(\delta^\star)$ is calculated from $\sigma^{\prime\star}$ by inverting $\sigma^\prime_{xx}\big(\phi_f(\delta^\star)\big)=\sigma^{\prime\star}$. We next derive an expression for $u_s(x)$ using Eq.~\eqref{eq:phi_to_u}, which can be rearranged using Eq.~\eqref{eq:proc_ss_q_dsigdx} and then integrated to give
\begin{equation}\label{eq:proc_ss_us_I2}
    \frac{u_s(x)}{L} =\frac{\delta^\star}{L}-\frac{k_0\mathcal{M}}{\mu{}q^\star{}L} \,\bigg[\mathcal{I}_2\big(\phi_f(x)\big)-\mathcal{I}_2\big(\phi_f(\delta^\star)\big)\bigg],
\end{equation}
where we have applied the boundary condition that $u_s(\delta^\star)=\delta^\star$. Finally, we evaluate Eqs.~\eqref{eq:proc_ss_q_I1} and \eqref{eq:proc_ss_us_I2} at $x=L$ by applying the boundary condition that $u_s(L)=0$, and rearranging to eliminate $\delta^\star$:
\begin{equation}\label{eq:I1_I2_q}
    \begin{split}
        \bigg[\mathcal{I}_2\big(\phi_f(&L)\big)-\mathcal{I}_2\big(\phi_f(\delta^\star)\big)\bigg] \\
        &-\bigg[\mathcal{I}_1\big(\phi_f(L)\big)-\mathcal{I}_1\big(\phi_f(\delta^\star)\big)\bigg] =\frac{\mu{}q^\star{}L}{k_0\mathcal{M}}.
    \end{split}
\end{equation}
If $\Delta{p}^\star$ is known, $\phi_f(L)$ can calculated by inverting $\sigma^\prime_{xx}\big(\phi_f(L)\big)=\sigma^{\prime\star}-\Delta{p}^\star$. Equation~\eqref{eq:I1_I2_q} then provides an explicit expression for $q^\star$. If $q^\star$ is known instead, Eq.~\eqref{eq:I1_I2_q} provides an implicit expression for $\phi_f(L)$, which can then be used to calculate $\Delta{p}^\star$ from $\Delta{p}^\star=\sigma^{\prime\star}-\sigma^\prime_{xx}\big(\phi_f(L)\big)$.

\medskip
\noindent\textbf{3.} Now that both $q^\star$ and $\Delta{p}^\star$ are known, $\delta^\star$ can be calculated explicitly from Eq.~\eqref{eq:proc_ss_us_I2} evaluated at $x=L$,
\begin{equation}\label{eq:I1_I2_delta}
    \frac{\delta^\star}{L} =\frac{k_0\mathcal{M}}{\mu{}q^\star{}L}\,\bigg[\mathcal{I}_2\big(\phi_f(L)\big)-\mathcal{I}_2\big(\phi_f(\delta^\star)\big)\bigg].
\end{equation}
    
\medskip
\noindent\textbf{4.} Equation~\eqref{eq:proc_ss_q_I1} now provides an implicit expression for $\phi_f(x)$ in terms of $q^\star$ and $\delta^\star$.
    
\medskip
\noindent\textbf{5.} Equation~\eqref{eq:proc_ss_us_I2} now provides an explicit expression for $u_s(x)$ in terms of $q^\star$, $\delta^\star$, and $\phi_f(x)$.
    
\medskip
\noindent\textbf{6.} Finally, the effective stress can be calculated from $\phi_f(x)$, and then the pressure from the effective stress,
\begin{subequations}
    \begin{align}
        \sigma^\prime_{xx}(x) &=\sigma^\prime_{xx}\big(\phi_f(x)\big), \\
        p(x) &=\sigma^\prime_{xx}(x)-(\sigma^{\prime\star}-\Delta{p}^\star),
    \end{align}
\end{subequations}
where the latter comes from integrating Eq.~\eqref{eq:equilibrium_rect} and applying the final boundary condition, $p(L)=0$.

\medskip
\noindent This procedure can be implemented analytically as long as the integrals can be evaluated exactly, although numerical root-finding is required in most cases. When the integrals cannot be evaluated exactly, it is straightforward to implement them numerically using standard quadrature techniques.

\section{Steady-state solutions: Maximum values}\label{app:s:Shutdown}

The effective stress is always largest in magnitude at the right boundary ($x=L$), so this is where the porosity and permeability are the smallest, and the flow must stop when these vanish. Provided that $\sigma^{\prime\star}$ is known, this condition allows for the direct calculation of the maximum achievable values $q^\star_\mathrm{max}$, $\Delta{p}^\star_\mathrm{max}$, and $\delta^\star_\mathrm{max}$ for which the porosity vanishes at $x=L$. The maximum flux $q^\star_\mathrm{max}$ can be evaluated directly from Eq.~\eqref{eq:I1_I2_q} by setting $\phi_f(L)=0$,
\begin{equation}
    \frac{\mu{}q^\star_\mathrm{max}L}{k_0\mathcal{M}} =\bigg[\mathcal{I}_2\big(0\big)-\mathcal{I}_2\big(\phi_f(\delta^\star)\big)\bigg] -\bigg[\mathcal{I}_1\big(0\big)-\mathcal{I}_1\big(\phi_f(\delta^\star)\big)\bigg].
\end{equation}
The maximum pressure drop $\Delta{p}^\star_\mathrm{max}$ can be evaluated directly from the elasticity law by setting $\phi_f(L)=0$,
\begin{equation}
    \Delta{p}^\star_\mathrm{max}=\sigma^{\prime\star}-\sigma^\prime_{xx}(\phi_f=0).
\end{equation}
The maximum deflection $\delta^\star_\mathrm{max}$ can be evaluated directly from Eq.~\eqref{eq:I1_I2_delta} by setting $q^\star=q^\star_\mathrm{max}$ and $\phi_f(L)=0$,
\begin{equation}
    \frac{\delta^\star_\mathrm{max}}{L} =\frac{k_0\mathcal{M}}{\mu{}q^\star_\mathrm{max}L}\,\bigg[\mathcal{I}_2\big(0\big)-\mathcal{I}_2\big(\phi_f(\delta^\star)\big)\bigg].
\end{equation}
These three values occur simultaneously for a given $\sigma^{\prime\star}$, and they are physical limits in the sense that it is not possible to drive a flux greater than $q^\star_\mathrm{max}$ or a deflection greater than $\delta^\star_\mathrm{max}$, or to apply a pressure drop greater than $\Delta{p}^\star_\mathrm{max}$, without producing a negative porosity at the right boundary. Although solutions may exist for larger values, they will be nonphysical.

\section{Steady-state solutions: Integrals for specific cases}\label{app:s:ss_Integrals}

Here, we evaluate the integrals $\mathcal{I}_1(\phi_f)$ and $\mathcal{I}_2(\phi_f)$ (Eq.~\ref{app:I1_I2}) for the scenarios considered in \S\ref{ss:flow_rect_ss} above: Linear elasticity (for use with the intermediate model) and Hencky elasticity (for use with the nonlinear model), each for both constant permeability ($k=k_0$) and deformation-dependent permeability ($k=k(\phi_f)$ via the normalized Kozeny-Carman formula, Eq.~\ref{eq:kozeny-carman-k0}). In each case, we first write the elasticity law in terms of the porosity and then evaluate the first derivative of this function. For linear elasticity, we have
\begin{equation}
    \frac{\sigma^{\prime}_{xx}(\phi_f)}{\mathcal{M}} =\frac{\phi_f-\phi_{f,0}}{1-\phi_{f,0}} \,\,\,\,\mathrm{and}\,\,\,\, \frac{1}{\mathcal{M}}\frac{\mathrm{d}\sigma_{xx}^\prime}{\mathrm{d}\phi_f} =\frac{1}{1-\phi_{f,0}}.
\end{equation}
For Hencky elasticity, we instead have
\begin{subequations}
    \begin{align}
        \frac{\sigma^{\prime}_{xx}(\phi_f)}{\mathcal{M}} &=\left(\frac{1-\phi_f}{1-\phi_{f,0}}\right)\ln\left(\frac{1-\phi_{f,0}}{1-\phi_f}\right) \quad\mathrm{and}\\ \noalign{\medskip}
        \frac{1}{\mathcal{M}}\frac{\mathrm{d}\sigma_{xx}^\prime}{\mathrm{d}\phi_f} &=\frac{1}{1-\phi_{f,0}}\left[1-\ln\left(\frac{1-\phi_{f,0}}{1-\phi_f}\right)\right].
    \end{align}
\end{subequations}

We now evaluate the two integrals. These expressions can then be used with the procedure described in Appendix~\ref{app:s:proc_steady_rect} to evaluate the full steady-state solutions for these constitutive models. Note that the integrals are indefinite, so arbitrary constants can be added or subtracted from the expressions below without loss of generality.

\begin{widetext}
\begin{itemize}
    
    \item intermediate-$k_0$ (linear elasticity with constant permeability)
    
\begin{equation}
    \mathcal{I}_1(\phi_f) =\frac{\phi_f-\phi_{f,0}}{1-\phi_{f,0}}  \quad\mathrm{and}\quad \mathcal{I}_2(\phi_f) =\frac{1}{2}\left(\frac{\phi_f-\phi_{f,0}}{1-\phi_{f,0}}\right)^2
\end{equation}
    
    \item intermediate-$k_\mathrm{KC}$ (linear elasticity with normalized Kozeny-Carman permeability)
        
\begin{subequations}
    \begin{align}
        \mathcal{I}_1(\phi_f) &= \frac{1-\phi_{f,0}}{\phi_{f,0}^3}\,\bigg[\frac{1}{2}\phi_f^2 +2\phi_f+\frac{1}{1-\phi_f}+3\ln(1-\phi_f)\bigg] \quad\mathrm{and}\\ \noalign{\medskip}
        \begin{split}
            \mathcal{I}_2(\phi_f) &=\frac{1}{\phi_{f,0}^3}\,\bigg[-\frac{1}{3}(1-\phi_f)^3 +\frac{1}{2}(4-\phi_{f,0})(1-\phi_f)^2 \\
            &\phantom{=\frac{1}{\phi_{f,0}^3}\,\bigg[} -3(2-\phi_{f,0})(1-\phi_f)+\frac{1-\phi_{f,0}}{1-\phi_f}+(4-3\phi_{f,0})\ln(1-\phi_f)\bigg]
        \end{split}
    \end{align}
\end{subequations}

    \item{nonlinear-$k_0$ (Hencky elasticity with constant permeability)}
    
\begin{subequations}
    \begin{align}
        \mathcal{I}_1(\phi_f) &=\left(\frac{1-\phi_f}{1-\phi_{f,0}}\right) \ln\left(\frac{1-\phi_{f,0}}{1-\phi_f}\right) \quad\mathrm{and}\\ \noalign{\medskip}
        \mathcal{I}_2(\phi_f) &=\frac{1}{4}\left(\frac{1-\phi_f}{1-\phi_{f,0}}\right)^2+\frac{1}{2}\left[1-\left(\frac{\phi_f-\phi_{f,0}}{1-\phi_{f,0}}\right)^2\right]\ln\left(\frac{1-\phi_{f,0}}{1-\phi_f}\right)
    \end{align}
\end{subequations}
    
    \item nonlinear-$k_\mathrm{KC}$ (Hencky elasticity with normalized Kozeny-Carman permeability)
    
\begin{subequations}
    \begin{align}
        \begin{split}
            \mathcal{I}_1(\phi_f) &=\frac{1-\phi_{f,0}}{4\phi_{f,0}^3}\,\bigg[ \phi_f^2 +2\left(\frac{\phi_f^3+3\phi_f^2-4\phi_f-2}{1-\phi_f}\right)\ln\left(\frac{1-\phi_{f,0}}{1-\phi_f}\right) \\
            &\phantom{=\frac{1-\phi_{f,0}}{4\phi_{f,0}^3}\,\bigg[} +6\ln^2\left(\frac{1-\phi_{f,0}}{1-\phi_f}\right) -2\phi_f+\frac{8}{1-\phi_f}+2\ln(1-\phi_f)\bigg]
        \end{split} \\ \noalign{\medskip} 
        \begin{split}
            \mathcal{I}_2(\phi_f) &=\frac{1}{36\phi_{f,0}^3}\,\bigg\{ 8\phi_f^3+3(4-3\phi_{f,0})\phi_f^2 -6(8-3\phi_{f,0})\phi_f-6(2+3\phi_{f,0})\ln(1-\phi_f) \\
            &\phantom{=\frac{1}{36\phi_{f,0}^3}} +6\left[\frac{2\phi_f^4+(4-3\phi_{f,0})(3+\phi_f)\phi_f^2 -6(3-2\phi_{f,0})\phi_f-6(1-\phi_{f,0})}{1-\phi_f}\right]\ln\left(\frac{1-\phi_{f,0}}{1-\phi_f}\right) \\
            &\phantom{=\frac{1}{36\phi_{f,0}^3}\,} +72\left(\frac{1-\phi_{f,0}}{1-\phi_f}\right) +18(4-3\phi_{f,0})\ln^2\left(\frac{1-\phi_{f,0}}{1-\phi_f}\right) \bigg\}
        \end{split}
    \end{align}
\end{subequations}

\end{itemize}
\end{widetext}



\begin{thebibliography}{88}%
\makeatletter
\providecommand \@ifxundefined [1]{%
 \@ifx{#1\undefined}
}%
\providecommand \@ifnum [1]{%
 \ifnum #1\expandafter \@firstoftwo
 \else \expandafter \@secondoftwo
 \fi
}%
\providecommand \@ifx [1]{%
 \ifx #1\expandafter \@firstoftwo
 \else \expandafter \@secondoftwo
 \fi
}%
\providecommand \natexlab [1]{#1}%
\providecommand \enquote  [1]{``#1''}%
\providecommand \bibnamefont  [1]{#1}%
\providecommand \bibfnamefont [1]{#1}%
\providecommand \citenamefont [1]{#1}%
\providecommand \href@noop [0]{\@secondoftwo}%
\providecommand \href [0]{\begingroup \@sanitize@url \@href}%
\providecommand \@href[1]{\@@startlink{#1}\@@href}%
\providecommand \@@href[1]{\endgroup#1\@@endlink}%
\providecommand \@sanitize@url [0]{\catcode `\\12\catcode `\$12\catcode
  `\&12\catcode `\#12\catcode `\^12\catcode `\_12\catcode `\%12\relax}%
\providecommand \@@startlink[1]{}%
\providecommand \@@endlink[0]{}%
\providecommand \url  [0]{\begingroup\@sanitize@url \@url }%
\providecommand \@url [1]{\endgroup\@href {#1}{\urlprefix }}%
\providecommand \urlprefix  [0]{URL }%
\providecommand \Eprint [0]{\href }%
\providecommand \doibase [0]{http://dx.doi.org/}%
\providecommand \selectlanguage [0]{\@gobble}%
\providecommand \bibinfo  [0]{\@secondoftwo}%
\providecommand \bibfield  [0]{\@secondoftwo}%
\providecommand \translation [1]{[#1]}%
\providecommand \BibitemOpen [0]{}%
\providecommand \bibitemStop [0]{}%
\providecommand \bibitemNoStop [0]{.\EOS\space}%
\providecommand \EOS [0]{\spacefactor3000\relax}%
\providecommand \BibitemShut  [1]{\csname bibitem#1\endcsname}%
\let\auto@bib@innerbib\@empty
\bibitem [{\citenamefont {Mow}\ \emph {et~al.}(1980)\citenamefont {Mow},
  \citenamefont {Kuei}, \citenamefont {Lai},\ and\ \citenamefont
  {Armstrong}}]{mow-jbiomecheng-1980}%
  \BibitemOpen
  \bibfield  {author} {\bibinfo {author} {\bibfnamefont {V.~C.}\ \bibnamefont
  {Mow}}, \bibinfo {author} {\bibfnamefont {S.~C.}\ \bibnamefont {Kuei}},
  \bibinfo {author} {\bibfnamefont {W.~M.}\ \bibnamefont {Lai}}, \ and\
  \bibinfo {author} {\bibfnamefont {C.~G.}\ \bibnamefont {Armstrong}},\
  }\bibfield  {title} {\enquote {\bibinfo {title} {Biphasic creep and stress
  relaxation of articular cartilage in compression: {Theory} and
  experiments},}\ }\href {\doibase 10.1115/1.3138202} {\bibfield  {journal}
  {\bibinfo  {journal} {Journal of Biomechanical Engineering}\ }\textbf
  {\bibinfo {volume} {102}},\ \bibinfo {pages} {73--84} (\bibinfo {year}
  {1980})}\BibitemShut {NoStop}%
\bibitem [{\citenamefont {Lai}\ \emph {et~al.}(1991)\citenamefont {Lai},
  \citenamefont {Hou},\ and\ \citenamefont {Mow}}]{lai-jbiomecheng-1991}%
  \BibitemOpen
  \bibfield  {author} {\bibinfo {author} {\bibfnamefont {W.~M.}\ \bibnamefont
  {Lai}}, \bibinfo {author} {\bibfnamefont {J.~S.}\ \bibnamefont {Hou}}, \ and\
  \bibinfo {author} {\bibfnamefont {V.~C.}\ \bibnamefont {Mow}},\ }\bibfield
  {title} {\enquote {\bibinfo {title} {A triphasic theory for the swelling and
  deformation behaviors of articular cartilage},}\ }\href {\doibase
  10.1115/1.2894880} {\bibfield  {journal} {\bibinfo  {journal} {Journal of
  Biomechanical Engineering}\ }\textbf {\bibinfo {volume} {113}},\ \bibinfo
  {pages} {245--258} (\bibinfo {year} {1991})}\BibitemShut {NoStop}%
\bibitem [{\citenamefont {Yang}\ and\ \citenamefont
  {Taber}(1991)}]{yang-jbiomech-1991}%
  \BibitemOpen
  \bibfield  {author} {\bibinfo {author} {\bibfnamefont {M.}~\bibnamefont
  {Yang}}\ and\ \bibinfo {author} {\bibfnamefont {L.~A.}\ \bibnamefont
  {Taber}},\ }\bibfield  {title} {\enquote {\bibinfo {title} {The possible role
  of poroelasticity in the apparent viscoelastic behavior of passive cardiac
  muscle},}\ }\href {\doibase 10.1016/0021-9290(91)90291-T} {\bibfield
  {journal} {\bibinfo  {journal} {Journal of Biomechanics}\ }\textbf {\bibinfo
  {volume} {24}},\ \bibinfo {pages} {587--597} (\bibinfo {year}
  {1991})}\BibitemShut {NoStop}%
\bibitem [{\citenamefont {Cowin}(1999)}]{cowin-jbiomech-1999}%
  \BibitemOpen
  \bibfield  {author} {\bibinfo {author} {\bibfnamefont {S.~C.}\ \bibnamefont
  {Cowin}},\ }\bibfield  {title} {\enquote {\bibinfo {title} {Bone
  poroelasticity},}\ }\href {\doibase 10.1016/S0021-9290(98)00161-4} {\bibfield
   {journal} {\bibinfo  {journal} {Journal of Biomechanics}\ }\textbf {\bibinfo
  {volume} {32}},\ \bibinfo {pages} {217--238} (\bibinfo {year}
  {1999})}\BibitemShut {NoStop}%
\bibitem [{\citenamefont {Franceschini}\ \emph {et~al.}(2006)\citenamefont
  {Franceschini}, \citenamefont {Bigoni}, \citenamefont {Regitnig},\ and\
  \citenamefont {Holzapfel}}]{franceschini-jmps-2006}%
  \BibitemOpen
  \bibfield  {author} {\bibinfo {author} {\bibfnamefont {G.}~\bibnamefont
  {Franceschini}}, \bibinfo {author} {\bibfnamefont {D.}~\bibnamefont
  {Bigoni}}, \bibinfo {author} {\bibfnamefont {P.}~\bibnamefont {Regitnig}}, \
  and\ \bibinfo {author} {\bibfnamefont {G.~A.}\ \bibnamefont {Holzapfel}},\
  }\bibfield  {title} {\enquote {\bibinfo {title} {Brain tissue deforms
  similarly to filled elastomers and follows consolidation theory},}\ }\href
  {\doibase 10.1016/j.jmps.2006.05.004} {\bibfield  {journal} {\bibinfo
  {journal} {Journal of the Mechanical and Physics of Solids}\ }\textbf
  {\bibinfo {volume} {54}},\ \bibinfo {pages} {2592--2620} (\bibinfo {year}
  {2006})}\BibitemShut {NoStop}%
\bibitem [{\citenamefont {Skotheim}\ and\ \citenamefont
  {Mahadevan}(2005)}]{skotheim-science-2005}%
  \BibitemOpen
  \bibfield  {author} {\bibinfo {author} {\bibfnamefont {J.~M.}\ \bibnamefont
  {Skotheim}}\ and\ \bibinfo {author} {\bibfnamefont {L.}~\bibnamefont
  {Mahadevan}},\ }\bibfield  {title} {\enquote {\bibinfo {title} {Physical
  limits and design principles for plant and fungal movements},}\ }\href
  {\doibase 10.1126/science.1107976} {\bibfield  {journal} {\bibinfo  {journal}
  {Science}\ }\textbf {\bibinfo {volume} {308}},\ \bibinfo {pages} {1308--1310}
  (\bibinfo {year} {2005})}\BibitemShut {NoStop}%
\bibitem [{\citenamefont {Charras}\ \emph {et~al.}(2005)\citenamefont
  {Charras}, \citenamefont {Yarrow}, \citenamefont {Horton}, \citenamefont
  {Mahadevan},\ and\ \citenamefont {Mitchison}}]{charras-nature-2005}%
  \BibitemOpen
  \bibfield  {author} {\bibinfo {author} {\bibfnamefont {G.~T.}\ \bibnamefont
  {Charras}}, \bibinfo {author} {\bibfnamefont {J.~C.}\ \bibnamefont {Yarrow}},
  \bibinfo {author} {\bibfnamefont {M.~A.}\ \bibnamefont {Horton}}, \bibinfo
  {author} {\bibfnamefont {L.}~\bibnamefont {Mahadevan}}, \ and\ \bibinfo
  {author} {\bibfnamefont {T.~J.}\ \bibnamefont {Mitchison}},\ }\bibfield
  {title} {\enquote {\bibinfo {title} {Non-equilibration of hydrostatic
  pressure in blebbing cells},}\ }\href {\doibase 10.1038/nature03550}
  {\bibfield  {journal} {\bibinfo  {journal} {Nature}\ }\textbf {\bibinfo
  {volume} {435}},\ \bibinfo {pages} {365--369} (\bibinfo {year}
  {2005})}\BibitemShut {NoStop}%
\bibitem [{\citenamefont {Dumais}\ and\ \citenamefont
  {Forterre}(2012)}]{dumais-annrevfluidmech-2012}%
  \BibitemOpen
  \bibfield  {author} {\bibinfo {author} {\bibfnamefont {J.}~\bibnamefont
  {Dumais}}\ and\ \bibinfo {author} {\bibfnamefont {Y.}~\bibnamefont
  {Forterre}},\ }\bibfield  {title} {\enquote {\bibinfo {title} {`{Vegetable
  Dynamicks}': {T}he role of water in plant movements},}\ }\href {\doibase
  10.1146/annurev-fluid-120710-101200} {\bibfield  {journal} {\bibinfo
  {journal} {Annual Review of Fluid Mechanics}\ }\textbf {\bibinfo {volume}
  {44}},\ \bibinfo {pages} {453--478} (\bibinfo {year} {2012})}\BibitemShut
  {NoStop}%
\bibitem [{\citenamefont {Moeendarbary}\ \emph {et~al.}(2013)\citenamefont
  {Moeendarbary}, \citenamefont {Valon}, \citenamefont {Fritzsche},
  \citenamefont {Harris}, \citenamefont {Moulding}, \citenamefont {Thrasher},
  \citenamefont {Stride}, \citenamefont {Mahadevan},\ and\ \citenamefont
  {Charras}}]{moeendarbary-natmaterials-2013}%
  \BibitemOpen
  \bibfield  {author} {\bibinfo {author} {\bibfnamefont {E.}~\bibnamefont
  {Moeendarbary}}, \bibinfo {author} {\bibfnamefont {L.}~\bibnamefont {Valon}},
  \bibinfo {author} {\bibfnamefont {M.}~\bibnamefont {Fritzsche}}, \bibinfo
  {author} {\bibfnamefont {A.~R.}\ \bibnamefont {Harris}}, \bibinfo {author}
  {\bibfnamefont {D.~A.}\ \bibnamefont {Moulding}}, \bibinfo {author}
  {\bibfnamefont {A.~J.}\ \bibnamefont {Thrasher}}, \bibinfo {author}
  {\bibfnamefont {E.}~\bibnamefont {Stride}}, \bibinfo {author} {\bibfnamefont
  {L.}~\bibnamefont {Mahadevan}}, \ and\ \bibinfo {author} {\bibfnamefont
  {G.~T.}\ \bibnamefont {Charras}},\ }\bibfield  {title} {\enquote {\bibinfo
  {title} {The cytoplasm of living cells behaves as a poroelastic material},}\
  }\href {\doibase 10.1038/nmat3517} {\bibfield  {journal} {\bibinfo  {journal}
  {Nature Materials}\ }\textbf {\bibinfo {volume} {12}},\ \bibinfo {pages}
  {253--261} (\bibinfo {year} {2013})}\BibitemShut {NoStop}%
\bibitem [{\citenamefont {{McKenzie}}(1984)}]{mckenzie-jpetrology-1984}%
  \BibitemOpen
  \bibfield  {author} {\bibinfo {author} {\bibfnamefont {D.}~\bibnamefont
  {{McKenzie}}},\ }\bibfield  {title} {\enquote {\bibinfo {title} {The
  generation and compaction of partially molten rock},}\ }\href {\doibase
  10.1093/petrology/25.3.713} {\bibfield  {journal} {\bibinfo  {journal}
  {Journal of Petrology}\ }\textbf {\bibinfo {volume} {25}},\ \bibinfo {pages}
  {713--765} (\bibinfo {year} {1984})}\BibitemShut {NoStop}%
\bibitem [{\citenamefont {Fowler}(1985)}]{fowler-geoastrofluiddynamics-1985}%
  \BibitemOpen
  \bibfield  {author} {\bibinfo {author} {\bibfnamefont {A.~C.}\ \bibnamefont
  {Fowler}},\ }\bibfield  {title} {\enquote {\bibinfo {title} {A mathematical
  model of magma transport in the asthenosphere},}\ }\href {\doibase
  10.1080/03091928508245423} {\bibfield  {journal} {\bibinfo  {journal}
  {Geophysical \& Astrophysical Fluid Dynamics}\ }\textbf {\bibinfo {volume}
  {33}},\ \bibinfo {pages} {63--96} (\bibinfo {year} {1985})}\BibitemShut
  {NoStop}%
\bibitem [{\citenamefont {Spiegelman}(1993)}]{spiegelman-jfm-1993a}%
  \BibitemOpen
  \bibfield  {author} {\bibinfo {author} {\bibfnamefont {M.}~\bibnamefont
  {Spiegelman}},\ }\bibfield  {title} {\enquote {\bibinfo {title} {Flow in
  deformable porous media. {P}art 1. {S}imple analysis},}\ }\href {\doibase
  10.1017/S0022112093000369} {\bibfield  {journal} {\bibinfo  {journal}
  {Journal of Fluid Mechanics}\ }\textbf {\bibinfo {volume} {247}},\ \bibinfo
  {pages} {17--38} (\bibinfo {year} {1993})}\BibitemShut {NoStop}%
\bibitem [{\citenamefont {Bercovici}\ \emph {et~al.}(2001)\citenamefont
  {Bercovici}, \citenamefont {Ricard},\ and\ \citenamefont
  {Schubert}}]{bercovici-jgr-2001a}%
  \BibitemOpen
  \bibfield  {author} {\bibinfo {author} {\bibfnamefont {D.}~\bibnamefont
  {Bercovici}}, \bibinfo {author} {\bibfnamefont {Y.}~\bibnamefont {Ricard}}, \
  and\ \bibinfo {author} {\bibfnamefont {G.}~\bibnamefont {Schubert}},\
  }\bibfield  {title} {\enquote {\bibinfo {title} {A two-phase model for
  compaction and damage 1. {General} theory},}\ }\href {\doibase
  10.1029/2000JB900430} {\bibfield  {journal} {\bibinfo  {journal} {Journal of
  Geophysical Research}\ }\textbf {\bibinfo {volume} {106}},\ \bibinfo {pages}
  {8887--8906} (\bibinfo {year} {2001})}\BibitemShut {NoStop}%
\bibitem [{\citenamefont {Katz}(2008)}]{katz-jpetrology-2008}%
  \BibitemOpen
  \bibfield  {author} {\bibinfo {author} {\bibfnamefont {R.~F.}\ \bibnamefont
  {Katz}},\ }\bibfield  {title} {\enquote {\bibinfo {title} {Magma dynamics
  with the enthalpy method: {Benchmark} solutions and magmatic focusing at
  mid-ocean ridges},}\ }\href {\doibase 10.1093/petrology/egn058} {\bibfield
  {journal} {\bibinfo  {journal} {Journal of Petrology}\ }\textbf {\bibinfo
  {volume} {49}},\ \bibinfo {pages} {2099--2121} (\bibinfo {year}
  {2008})}\BibitemShut {NoStop}%
\bibitem [{\citenamefont {Hesse}\ \emph {et~al.}(2011)\citenamefont {Hesse},
  \citenamefont {Schiemenz}, \citenamefont {Liang},\ and\ \citenamefont
  {Parmentier}}]{hesse-geophysjint-2011}%
  \BibitemOpen
  \bibfield  {author} {\bibinfo {author} {\bibfnamefont {M.~A.}\ \bibnamefont
  {Hesse}}, \bibinfo {author} {\bibfnamefont {A.~R.}\ \bibnamefont
  {Schiemenz}}, \bibinfo {author} {\bibfnamefont {Y.}~\bibnamefont {Liang}}, \
  and\ \bibinfo {author} {\bibfnamefont {E.~M.}\ \bibnamefont {Parmentier}},\
  }\bibfield  {title} {\enquote {\bibinfo {title} {Compaction-dissolution waves
  in an upwelling mantle column},}\ }\href {\doibase
  10.1111/j.1365-246X.2011.05177.x} {\bibfield  {journal} {\bibinfo  {journal}
  {Geophysical Journal International}\ }\textbf {\bibinfo {volume} {187}},\
  \bibinfo {pages} {1057--1075} (\bibinfo {year} {2011})}\BibitemShut {NoStop}%
\bibitem [{\citenamefont {Bear}\ and\ \citenamefont
  {Corapcioglu}(1981)}]{bear-wrr-1981a}%
  \BibitemOpen
  \bibfield  {author} {\bibinfo {author} {\bibfnamefont {J.}~\bibnamefont
  {Bear}}\ and\ \bibinfo {author} {\bibfnamefont {M.~Y.}\ \bibnamefont
  {Corapcioglu}},\ }\bibfield  {title} {\enquote {\bibinfo {title}
  {Mathematical model for regional land subsidence due to pumping:
  1.~{Integrated} aquifer subsidence equations based on vertical displacement
  only},}\ }\href {\doibase 10.1029/WR017i004p00937} {\bibfield  {journal}
  {\bibinfo  {journal} {Water Resources Research}\ }\textbf {\bibinfo {volume}
  {17}},\ \bibinfo {pages} {937--946} (\bibinfo {year} {1981})}\BibitemShut
  {NoStop}%
\bibitem [{\citenamefont {Wang}(2000)}]{wang-princeton-2000}%
  \BibitemOpen
  \bibfield  {author} {\bibinfo {author} {\bibfnamefont {H.~F.}\ \bibnamefont
  {Wang}},\ }\href@noop {} {\emph {\bibinfo {title} {Theory of Linear
  Poroelasticity}}}\ (\bibinfo  {publisher} {Princeton University Press},\
  \bibinfo {address} {Princeton NJ},\ \bibinfo {year} {2000})\BibitemShut
  {NoStop}%
\bibitem [{\citenamefont {Szulczewski}\ \emph {et~al.}(2012)\citenamefont
  {Szulczewski}, \citenamefont {MacMinn}, \citenamefont {Herzog},\ and\
  \citenamefont {Juanes}}]{szulczewski-pnas-2012}%
  \BibitemOpen
  \bibfield  {author} {\bibinfo {author} {\bibfnamefont {M.~L.}\ \bibnamefont
  {Szulczewski}}, \bibinfo {author} {\bibfnamefont {C.~W.}\ \bibnamefont
  {MacMinn}}, \bibinfo {author} {\bibfnamefont {H.~J.}\ \bibnamefont {Herzog}},
  \ and\ \bibinfo {author} {\bibfnamefont {R.}~\bibnamefont {Juanes}},\
  }\bibfield  {title} {\enquote {\bibinfo {title} {Lifetime of carbon capture
  and storage as a climate-change mitigation technology},}\ }\href {\doibase
  10.1073/pnas.1115347109} {\bibfield  {journal} {\bibinfo  {journal}
  {Proceedings of the National Academy of Sciences of the United States of
  America}\ }\textbf {\bibinfo {volume} {109}},\ \bibinfo {pages} {5185--5189}
  (\bibinfo {year} {2012})}\BibitemShut {NoStop}%
\bibitem [{\citenamefont {Rutqvist}(2012)}]{rutqvist-geotechgeologeng-2012}%
  \BibitemOpen
  \bibfield  {author} {\bibinfo {author} {\bibfnamefont {J.}~\bibnamefont
  {Rutqvist}},\ }\bibfield  {title} {\enquote {\bibinfo {title} {The
  geomechanics of {CO}$_2$ storage in deep sedimentary formations},}\ }\href
  {\doibase 10.1007/s10706-011-9491-0} {\bibfield  {journal} {\bibinfo
  {journal} {Geotechnical and Geological Engineering}\ }\textbf {\bibinfo
  {volume} {30}},\ \bibinfo {pages} {525--551} (\bibinfo {year}
  {2012})}\BibitemShut {NoStop}%
\bibitem [{\citenamefont {Chang}\ \emph {et~al.}(2013)\citenamefont {Chang},
  \citenamefont {Hesse},\ and\ \citenamefont {Nicot}}]{chang-wrr-2013}%
  \BibitemOpen
  \bibfield  {author} {\bibinfo {author} {\bibfnamefont {K.~W.}\ \bibnamefont
  {Chang}}, \bibinfo {author} {\bibfnamefont {M.~A.}\ \bibnamefont {Hesse}}, \
  and\ \bibinfo {author} {\bibfnamefont {J.-P.}\ \bibnamefont {Nicot}},\
  }\bibfield  {title} {\enquote {\bibinfo {title} {Reduction in lateral
  pressure propagation due to dissipation into ambient mudrocks during
  geological carbon dioxide storage},}\ }\href {\doibase 10.1002/wrcr.20197}
  {\bibfield  {journal} {\bibinfo  {journal} {Water Resources Research}\
  }\textbf {\bibinfo {volume} {49}},\ \bibinfo {pages} {2573--2588} (\bibinfo
  {year} {2013})}\BibitemShut {NoStop}%
\bibitem [{\citenamefont {Verdon}\ \emph {et~al.}(2013)\citenamefont {Verdon},
  \citenamefont {Kendall}, \citenamefont {Stork}, \citenamefont {Chadwick},
  \citenamefont {White},\ and\ \citenamefont {Bissell}}]{verdon-pnas-2013}%
  \BibitemOpen
  \bibfield  {author} {\bibinfo {author} {\bibfnamefont {J.~P.}\ \bibnamefont
  {Verdon}}, \bibinfo {author} {\bibfnamefont {J.-M.}\ \bibnamefont {Kendall}},
  \bibinfo {author} {\bibfnamefont {A.~L.}\ \bibnamefont {Stork}}, \bibinfo
  {author} {\bibfnamefont {R.~A.}\ \bibnamefont {Chadwick}}, \bibinfo {author}
  {\bibfnamefont {D.~J.}\ \bibnamefont {White}}, \ and\ \bibinfo {author}
  {\bibfnamefont {R.~C.}\ \bibnamefont {Bissell}},\ }\bibfield  {title}
  {\enquote {\bibinfo {title} {Comparison of geomechanical deformation induced
  by megatonne-scale {CO}$_2$ storage at {Sleipner}, {Weyburn}, and {In
  Salah}},}\ }\href {\doibase 10.1073/pnas.1302156110} {\bibfield  {journal}
  {\bibinfo  {journal} {Proceedings of the National Academy of Sciences of the
  United States of America}\ }\textbf {\bibinfo {volume} {110}},\ \bibinfo
  {pages} {E2762--71} (\bibinfo {year} {2013})}\BibitemShut {NoStop}%
\bibitem [{\citenamefont {Yarushina}\ \emph {et~al.}(2013)\citenamefont
  {Yarushina}, \citenamefont {Bercovici},\ and\ \citenamefont
  {Oristaglio}}]{yarushina-geophysjint-2013}%
  \BibitemOpen
  \bibfield  {author} {\bibinfo {author} {\bibfnamefont {V.~M.}\ \bibnamefont
  {Yarushina}}, \bibinfo {author} {\bibfnamefont {D.}~\bibnamefont
  {Bercovici}}, \ and\ \bibinfo {author} {\bibfnamefont {M.~L.}\ \bibnamefont
  {Oristaglio}},\ }\bibfield  {title} {\enquote {\bibinfo {title} {Rock
  deformation models and fluid leak-off in hydraulic fracturing},}\ }\href
  {\doibase 10.1093/gji/ggt199} {\bibfield  {journal} {\bibinfo  {journal}
  {Geophysical Journal International}\ }\textbf {\bibinfo {volume} {194}},\
  \bibinfo {pages} {1514--1526} (\bibinfo {year} {2013})}\BibitemShut {NoStop}%
\bibitem [{\citenamefont {Jha}\ and\ \citenamefont
  {Juanes}(2014)}]{jha-wrr-2014}%
  \BibitemOpen
  \bibfield  {author} {\bibinfo {author} {\bibfnamefont {B.}~\bibnamefont
  {Jha}}\ and\ \bibinfo {author} {\bibfnamefont {R.}~\bibnamefont {Juanes}},\
  }\bibfield  {title} {\enquote {\bibinfo {title} {Coupled multiphase flow and
  poromechanics: {A} computational model of pore-pressure effects on fault slip
  and earthquake triggering},}\ }\href {\doibase 10.1002/2013WR015175}
  {\bibfield  {journal} {\bibinfo  {journal} {Water Resources Research}\
  }\textbf {\bibinfo {volume} {50}},\ \bibinfo {pages} {3776--3808} (\bibinfo
  {year} {2014})}\BibitemShut {NoStop}%
\bibitem [{\citenamefont {Cai}\ and\ \citenamefont
  {Bercovici}(2014)}]{cai-geophysjint-2014}%
  \BibitemOpen
  \bibfield  {author} {\bibinfo {author} {\bibfnamefont {Z.}~\bibnamefont
  {Cai}}\ and\ \bibinfo {author} {\bibfnamefont {D.}~\bibnamefont
  {Bercovici}},\ }\bibfield  {title} {\enquote {\bibinfo {title} {Two-phase
  viscoelastic damage theory, with applications to subsurface fluid
  injection},}\ }\href {\doibase 10.1093/gji/ggu344} {\bibfield  {journal}
  {\bibinfo  {journal} {Geophysical Journal International}\ }\textbf {\bibinfo
  {volume} {199}},\ \bibinfo {pages} {1481--1496} (\bibinfo {year}
  {2014})}\BibitemShut {NoStop}%
\bibitem [{\citenamefont {Hewitt}\ \emph {et~al.}(2015)\citenamefont {Hewitt},
  \citenamefont {Neufeld},\ and\ \citenamefont {Balmforth}}]{hewitt-jfm-2015}%
  \BibitemOpen
  \bibfield  {author} {\bibinfo {author} {\bibfnamefont {D.~R.}\ \bibnamefont
  {Hewitt}}, \bibinfo {author} {\bibfnamefont {J.~A.}\ \bibnamefont {Neufeld}},
  \ and\ \bibinfo {author} {\bibfnamefont {N.~J.}\ \bibnamefont {Balmforth}},\
  }\bibfield  {title} {\enquote {\bibinfo {title} {Shallow, gravity-driven flow
  in a poro-elastic layer},}\ }\href {\doibase 10.1017/jfm.2015.361} {\bibfield
   {journal} {\bibinfo  {journal} {Journal of Fluid Mechanics}\ }\textbf
  {\bibinfo {volume} {778}},\ \bibinfo {pages} {335--360} (\bibinfo {year}
  {2015})}\BibitemShut {NoStop}%
\bibitem [{\citenamefont {{de Boer}}(1996)}]{deboer-applmechrev-1996}%
  \BibitemOpen
  \bibfield  {author} {\bibinfo {author} {\bibfnamefont {R.}~\bibnamefont {{de
  Boer}}},\ }\bibfield  {title} {\enquote {\bibinfo {title} {Highlights in the
  historical development of the porous media theory: {Toward} a consistent
  macroscopic theory},}\ }\href {\doibase 10.1115/1.3101926} {\bibfield
  {journal} {\bibinfo  {journal} {Applied Mechanics Reviews}\ }\textbf
  {\bibinfo {volume} {49}},\ \bibinfo {pages} {201--262} (\bibinfo {year}
  {1996})}\BibitemShut {NoStop}%
\bibitem [{\citenamefont {Biot}(1941)}]{biot-japplphys-1941}%
  \BibitemOpen
  \bibfield  {author} {\bibinfo {author} {\bibfnamefont {M.~A.}\ \bibnamefont
  {Biot}},\ }\bibfield  {title} {\enquote {\bibinfo {title} {General theory of
  three-dimensional consolidation},}\ }\href@noop {} {\bibfield  {journal}
  {\bibinfo  {journal} {Journal of Applied Physics}\ }\textbf {\bibinfo
  {volume} {12}},\ \bibinfo {pages} {155--164} (\bibinfo {year}
  {1941})}\BibitemShut {NoStop}%
\bibitem [{\citenamefont {Biot}(1956)}]{biot-jacoustsocam-1956a}%
  \BibitemOpen
  \bibfield  {author} {\bibinfo {author} {\bibfnamefont {M.~A.}\ \bibnamefont
  {Biot}},\ }\bibfield  {title} {\enquote {\bibinfo {title} {Theory of
  propagation of elastic waves in a fluid-saturated porous solid. {I}.
  {Low}-frequency range},}\ }\href {\doibase 10.1121/1.1908239} {\bibfield
  {journal} {\bibinfo  {journal} {Journal of the Acoustical Society of
  America}\ }\textbf {\bibinfo {volume} {28}},\ \bibinfo {pages} {168--178}
  (\bibinfo {year} {1956})}\BibitemShut {NoStop}%
\bibitem [{\citenamefont {Biot}(1962)}]{biot-japplphys-1962}%
  \BibitemOpen
  \bibfield  {author} {\bibinfo {author} {\bibfnamefont {M.~A.}\ \bibnamefont
  {Biot}},\ }\bibfield  {title} {\enquote {\bibinfo {title} {Mechanics of
  deformation and acoustic propagation in porous media},}\ }\href {\doibase
  10.1063/1.1728759} {\bibfield  {journal} {\bibinfo  {journal} {Journal of
  Applied Physics}\ }\textbf {\bibinfo {volume} {33}},\ \bibinfo {pages} {1482}
  (\bibinfo {year} {1962})}\BibitemShut {NoStop}%
\bibitem [{\citenamefont {Rice}\ and\ \citenamefont
  {Cleary}(1976)}]{rice-revgeophysspacephys-1976}%
  \BibitemOpen
  \bibfield  {author} {\bibinfo {author} {\bibfnamefont {J.~R.}\ \bibnamefont
  {Rice}}\ and\ \bibinfo {author} {\bibfnamefont {M.~P.}\ \bibnamefont
  {Cleary}},\ }\bibfield  {title} {\enquote {\bibinfo {title} {Some basic
  stress diffusion solutions for fluid-saturated elastic porous media with
  compressible constituents},}\ }\href@noop {} {\bibfield  {journal} {\bibinfo
  {journal} {Reviews of Geophysics and Space Physics}\ }\textbf {\bibinfo
  {volume} {14}},\ \bibinfo {pages} {227--241} (\bibinfo {year}
  {1976})}\BibitemShut {NoStop}%
\bibitem [{\citenamefont {Biot}(1972)}]{biot-indianaumathj-1972}%
  \BibitemOpen
  \bibfield  {author} {\bibinfo {author} {\bibfnamefont {M.~A.}\ \bibnamefont
  {Biot}},\ }\bibfield  {title} {\enquote {\bibinfo {title} {Theory of finite
  deformations of porous solids},}\ }\href@noop {} {\bibfield  {journal}
  {\bibinfo  {journal} {Indiana University Mathematics Journal}\ }\textbf
  {\bibinfo {volume} {21}},\ \bibinfo {pages} {597--620} (\bibinfo {year}
  {1972})}\BibitemShut {NoStop}%
\bibitem [{\citenamefont {Coussy}(2004)}]{coussy-wiley-2004}%
  \BibitemOpen
  \bibfield  {author} {\bibinfo {author} {\bibfnamefont {O.}~\bibnamefont
  {Coussy}},\ }\href@noop {} {\emph {\bibinfo {title} {Poromechanics}}}\
  (\bibinfo  {publisher} {Wiley},\ \bibinfo {year} {2004})\BibitemShut
  {NoStop}%
\bibitem [{\citenamefont {Holmes}\ and\ \citenamefont
  {Mow}(1990)}]{holmes-jbiomech-1990}%
  \BibitemOpen
  \bibfield  {author} {\bibinfo {author} {\bibfnamefont {M.~H.}\ \bibnamefont
  {Holmes}}\ and\ \bibinfo {author} {\bibfnamefont {V.~C.}\ \bibnamefont
  {Mow}},\ }\bibfield  {title} {\enquote {\bibinfo {title} {The nonlinear
  characteristics of soft gels and hydrated connective tissues in
  ultrafiltration},}\ }\href {\doibase 10.1016/0021-9290(90)90007-P} {\bibfield
   {journal} {\bibinfo  {journal} {Journal of Biomechanics}\ }\textbf {\bibinfo
  {volume} {23}},\ \bibinfo {pages} {1145--1156} (\bibinfo {year}
  {1990})}\BibitemShut {NoStop}%
\bibitem [{\citenamefont {Kwan}\ \emph {et~al.}(1990)\citenamefont {Kwan},
  \citenamefont {Lai},\ and\ \citenamefont {Mow}}]{kwan-jbiomechanics-1990}%
  \BibitemOpen
  \bibfield  {author} {\bibinfo {author} {\bibfnamefont {M.~K.}\ \bibnamefont
  {Kwan}}, \bibinfo {author} {\bibfnamefont {W.~M.}\ \bibnamefont {Lai}}, \
  and\ \bibinfo {author} {\bibfnamefont {V.~C.}\ \bibnamefont {Mow}},\
  }\bibfield  {title} {\enquote {\bibinfo {title} {A finite deformation theory
  for cartilage and other soft hydrated connective tissues---{I.} Efquilibrium
  results},}\ }\href {\doibase 10.1016/0021-9290(90)90348-7} {\bibfield
  {journal} {\bibinfo  {journal} {Journal of Biomechanics}\ }\textbf {\bibinfo
  {volume} {23}},\ \bibinfo {pages} {145--155} (\bibinfo {year}
  {1990})}\BibitemShut {NoStop}%
\bibitem [{\citenamefont {Simon}(1992)}]{simon-applmechrev-1992}%
  \BibitemOpen
  \bibfield  {author} {\bibinfo {author} {\bibfnamefont {B.~R.}\ \bibnamefont
  {Simon}},\ }\bibfield  {title} {\enquote {\bibinfo {title} {Multiphase
  poroelastic finite element models for soft tissue structures},}\ }\href {\doibase
  10.1115/1.3121397} {\bibfield  {journal} {\bibinfo  {journal} {Applied
  Mechanics Reviews}\ }\textbf {\bibinfo {volume} {45}},\ \bibinfo {pages}
  {191--218} (\bibinfo {year} {1992})}\BibitemShut {NoStop}%
\bibitem [{\citenamefont {Argoubi}\ and\ \citenamefont
  {Shirazi-Adl}(1996)}]{argoubi-jbiomech-1996}%
  \BibitemOpen
  \bibfield  {author} {\bibinfo {author} {\bibfnamefont {M}~\bibnamefont
  {Argoubi}}\ and\ \bibinfo {author} {\bibfnamefont {A}~\bibnamefont
  {Shirazi-Adl}},\ }\bibfield  {title} {\enquote {\bibinfo {title} {Poroelastic
  creep response analysis of a lumbar motion segment in compression},}\ }\href
  {\doibase 10.1016/0021-9290(96)00035-8} {\bibfield  {journal} {\bibinfo
  {journal} {Journal of Biomechanics}\ }\textbf {\bibinfo {volume} {29}},\
  \bibinfo {pages} {1331--1339} (\bibinfo {year} {1996})}\BibitemShut {NoStop}%
\bibitem [{\citenamefont {Federico}\ and\ \citenamefont
  {Grillo}(2012)}]{federico-mechmater-2012}%
  \BibitemOpen
  \bibfield  {author} {\bibinfo {author} {\bibfnamefont {S.}~\bibnamefont
  {Federico}}\ and\ \bibinfo {author} {\bibfnamefont {A.}~\bibnamefont
  {Grillo}},\ }\bibfield  {title} {\enquote {\bibinfo {title} {Elasticity and
  permeability of porous fibre-reinforced materials under large
  deformations},}\ }\href {\doibase 10.1016/j.mechmat.2011.07.010} {\bibfield
  {journal} {\bibinfo  {journal} {Mechanics of Materials}\ }\textbf {\bibinfo
  {volume} {44}},\ \bibinfo {pages} {58--71} (\bibinfo {year}
  {2012})}\BibitemShut {NoStop}%
\bibitem [{\citenamefont {Tomic}\ \emph {et~al.}(2014)\citenamefont {Tomic},
  \citenamefont {Grillo},\ and\ \citenamefont
  {Federico}}]{tomic-imajapplmath-2014}%
  \BibitemOpen
  \bibfield  {author} {\bibinfo {author} {\bibfnamefont {A.}~\bibnamefont
  {Tomic}}, \bibinfo {author} {\bibfnamefont {A.}~\bibnamefont {Grillo}}, \
  and\ \bibinfo {author} {\bibfnamefont {S.}~\bibnamefont {Federico}},\
  }\bibfield  {title} {\enquote {\bibinfo {title} {Poroelastic materials
  reinforced by statistically oriented fibers---numerical implementation and
  applications to articular cartilage},}\ }\href {\doibase
  10.1093/imamat/hxu039} {\bibfield  {journal} {\bibinfo  {journal} {{IMA}
  Journal of Applied Mathematics}\ }\textbf {\bibinfo {volume} {79}},\ \bibinfo
  {pages} {1027--1059} (\bibinfo {year} {2014})}\BibitemShut {NoStop}%
\bibitem [{\citenamefont {Vuong}\ \emph {et~al.}(2015)\citenamefont {Vuong},
  \citenamefont {Yoshihara},\ and\ \citenamefont
  {Wall}}]{vuong-compmethapplmecheng-2015}%
  \BibitemOpen
  \bibfield  {author} {\bibinfo {author} {\bibfnamefont {A.-T.}\ \bibnamefont
  {Vuong}}, \bibinfo {author} {\bibfnamefont {L.}~\bibnamefont {Yoshihara}}, \
  and\ \bibinfo {author} {\bibfnamefont {W.~A.}\ \bibnamefont {Wall}},\
  }\bibfield  {title} {\enquote {\bibinfo {title} {A general approach for
  modeling interacting flow through porous media under finite deformations},}\
  }\href {\doibase 10.1016/j.cma.2014.08.018} {\bibfield  {journal} {\bibinfo
  {journal} {Computer Methods in Applied Mechanics and Engineering}\ }\textbf
  {\bibinfo {volume} {283}},\ \bibinfo {pages} {1240--1259} (\bibinfo {year}
  {2015})}\BibitemShut {NoStop}%
\bibitem [{\citenamefont {Carter}\ \emph {et~al.}(1979)\citenamefont {Carter},
  \citenamefont {Booker},\ and\ \citenamefont {Small}}]{carter-intjnag-1979a}%
  \BibitemOpen
  \bibfield  {author} {\bibinfo {author} {\bibfnamefont {J.~P.}\ \bibnamefont
  {Carter}}, \bibinfo {author} {\bibfnamefont {J.~R.}\ \bibnamefont {Booker}},
  \ and\ \bibinfo {author} {\bibfnamefont {J.~C.}\ \bibnamefont {Small}},\
  }\bibfield  {title} {\enquote {\bibinfo {title} {The analysis of finite
  elasto-plastic consolidation},}\ }\href {\doibase 10.1002/nag.1610030202}
  {\bibfield  {journal} {\bibinfo  {journal} {International Journal for
  Numerical and Analytical Methods in Geomechanics}\ }\textbf {\bibinfo
  {volume} {3}},\ \bibinfo {pages} {107--129} (\bibinfo {year}
  {1979})}\BibitemShut {NoStop}%
\bibitem [{\citenamefont {Borja}\ and\ \citenamefont
  {Alarc{\'{o}}n}(1995)}]{borja-compmethapplmecheng-1995}%
  \BibitemOpen
  \bibfield  {author} {\bibinfo {author} {\bibfnamefont {R.~I.}\ \bibnamefont
  {Borja}}\ and\ \bibinfo {author} {\bibfnamefont {E.}~\bibnamefont
  {Alarc{\'{o}}n}},\ }\bibfield  {title} {\enquote {\bibinfo {title} {A
  mathematical framework for finite strain elastoplastic consolidation. Part 1:
  Balance laws, variational formulation, and linearization},}\ }\href {\doibase
  10.1016/0045-7825(94)00720-8} {\bibfield  {journal} {\bibinfo  {journal}
  {Computer Methods in Applied Mechanics and Engineering}\ }\textbf {\bibinfo
  {volume} {122}},\ \bibinfo {pages} {145--171} (\bibinfo {year}
  {1995})}\BibitemShut {NoStop}%
\bibitem [{\citenamefont {Li}\ \emph {et~al.}(2004)\citenamefont {Li},
  \citenamefont {Borja},\ and\ \citenamefont
  {Regueiro}}]{li-compmethapplmecheng-2004}%
  \BibitemOpen
  \bibfield  {author} {\bibinfo {author} {\bibfnamefont {C.}~\bibnamefont
  {Li}}, \bibinfo {author} {\bibfnamefont {R.~I.}\ \bibnamefont {Borja}}, \
  and\ \bibinfo {author} {\bibfnamefont {R.~A.}\ \bibnamefont {Regueiro}},\
  }\bibfield  {title} {\enquote {\bibinfo {title} {Dynamics of porous media at
  finite strain},}\ }\href {\doibase 10.1016/j.cma.2004.02.014} {\bibfield
  {journal} {\bibinfo  {journal} {Computer Methods in Applied Mechanics and
  Engineering}\ }\textbf {\bibinfo {volume} {193}},\ \bibinfo {pages}
  {3837--3870} (\bibinfo {year} {2004})}\BibitemShut {NoStop}%
\bibitem [{\citenamefont {Uzuoka}\ and\ \citenamefont
  {Borja}(2011)}]{uzuoka-intjnag-2011}%
  \BibitemOpen
  \bibfield  {author} {\bibinfo {author} {\bibfnamefont {R.}~\bibnamefont
  {Uzuoka}}\ and\ \bibinfo {author} {\bibfnamefont {R.~I.}\ \bibnamefont
  {Borja}},\ }\bibfield  {title} {\enquote {\bibinfo {title} {Dynamics of
  unsaturated poroelastic solids at finite strain},}\ }\href {\doibase
  10.1002/nag.1061} {\bibfield  {journal} {\bibinfo  {journal} {International
  Journal for Numerical and Analytical Methods in Geomechanics}\ }\textbf
  {\bibinfo {volume} {36}},\ \bibinfo {pages} {1535--1573} (\bibinfo {year}
  {2011})}\BibitemShut {NoStop}%
\bibitem [{\citenamefont {Bennethum}\ and\ \citenamefont
  {Cushman}(1996)}]{bennethum-intjengsci-1996}%
  \BibitemOpen
  \bibfield  {author} {\bibinfo {author} {\bibfnamefont {L.~S.}\ \bibnamefont
  {Bennethum}}\ and\ \bibinfo {author} {\bibfnamefont {J.~H.}\ \bibnamefont
  {Cushman}},\ }\bibfield  {title} {\enquote {\bibinfo {title} {Multiscale,
  hybrid mixture theory for swelling systems---{I}: Balance laws},}\ }\href
  {\doibase 10.1016/0020-7225(95)00089-5} {\bibfield  {journal} {\bibinfo
  {journal} {International Journal of Engineering Science}\ }\textbf {\bibinfo
  {volume} {34}},\ \bibinfo {pages} {125--145} (\bibinfo {year}
  {1996})}\BibitemShut {NoStop}%
\bibitem [{\citenamefont {Huyghe}\ and\ \citenamefont
  {Janssen}(1997)}]{huyghe-intjengsci-1997}%
  \BibitemOpen
  \bibfield  {author} {\bibinfo {author} {\bibfnamefont {J.~M.}\ \bibnamefont
  {Huyghe}}\ and\ \bibinfo {author} {\bibfnamefont {J.~D.}\ \bibnamefont
  {Janssen}},\ }\bibfield  {title} {\enquote {\bibinfo {title} {Quadriphasic
  mechanics of swelling incompressible porous media},}\ }\href {\doibase
  10.1016/S0020-7225(96)00119-X} {\bibfield  {journal} {\bibinfo  {journal}
  {International Journal of Engineering Science}\ }\textbf {\bibinfo {volume}
  {35}},\ \bibinfo {pages} {793--802} (\bibinfo {year} {1997})}\BibitemShut
  {NoStop}%
\bibitem [{\citenamefont {Hong}\ \emph {et~al.}(2008)\citenamefont {Hong},
  \citenamefont {Zhao}, \citenamefont {Zhou},\ and\ \citenamefont
  {Suo}}]{hong-jmps-2008}%
  \BibitemOpen
  \bibfield  {author} {\bibinfo {author} {\bibfnamefont {W.}~\bibnamefont
  {Hong}}, \bibinfo {author} {\bibfnamefont {X.}~\bibnamefont {Zhao}}, \bibinfo
  {author} {\bibfnamefont {J.}~\bibnamefont {Zhou}}, \ and\ \bibinfo {author}
  {\bibfnamefont {Z.}~\bibnamefont {Suo}},\ }\bibfield  {title} {\enquote
  {\bibinfo {title} {A theory of coupled diffusion and large deformation in
  polymeric gels},}\ }\href {\doibase 10.1016/j.jmps.2007.11.010} {\bibfield
  {journal} {\bibinfo  {journal} {Journal of the Mechanics and Physics of
  Solids}\ }\textbf {\bibinfo {volume} {56}},\ \bibinfo {pages} {1779--1793}
  (\bibinfo {year} {2008})}\BibitemShut {NoStop}%
\bibitem [{\citenamefont {Duda}\ \emph {et~al.}(2010)\citenamefont {Duda},
  \citenamefont {Souza},\ and\ \citenamefont {Fried}}]{duda-jmps-2010}%
  \BibitemOpen
  \bibfield  {author} {\bibinfo {author} {\bibfnamefont {F.~P.}\ \bibnamefont
  {Duda}}, \bibinfo {author} {\bibfnamefont {A.~C.}\ \bibnamefont {Souza}}, \
  and\ \bibinfo {author} {\bibfnamefont {E.}~\bibnamefont {Fried}},\ }\bibfield
   {title} {\enquote {\bibinfo {title} {A theory for species migration in a
  finitely strained solid with application to polymer network swelling},}\
  }\href {\doibase 10.1016/j.jmps.2010.01.009} {\bibfield  {journal} {\bibinfo
  {journal} {Journal of the Mechanical and Physics of Solids}\ }\textbf
  {\bibinfo {volume} {58}},\ \bibinfo {pages} {515--529} (\bibinfo {year}
  {2010})}\BibitemShut {NoStop}%
\bibitem [{\citenamefont {Chester}\ and\ \citenamefont
  {Anand}(2010)}]{chester-jmps-2010}%
  \BibitemOpen
  \bibfield  {author} {\bibinfo {author} {\bibfnamefont {S.~A.}\ \bibnamefont
  {Chester}}\ and\ \bibinfo {author} {\bibfnamefont {L.}~\bibnamefont
  {Anand}},\ }\bibfield  {title} {\enquote {\bibinfo {title} {A coupled theory
  of fluid permeation and large deformations for elastomeric materials},}\
  }\href {\doibase 10.1016/j.jmps.2010.07.020} {\bibfield  {journal} {\bibinfo
  {journal} {Journal of the Mechanics and Physics of Solids}\ }\textbf
  {\bibinfo {volume} {58}},\ \bibinfo {pages} {1879--1906} (\bibinfo {year}
  {2010})}\BibitemShut {NoStop}%
\bibitem [{\citenamefont {Hassanizadeh}\ and\ \citenamefont
  {Gray}(1979)}]{hassanizadeh-awr-1979}%
  \BibitemOpen
  \bibfield  {author} {\bibinfo {author} {\bibfnamefont {S.~M.}\ \bibnamefont
  {Hassanizadeh}}\ and\ \bibinfo {author} {\bibfnamefont {W.~G.}\ \bibnamefont
  {Gray}},\ }\bibfield  {title} {\enquote {\bibinfo {title} {General
  conservation equations for multi-phase systems: 1. {A}veraging procedure},}\
  }\href@noop {} {\bibfield  {journal} {\bibinfo  {journal} {Advances in Water
  Resources}\ }\textbf {\bibinfo {volume} {2}},\ \bibinfo {pages} {131--144}
  (\bibinfo {year} {1979})}\BibitemShut {NoStop}%
\bibitem [{\citenamefont {Bennethum}(2006)}]{bennethum-jengmech-2006}%
  \BibitemOpen
  \bibfield  {author} {\bibinfo {author} {\bibfnamefont {L.~S.}\ \bibnamefont
  {Bennethum}},\ }\bibfield  {title} {\enquote {\bibinfo {title}
  {Compressibility moduli for porous materials incorporating volume
  fraction},}\ }\href {\doibase 10.1061/(ASCE)0733-9399(2006)132:11(1205)}
  {\bibfield  {journal} {\bibinfo  {journal} {Journal of Engineering
  Mechanics}\ }\textbf {\bibinfo {volume} {132}},\ \bibinfo {pages}
  {1205--1214} (\bibinfo {year} {2006})}\BibitemShut {NoStop}%
\bibitem [{\citenamefont {Jaeger}\ \emph {et~al.}(2007)\citenamefont {Jaeger},
  \citenamefont {Cook},\ and\ \citenamefont {Zimmerman}}]{jaeger-wiley-2007}%
  \BibitemOpen
  \bibfield  {author} {\bibinfo {author} {\bibfnamefont {J.~C.}\ \bibnamefont
  {Jaeger}}, \bibinfo {author} {\bibfnamefont {N.~G.~W.}\ \bibnamefont {Cook}},
  \ and\ \bibinfo {author} {\bibfnamefont {R.}~\bibnamefont {Zimmerman}},\
  }\href@noop {} {\emph {\bibinfo {title} {Fundamentals of Rock Mechanics}}},\
  \bibinfo {edition} {4th}\ ed.\ (\bibinfo  {publisher} {Wiley-Blackwell},\
  \bibinfo {year} {2007})\ \bibinfo {note} {ISBN~978-0-632-05759-7}\BibitemShut {NoStop}%
\bibitem [{\citenamefont {Gajo}(2010)}]{gajo-rspa-2010}%
  \BibitemOpen
  \bibfield  {author} {\bibinfo {author} {\bibfnamefont {A.}~\bibnamefont
  {Gajo}},\ }\bibfield  {title} {\enquote {\bibinfo {title} {A general approach
  to isothermal hyperelastic modelling of saturated porous media at finite
  strains with compressible solid constituents},}\ }\href {\doibase
  10.1098/rspa.2010.0018} {\bibfield  {journal} {\bibinfo  {journal}
  {Proceedings of the Royal Society A}\ }\textbf {\bibinfo {volume} {466}},\
  \bibinfo {pages} {3061--3087} (\bibinfo {year} {2010})}\BibitemShut {NoStop}%
\bibitem [{\citenamefont {Marsden}\ and\ \citenamefont
  {Hughes}(1983)}]{marsden-dover-1983}%
  \BibitemOpen
  \bibfield  {author} {\bibinfo {author} {\bibfnamefont {J.~E.}\ \bibnamefont
  {Marsden}}\ and\ \bibinfo {author} {\bibfnamefont {T.~J.~R.}\ \bibnamefont
  {Hughes}},\ }\href@noop {} {\emph {\bibinfo {title} {Mathematical foundations
  of elasticity}}}\ (\bibinfo  {publisher} {Dover Publications, Inc.},\
  \bibinfo {address} {New York},\ \bibinfo {year} {1983})\ \bibinfo {note}
  {ISBN~0-486-67865-2}\BibitemShut {NoStop}%
\bibitem [{\citenamefont {Gurtin}\ \emph {et~al.}(2010)\citenamefont {Gurtin},
  \citenamefont {Fried},\ and\ \citenamefont {Anand}}]{gurtin-cambridge-2010}%
  \BibitemOpen
  \bibfield  {author} {\bibinfo {author} {\bibfnamefont {M.~E.}\ \bibnamefont
  {Gurtin}}, \bibinfo {author} {\bibfnamefont {E.}~\bibnamefont {Fried}}, \
  and\ \bibinfo {author} {\bibfnamefont {L.}~\bibnamefont {Anand}},\
  }\href@noop {} {\emph {\bibinfo {title} {The Mechanics and Thermodynamics of
  Continua}}}\ (\bibinfo  {publisher} {Cambridge University Press},\ \bibinfo
  {year} {2010})\BibitemShut {NoStop}%
\bibitem [{\citenamefont {Bird}\ \emph
  {et~al.}(1987{\natexlab{a}})\citenamefont {Bird}, \citenamefont {Armstrong},\
  and\ \citenamefont {Hassager}}]{bird-wiley-1987a}%
  \BibitemOpen
  \bibfield  {author} {\bibinfo {author} {\bibfnamefont {R.~B.}\ \bibnamefont
  {Bird}}, \bibinfo {author} {\bibfnamefont {R.~C.}\ \bibnamefont {Armstrong}},
  \ and\ \bibinfo {author} {\bibfnamefont {O.}~\bibnamefont {Hassager}},\
  }\href@noop {} {\emph {\bibinfo {title} {Dynamics of Polymer Liquids, Volume
  1: Fluid Mechanics}}}\ (\bibinfo  {publisher} {Wiley-Interscience},\ \bibinfo
  {year} {1987})\BibitemShut {NoStop}%
\bibitem [{\citenamefont {Bird}\ \emph
  {et~al.}(1987{\natexlab{b}})\citenamefont {Bird}, \citenamefont {Curtiss},
  \citenamefont {Armstrong},\ and\ \citenamefont
  {Hassager}}]{bird-wiley-1987b}%
  \BibitemOpen
  \bibfield  {author} {\bibinfo {author} {\bibfnamefont {R.~B.}\ \bibnamefont
  {Bird}}, \bibinfo {author} {\bibfnamefont {C.~F.}\ \bibnamefont {Curtiss}},
  \bibinfo {author} {\bibfnamefont {R.~C.}\ \bibnamefont {Armstrong}}, \ and\
  \bibinfo {author} {\bibfnamefont {O.}~\bibnamefont {Hassager}},\ }\href@noop
  {} {\emph {\bibinfo {title} {Dynamics of Polymer Liquids, Volume 2: Kinetic
  Theory}}}\ (\bibinfo  {publisher} {Wiley-Interscience},\ \bibinfo {year}
  {1987})\BibitemShut {NoStop}%
\bibitem [{\citenamefont {Pinder}\ and\ \citenamefont
  {Gray}(2008)}]{pinder-wiley-2008}%
  \BibitemOpen
  \bibfield  {author} {\bibinfo {author} {\bibfnamefont {G.~F.}\ \bibnamefont
  {Pinder}}\ and\ \bibinfo {author} {\bibfnamefont {W.~G.}\ \bibnamefont
  {Gray}},\ }\href@noop {} {\emph {\bibinfo {title} {Essentials of multiphase
  flow in porous media}}}\ (\bibinfo  {publisher} {Wiley-Interscience},\
  \bibinfo {address} {Hoboken, NJ, USA},\ \bibinfo {year} {2008})\BibitemShut
  {NoStop}%
\bibitem [{\citenamefont {Wu}\ and\ \citenamefont
  {Pruess}(1996)}]{wu-advporousmed-1996}%
  \BibitemOpen
  \bibfield  {author} {\bibinfo {author} {\bibfnamefont {Y.-S.}\ \bibnamefont
  {Wu}}\ and\ \bibinfo {author} {\bibfnamefont {K.}~\bibnamefont {Pruess}},\
  }\bibfield  {title} {\enquote {\bibinfo {title} {Flow of non-{Newtonian}
  fluids in porous media},}\ }in\ \href {\doibase
  10.1016/S1873-975X(96)80004-7} {\emph {\bibinfo {booktitle} {Advances in
  Porous Media}}},\ Vol.~\bibinfo {volume} {3}\ (\bibinfo  {publisher}
  {Elsevier},\ \bibinfo {year} {1996})\ Chap.~\bibinfo {chapter} {2}, pp.\
  \bibinfo {pages} {87--184}\BibitemShut {NoStop}%
\bibitem [{\citenamefont {{von Terzaghi}}(1936)}]{terzaghi-procsmfe-1936}%
  \BibitemOpen
  \bibfield  {author} {\bibinfo {author} {\bibfnamefont {K.}~\bibnamefont {{von
  Terzaghi}}},\ }\bibfield  {title} {\enquote {\bibinfo {title} {The shearing
  resistance of saturated soil and the angle between the planes of shear},}\
  }in\ \href@noop {} {\emph {\bibinfo {booktitle} {Proceedings of the
  International Conference on Soil Mechanics and Foundation Engineering, June
  22 to 26}}},\ Vol.~\bibinfo {volume} {1}\ (\bibinfo {year} {1936})\ pp.\
  \bibinfo {pages} {54--56}\BibitemShut {NoStop}%
\bibitem [{\citenamefont {Nur}\ and\ \citenamefont
  {Byerlee}(1971)}]{nur-jgr-1971}%
  \BibitemOpen
  \bibfield  {author} {\bibinfo {author} {\bibfnamefont {A.}~\bibnamefont
  {Nur}}\ and\ \bibinfo {author} {\bibfnamefont {J.~D.}\ \bibnamefont
  {Byerlee}},\ }\bibfield  {title} {\enquote {\bibinfo {title} {An exact
  effective stress law for elastic deformation of rock with fluids},}\ }\href
  {\doibase 10.1029/JB076i026p06414} {\bibfield  {journal} {\bibinfo  {journal}
  {Journal of Geophysical Research}\ }\textbf {\bibinfo {volume} {76}},\
  \bibinfo {pages} {6414--6419} (\bibinfo {year} {1971})}\BibitemShut {NoStop}%
\bibitem [{\citenamefont {Anand}(1979)}]{anand-japplmech-1979}%
  \BibitemOpen
  \bibfield  {author} {\bibinfo {author} {\bibfnamefont {L.}~\bibnamefont
  {Anand}},\ }\bibfield  {title} {\enquote {\bibinfo {title} {On {H. Hencky's}
  approximate strain-energy function for moderate deformations},}\ }\href
  {\doibase 10.1115/1.3424532} {\bibfield  {journal} {\bibinfo  {journal}
  {Journal of Applied Mechanics}\ }\textbf {\bibinfo {volume} {46}},\ \bibinfo
  {pages} {78--82} (\bibinfo {year} {1979})}\BibitemShut {NoStop}%
\bibitem [{\citenamefont {Xiao}\ and\ \citenamefont
  {Chen}(2002)}]{xiao-actamechanica-2002}%
  \BibitemOpen
  \bibfield  {author} {\bibinfo {author} {\bibfnamefont {H.}~\bibnamefont
  {Xiao}}\ and\ \bibinfo {author} {\bibfnamefont {L.~S.}\ \bibnamefont
  {Chen}},\ }\bibfield  {title} {\enquote {\bibinfo {title} {Hencky's
  elasticity model and linear stress-strain relations in isotropic finite
  hyperelasticity},}\ }\href {\doibase 10.1007/BF01182154} {\bibfield
  {journal} {\bibinfo  {journal} {Acta Mechanica}\ }\textbf {\bibinfo {volume}
  {157}},\ \bibinfo {pages} {51--60} (\bibinfo {year} {2002})}\BibitemShut
  {NoStop}%
\bibitem [{\citenamefont {Anand}(1986)}]{anand-jmps-1986}%
  \BibitemOpen
  \bibfield  {author} {\bibinfo {author} {\bibfnamefont {L.}~\bibnamefont
  {Anand}},\ }\bibfield  {title} {\enquote {\bibinfo {title} {Moderate
  deformations in extension-torsion of incompressible isotropic elastic
  materials},}\ }\href {\doibase 10.1016/0022-5096(86)90021-9} {\bibfield
  {journal} {\bibinfo  {journal} {Journal of the Mechanics and Physics of
  Solids}\ }\textbf {\bibinfo {volume} {34}},\ \bibinfo {pages} {293--304}
  (\bibinfo {year} {1986})}\BibitemShut {NoStop}%
\bibitem [{\citenamefont {Ba\u{z}ant}(1998)}]{bazant-jengmatertechnol-1998}%
  \BibitemOpen
  \bibfield  {author} {\bibinfo {author} {\bibfnamefont {Z.~P.}\ \bibnamefont
  {Ba\u{z}ant}},\ }\bibfield  {title} {\enquote {\bibinfo {title}
  {Easy-to-compute tensors with symmetric inverse approximating {Hencky} finite
  strain and its rate},}\ }\href {\doibase 10.1115/1.2807001} {\bibfield
  {journal} {\bibinfo  {journal} {Journal of Engineering Materials and
  Technology}\ }\textbf {\bibinfo {volume} {120}},\ \bibinfo {pages} {131--136}
  (\bibinfo {year} {1998})}\BibitemShut {NoStop}%
\bibitem [{\citenamefont {Beavers}\ and\ \citenamefont
  {Joseph}(1967)}]{beavers-jfm-1967}%
  \BibitemOpen
  \bibfield  {author} {\bibinfo {author} {\bibfnamefont {G.~S.}\ \bibnamefont
  {Beavers}}\ and\ \bibinfo {author} {\bibfnamefont {D.~D.}\ \bibnamefont
  {Joseph}},\ }\bibfield  {title} {\enquote {\bibinfo {title} {Boundary
  conditions at a naturally permeable wall},}\ }\href {\doibase
  10.1017/S0022112067001375} {\bibfield  {journal} {\bibinfo  {journal}
  {Journal of Fluid Mechanics}\ }\textbf {\bibinfo {volume} {30}},\ \bibinfo
  {pages} {197--207} (\bibinfo {year} {1967})}\BibitemShut {NoStop}%
\bibitem [{\citenamefont {Shavit}(2009)}]{shavit-tipm-2009}%
  \BibitemOpen
  \bibfield  {author} {\bibinfo {author} {\bibfnamefont {U.}~\bibnamefont
  {Shavit}},\ }\bibfield  {title} {\enquote {\bibinfo {title} {{Special Issue}
  on `{Transport} phenomena at the interface between fluid and porous
  domains'},}\ }\href {\doibase 10.1007/s11242-009-9414-1} {\bibfield
  {journal} {\bibinfo  {journal} {Transport in Porous Media}\ }\textbf
  {\bibinfo {volume} {78}},\ \bibinfo {pages} {327--330} (\bibinfo {year}
  {2009})}\BibitemShut {NoStop}%
\bibitem [{\citenamefont {Mosthaf}\ \emph {et~al.}(2011)\citenamefont
  {Mosthaf}, \citenamefont {Baber}, \citenamefont {Flemisch}, \citenamefont
  {Helmig}, \citenamefont {Leijnse}, \citenamefont {Rybak},\ and\ \citenamefont
  {Wohlmuth}}]{mosthaf-wrr-2011}%
  \BibitemOpen
  \bibfield  {author} {\bibinfo {author} {\bibfnamefont {K.}~\bibnamefont
  {Mosthaf}}, \bibinfo {author} {\bibfnamefont {K.}~\bibnamefont {Baber}},
  \bibinfo {author} {\bibfnamefont {B.}~\bibnamefont {Flemisch}}, \bibinfo
  {author} {\bibfnamefont {R.}~\bibnamefont {Helmig}}, \bibinfo {author}
  {\bibfnamefont {A.}~\bibnamefont {Leijnse}}, \bibinfo {author} {\bibfnamefont
  {I.}~\bibnamefont {Rybak}}, \ and\ \bibinfo {author} {\bibfnamefont
  {B.}~\bibnamefont {Wohlmuth}},\ }\bibfield  {title} {\enquote {\bibinfo
  {title} {A coupling concept for two-phase compositional porous-medium and
  single-phase compositional free flow},}\ }\href {\doibase
  10.1029/2011WR010685} {\bibfield  {journal} {\bibinfo  {journal} {Water
  Resources Research}\ }\textbf {\bibinfo {volume} {47}},\ \bibinfo {pages}
  {W10522} (\bibinfo {year} {2011})}\BibitemShut {NoStop}%
\bibitem [{\citenamefont {Barry}\ and\ \citenamefont
  {Aldis}(1990)}]{barry-jbiomech-1990}%
  \BibitemOpen
  \bibfield  {author} {\bibinfo {author} {\bibfnamefont {S.~I.}\ \bibnamefont
  {Barry}}\ and\ \bibinfo {author} {\bibfnamefont {G.~K.}\ \bibnamefont
  {Aldis}},\ }\bibfield  {title} {\enquote {\bibinfo {title} {Comparison of
  models for flow induced deformation of soft biological tissue},}\ }\href
  {\doibase 10.1016/0021-9290(90)90164-X} {\bibfield  {journal} {\bibinfo
  {journal} {Journal of Biomechanics}\ }\textbf {\bibinfo {volume} {23}},\
  \bibinfo {pages} {647--654} (\bibinfo {year} {1990})}\BibitemShut {NoStop}%
\bibitem [{\citenamefont {Barry}\ and\ \citenamefont
  {Aldis}(1991)}]{barry-intjnonlinmech-1991}%
  \BibitemOpen
  \bibfield  {author} {\bibinfo {author} {\bibfnamefont {S.~I.}\ \bibnamefont
  {Barry}}\ and\ \bibinfo {author} {\bibfnamefont {G.~K.}\ \bibnamefont
  {Aldis}},\ }\bibfield  {title} {\enquote {\bibinfo {title} {Unsteady flow
  induced deformation of porous materials},}\ }\href {\doibase
  10.1016/0020-7462(91)90020-T} {\bibfield  {journal} {\bibinfo  {journal}
  {International Journal of Non-Linear Mechanics}\ }\textbf {\bibinfo {volume}
  {26}},\ \bibinfo {pages} {687--699} (\bibinfo {year} {1991})}\BibitemShut
  {NoStop}%
\bibitem [{\citenamefont {Preziosi}\ \emph {et~al.}(1996)\citenamefont
  {Preziosi}, \citenamefont {Joseph},\ and\ \citenamefont
  {Beavers}}]{preziosi-intjmultiphaseflow-1996}%
  \BibitemOpen
  \bibfield  {author} {\bibinfo {author} {\bibfnamefont {L.}~\bibnamefont
  {Preziosi}}, \bibinfo {author} {\bibfnamefont {D.~D.}\ \bibnamefont
  {Joseph}}, \ and\ \bibinfo {author} {\bibfnamefont {G.~S.}\ \bibnamefont
  {Beavers}},\ }\bibfield  {title} {\enquote {\bibinfo {title} {Infiltration of
  initially dry, deformable porous media},}\ }\href {\doibase
  10.1016/0301-9322(96)00035-3} {\bibfield  {journal} {\bibinfo  {journal}
  {International Journal of Multiphase Flow}\ }\textbf {\bibinfo {volume}
  {22}},\ \bibinfo {pages} {1205--1222} (\bibinfo {year} {1996})}\BibitemShut
  {NoStop}%
\bibitem [{\citenamefont {Holmes}(1986)}]{holmes-jbiomecheng-1986}%
  \BibitemOpen
  \bibfield  {author} {\bibinfo {author} {\bibfnamefont {M.~H.}\ \bibnamefont
  {Holmes}},\ }\bibfield  {title} {\enquote {\bibinfo {title} {Finite
  deformation of soft tissue: {Analysis} of a mixture model in uni-axial
  compression},}\ }\href {\doibase 10.1115/1.3138633} {\bibfield  {journal}
  {\bibinfo  {journal} {Journal of Biomechanical Engineering}\ }\textbf
  {\bibinfo {volume} {108}},\ \bibinfo {pages} {372--381} (\bibinfo {year}
  {1986})}\BibitemShut {NoStop}%
\bibitem [{\citenamefont {Dawson}\ \emph {et~al.}(2008)\citenamefont {Dawson},
  \citenamefont {Mc{K}inley},\ and\ \citenamefont
  {Gibson}}]{dawson-japplmech-2008}%
  \BibitemOpen
  \bibfield  {author} {\bibinfo {author} {\bibfnamefont {M.~A.}\ \bibnamefont
  {Dawson}}, \bibinfo {author} {\bibfnamefont {G.~H.}\ \bibnamefont
  {Mc{K}inley}}, \ and\ \bibinfo {author} {\bibfnamefont {L.~J.}\ \bibnamefont
  {Gibson}},\ }\bibfield  {title} {\enquote {\bibinfo {title} {The dynamic
  compressive response of an open-cell foam impregnated with a {Newtonian}
  fluid},}\ }\href {\doibase 10.1115/1.2912940} {\bibfield  {journal} {\bibinfo
   {journal} {Journal of Applied Mechanics}\ }\textbf {\bibinfo {volume}
  {75}},\ \bibinfo {pages} {041015} (\bibinfo {year} {2008})}\BibitemShut
  {NoStop}%
\bibitem [{\citenamefont {Sommer}\ and\ \citenamefont
  {Mortensen}(1996)}]{sommer-jfm-1996}%
  \BibitemOpen
  \bibfield  {author} {\bibinfo {author} {\bibfnamefont {J.~L.}\ \bibnamefont
  {Sommer}}\ and\ \bibinfo {author} {\bibfnamefont {A.}~\bibnamefont
  {Mortensen}},\ }\bibfield  {title} {\enquote {\bibinfo {title} {Forced
  unidirectional infiltration of deformable porous media},}\ }\href {\doibase
  10.1017/S002211209600256X} {\bibfield  {journal} {\bibinfo  {journal}
  {Journal of Fluid Mechanics}\ }\textbf {\bibinfo {volume} {311}},\ \bibinfo
  {pages} {193--217} (\bibinfo {year} {1996})}\BibitemShut {NoStop}%
\bibitem [{\citenamefont {Anderson}(2005)}]{anderson-physfluids-2005}%
  \BibitemOpen
  \bibfield  {author} {\bibinfo {author} {\bibfnamefont {D.~M.}\ \bibnamefont
  {Anderson}},\ }\bibfield  {title} {\enquote {\bibinfo {title} {Imbibition of
  a liquid droplet on a deformable porous substrate},}\ }\href {\doibase
  10.1063/1.2000247} {\bibfield  {journal} {\bibinfo  {journal} {Physics of
  Fluids}\ }\textbf {\bibinfo {volume} {17}},\ \bibinfo {pages} {087104}
  (\bibinfo {year} {2005})}\BibitemShut {NoStop}%
\bibitem [{\citenamefont {Siddique}\ \emph {et~al.}(2009)\citenamefont
  {Siddique}, \citenamefont {Anderson},\ and\ \citenamefont
  {Bondarev}}]{siddique-physfluids-2009}%
  \BibitemOpen
  \bibfield  {author} {\bibinfo {author} {\bibfnamefont {J.~I.}\ \bibnamefont
  {Siddique}}, \bibinfo {author} {\bibfnamefont {D.~M.}\ \bibnamefont
  {Anderson}}, \ and\ \bibinfo {author} {\bibfnamefont {A.}~\bibnamefont
  {Bondarev}},\ }\bibfield  {title} {\enquote {\bibinfo {title} {Capillary rise
  of a liquid into a deformable porous material},}\ }\href {\doibase
  10.1063/1.3068194} {\bibfield  {journal} {\bibinfo  {journal} {Physics of
  Fluids}\ }\textbf {\bibinfo {volume} {21}},\ \bibinfo {pages} {013106}
  (\bibinfo {year} {2009})}\BibitemShut {NoStop}%
\bibitem [{\citenamefont {Beavers}\ \emph
  {et~al.}(1981{\natexlab{a}})\citenamefont {Beavers}, \citenamefont {Hajji},\
  and\ \citenamefont {Sparrow}}]{beavers-jfluidseng-1981a}%
  \BibitemOpen
  \bibfield  {author} {\bibinfo {author} {\bibfnamefont {G.~S.}\ \bibnamefont
  {Beavers}}, \bibinfo {author} {\bibfnamefont {A.}~\bibnamefont {Hajji}}, \
  and\ \bibinfo {author} {\bibfnamefont {E.~M.}\ \bibnamefont {Sparrow}},\
  }\bibfield  {title} {\enquote {\bibinfo {title} {Fluid flow through a class
  of highly-deformable porous media. {P}art {I}: {E}xperiments with air},}\
  }\href {\doibase 10.1115/1.3240807} {\bibfield  {journal} {\bibinfo
  {journal} {Journal of Fluids Engineering}\ }\textbf {\bibinfo {volume}
  {103}},\ \bibinfo {pages} {432} (\bibinfo {year}
  {1981}{\natexlab{a}})}\BibitemShut {NoStop}%
\bibitem [{\citenamefont {Beavers}\ \emph
  {et~al.}(1981{\natexlab{b}})\citenamefont {Beavers}, \citenamefont
  {Wittenberg},\ and\ \citenamefont {Sparrow}}]{beavers-jfluidseng-1981b}%
  \BibitemOpen
  \bibfield  {author} {\bibinfo {author} {\bibfnamefont {G.~S.}\ \bibnamefont
  {Beavers}}, \bibinfo {author} {\bibfnamefont {K.}~\bibnamefont {Wittenberg}},
  \ and\ \bibinfo {author} {\bibfnamefont {E.~M.}\ \bibnamefont {Sparrow}},\
  }\bibfield  {title} {\enquote {\bibinfo {title} {Fluid flow through a class
  of highly-deformable porous media. {P}art {II}: {E}xperiments with water},}\
  }\href {\doibase 10.1115/1.3240810} {\bibfield  {journal} {\bibinfo
  {journal} {Journal of Fluids Engineering}\ }\textbf {\bibinfo {volume}
  {103}},\ \bibinfo {pages} {440} (\bibinfo {year}
  {1981}{\natexlab{b}})}\BibitemShut {NoStop}%
\bibitem [{\citenamefont {Parker}\ \emph {et~al.}(1987)\citenamefont {Parker},
  \citenamefont {Mehta},\ and\ \citenamefont {Caro}}]{parker-japplmech-1987}%
  \BibitemOpen
  \bibfield  {author} {\bibinfo {author} {\bibfnamefont {K.~H.}\ \bibnamefont
  {Parker}}, \bibinfo {author} {\bibfnamefont {R.~V.}\ \bibnamefont {Mehta}}, \
  and\ \bibinfo {author} {\bibfnamefont {C.~G.}\ \bibnamefont {Caro}},\
  }\bibfield  {title} {\enquote {\bibinfo {title} {Steady flow in porous,
  elastically deformable materials},}\ }\href {\doibase 10.1115/1.3173119}
  {\bibfield  {journal} {\bibinfo  {journal} {Journal of Applied Mechanics}\
  }\textbf {\bibinfo {volume} {54}},\ \bibinfo {pages} {794} (\bibinfo {year}
  {1987})}\BibitemShut {NoStop}%
\bibitem [{}]{see_supp}%
  \BibitemOpen
  \href@noop {} {}\bibinfo {note} {See Supplemental Material for a \texttt{MATLAB} implementation of the models, solutions, and results presented in \S\ref{s:rect}--\ref{s:flow_rect}.}\BibitemShut
  {Stop}%
\bibitem [{\citenamefont {Beavers}\ \emph {et~al.}(1975)\citenamefont
  {Beavers}, \citenamefont {Wilson},\ and\ \citenamefont
  {Masha}}]{beavers-japplmech-1975}%
  \BibitemOpen
  \bibfield  {author} {\bibinfo {author} {\bibfnamefont {G.~S.}\ \bibnamefont
  {Beavers}}, \bibinfo {author} {\bibfnamefont {T.~A.}\ \bibnamefont {Wilson}},
  \ and\ \bibinfo {author} {\bibfnamefont {B.~A.}\ \bibnamefont {Masha}},\
  }\bibfield  {title} {\enquote {\bibinfo {title} {Flow through a deformable
  porous material},}\ }\href {\doibase 10.1115/1.3423648} {\bibfield  {journal}
  {\bibinfo  {journal} {Journal of Applied Mechanics}\ }\textbf {\bibinfo
  {volume} {42}},\ \bibinfo {pages} {598--602} (\bibinfo {year}
  {1975})}\BibitemShut {NoStop}%
\bibitem [{\citenamefont {Oloyede}\ and\ \citenamefont
  {Broom}(1991)}]{oloyede-clinicalbiomech-1991}%
  \BibitemOpen
  \bibfield  {author} {\bibinfo {author} {\bibfnamefont {A.}~\bibnamefont
  {Oloyede}}\ and\ \bibinfo {author} {\bibfnamefont {N.~D.}\ \bibnamefont
  {Broom}},\ }\bibfield  {title} {\enquote {\bibinfo {title} {Is classical
  consolidation theory applicable to articular cartilage deformation?}}\ }\href
  {\doibase 10.1016/0268-0033(91)90048-U} {\bibfield  {journal} {\bibinfo
  {journal} {Clinical Biomechanics}\ }\textbf {\bibinfo {volume} {6}},\
  \bibinfo {pages} {206--212} (\bibinfo {year} {1991})}\BibitemShut {NoStop}%
\bibitem [{\citenamefont {Murad}\ \emph {et~al.}(1995)\citenamefont {Murad},
  \citenamefont {Bennethum},\ and\ \citenamefont {Cushman}}]{murad-tipm-1995}%
  \BibitemOpen
  \bibfield  {author} {\bibinfo {author} {\bibfnamefont {M.~A.}\ \bibnamefont
  {Murad}}, \bibinfo {author} {\bibfnamefont {L.~S.}\ \bibnamefont
  {Bennethum}}, \ and\ \bibinfo {author} {\bibfnamefont {J.~H.}\ \bibnamefont
  {Cushman}},\ }\bibfield  {title} {\enquote {\bibinfo {title} {A multi-scale
  theory of swelling porous media: {I}. {Application} to one-dimensional
  consolidation},}\ }\href {\doibase 10.1007/BF00626661} {\bibfield  {journal}
  {\bibinfo  {journal} {Transport in Porous Media}\ }\textbf {\bibinfo {volume}
  {19}},\ \bibinfo {pages} {93--122} (\bibinfo {year} {1995})}\BibitemShut
  {NoStop}%
\bibitem [{\citenamefont {Lancellotta}\ and\ \citenamefont
  {Preziosi}(1997)}]{lancellotta-intjengsci-1997}%
  \BibitemOpen
  \bibfield  {author} {\bibinfo {author} {\bibfnamefont {R.}~\bibnamefont
  {Lancellotta}}\ and\ \bibinfo {author} {\bibfnamefont {L.}~\bibnamefont
  {Preziosi}},\ }\bibfield  {title} {\enquote {\bibinfo {title} {A general
  nonlinear mathematical model for soil consolidation problems},}\ }\href
  {\doibase 10.1016/S0020-7225(97)00024-4} {\bibfield  {journal} {\bibinfo
  {journal} {International Journal of Engineering Science}\ }\textbf {\bibinfo
  {volume} {35}},\ \bibinfo {pages} {1045--1063} (\bibinfo {year}
  {1997})}\BibitemShut {NoStop}%
\bibitem [{\citenamefont {Lanir}\ \emph {et~al.}(1990)\citenamefont {Lanir},
  \citenamefont {Sauob},\ and\ \citenamefont
  {Maretsky}}]{lanir-japplmech-1990}%
  \BibitemOpen
  \bibfield  {author} {\bibinfo {author} {\bibfnamefont {Y.}~\bibnamefont
  {Lanir}}, \bibinfo {author} {\bibfnamefont {S.}~\bibnamefont {Sauob}}, \ and\
  \bibinfo {author} {\bibfnamefont {P.}~\bibnamefont {Maretsky}},\ }\bibfield
  {title} {\enquote {\bibinfo {title} {Nonlinear finite deformation response of
  open cell polyurethane sponge to fluid filtration},}\ }\href {\doibase
  10.1115/1.2892010} {\bibfield  {journal} {\bibinfo  {journal} {Journal of
  Applied Mechanics}\ }\textbf {\bibinfo {volume} {57}},\ \bibinfo {pages}
  {449} (\bibinfo {year} {1990})}\BibitemShut {NoStop}%
\bibitem [{\citenamefont {Sobac}\ \emph {et~al.}(2011)\citenamefont {Sobac},
  \citenamefont {Colombani},\ and\ \citenamefont
  {Forterre}}]{sobac-mecind-2011}%
  \BibitemOpen
  \bibfield  {author} {\bibinfo {author} {\bibfnamefont {B.}~\bibnamefont
  {Sobac}}, \bibinfo {author} {\bibfnamefont {M.}~\bibnamefont {Colombani}}, \
  and\ \bibinfo {author} {\bibfnamefont {Y.}~\bibnamefont {Forterre}},\
  }\bibfield  {title} {\enquote {\bibinfo {title} {On the dynamics of
  poroelastic foams (\emph{in French})},}\ }\href {\doibase
  10.1051/meca/2011115} {\bibfield  {journal} {\bibinfo  {journal}
  {M\'{e}canique \& Industries}\ }\textbf {\bibinfo {volume} {12}},\ \bibinfo
  {pages} {231--238} (\bibinfo {year} {2011})}\BibitemShut {NoStop}%
\bibitem [{\citenamefont {Hewitt}\ \emph {et~al.}(2016)\citenamefont {Hewitt},
  \citenamefont {Nijjer}, \citenamefont {Worster},\ and\ \citenamefont
  {Neufeld}}]{hewitt-pre-2016}%
  \BibitemOpen
  \bibfield  {author} {\bibinfo {author} {\bibfnamefont {D.~R.}\ \bibnamefont
  {Hewitt}}, \bibinfo {author} {\bibfnamefont {J.~S.}\ \bibnamefont {Nijjer}},
  \bibinfo {author} {\bibfnamefont {M.~G.}\ \bibnamefont {Worster}}, \ and\
  \bibinfo {author} {\bibfnamefont {J.~A.}\ \bibnamefont {Neufeld}},\
  }\bibfield  {title} {\enquote {\bibinfo {title} {Flow-induced compaction of a
  deformable porous medium},}\ }\href {\doibase 10.1103/PhysRevE.93.023116}
  {\bibfield  {journal} {\bibinfo  {journal} {Physical Review E}\ }\textbf
  {\bibinfo {volume} {93}},\ \bibinfo {pages} {023116} (\bibinfo {year}
  {2016})}\BibitemShut {NoStop}%
\bibitem [{\citenamefont {{LeVeque}}(2004)}]{leveque-cambridge-2004}%
  \BibitemOpen
  \bibfield  {author} {\bibinfo {author} {\bibfnamefont {R.~J.}\ \bibnamefont
  {{LeVeque}}},\ }\href@noop {} {\emph {\bibinfo {title} {Finite-volume methods
  for hyperbolic problems}}}\ (\bibinfo  {publisher} {Cambridge University
  Press},\ \bibinfo {year} {2004})\BibitemShut {NoStop}%
\bibitem [{\citenamefont {Barry}\ and\ \citenamefont
  {Aldis}(1993)}]{barry-jaustralmathsocb-1993}%
  \BibitemOpen
  \bibfield  {author} {\bibinfo {author} {\bibfnamefont {S.~I.}\ \bibnamefont
  {Barry}}\ and\ \bibinfo {author} {\bibfnamefont {G.~K.}\ \bibnamefont
  {Aldis}},\ }\bibfield  {title} {\enquote {\bibinfo {title} {Radial flow
  through deformable porous shells},}\ }\href {\doibase
  10.1017/S0334270000008936} {\bibfield  {journal} {\bibinfo  {journal}
  {Journal of the Australian Mathematical Society. Series B. Applied
  Mathematics}\ }\textbf {\bibinfo {volume} {34}},\ \bibinfo {pages} {333--354}
  (\bibinfo {year} {1993})}\BibitemShut {NoStop}%
\end{thebibliography}

%

\end{document}